\documentclass[12pt]{article}
\usepackage[T1]{fontenc}
\usepackage{textcomp}
\usepackage[utf8]{inputenc}
\usepackage{array}
\usepackage{verbatim}
\usepackage{float}
\usepackage{varwidth}
\usepackage{amsmath}
\usepackage{amsthm}
\usepackage{amssymb}
\usepackage{graphicx}
\usepackage{setspace}
\usepackage[authoryear]{natbib}
\makeatletter

\providecommand{\tabularnewline}{\\}
\newenvironment{cellvarwidth}[1][t]
    {\begin{varwidth}[#1]{\linewidth}}
    {\@finalstrut\@arstrutbox\end{varwidth}}
\floatstyle{ruled}
\newfloat{algorithm}{tbp}{loa}
\providecommand{\algorithmname}{Algorithm}
\floatname{algorithm}{\protect\algorithmname}

\theoremstyle{plain}
\newtheorem{thm}{\protect\theoremname}
\theoremstyle{plain}
\newtheorem{prop}[thm]{\protect\propositionname}

\usepackage{iftex}
\ifPDFTeX
  \usepackage[utf8]{inputenc}
  \usepackage[T1]{fontenc}
\fi
\usepackage[english]{babel}

\usepackage[
  top=0.95in,
  bottom=0.95in,
  left=0.95in,
  right=0.95in
]{geometry}
\usepackage{setspace}

\usepackage{amsmath,amsthm,amssymb}
\usepackage{mathrsfs}
\usepackage{bbm}

\usepackage{graphicx}
\usepackage{float}
\usepackage{placeins}
\usepackage{pdflscape}

\usepackage{array}
\usepackage{booktabs}
\usepackage{longtable}
\usepackage{multirow}
\usepackage{threeparttable}
\usepackage{tablefootnote}
\usepackage{adjustbox}

\newcolumntype{x}[1]{>{\centering\hspace{0pt}}p{#1}}

\usepackage[justification=centering]{caption}
\usepackage{subfig}
\usepackage{algorithm}
\usepackage{algorithmic}

\usepackage{natbib}
\bibpunct[, ]{(}{)}{;}{a}{}{,}%
\usepackage{titlesec}

\titleformat{\section}
  {\normalfont\fontsize{15}{15}\bfseries}
  {\thesection}{1em}{}

\titleformat{\subsection}
  {\normalfont\fontsize{14}{14}\bfseries}
  {\thesubsection}{1em}{}

\titleformat{\subsubsection}[runin]
  {\normalfont\fontsize{13}{13}\bfseries}
  {\thesubsubsection}{0.5em}{}[.]

\titleformat{\paragraph}[runin]
  {\normalfont\fontsize{12}{12}\bfseries}
  {\theparagraph}{0.5em}{}

\titlespacing*{\section}{0pt}{1.2ex}{0.6ex}
\titlespacing*{\subsection}{0pt}{1.0ex}{0.4ex}
\titlespacing*{\subsubsection}{0pt}{0.8ex}{0.8em}
\titlespacing*{\paragraph}{0pt}{0.6ex}{0.8em}

\usepackage{etoolbox}
\newcommand{\webappendixsetup}{%
  \titleformat{\section}
    {\normalfont\fontsize{15}{15}\bfseries}
    {Web Appendix~\thesection:}{1em}{}%
  \titleformat{\subsection}
    {\normalfont\fontsize{14}{14}\bfseries}
    {\thesubsection}{1em}{}%
  \titleformat{\subsubsection}[runin]
    {\normalfont\fontsize{13}{13}\bfseries}
    {\thesubsubsection}{0.5em}{}[.]%
}

\pretocmd{\appendix}{\webappendixsetup}{}{}

\usepackage{footmisc}

\usepackage[dvipsnames]{xcolor}
\definecolor{darkblue}{RGB}{0,0,120}

\usepackage{hyperref}
\hypersetup{
  colorlinks=true,
  linkcolor=darkblue,
  filecolor=darkblue,
  citecolor=darkblue,
  urlcolor=darkblue,
}

\usepackage{cleveref}

\ifdefined\showcaptionsetup
 \PassOptionsToPackage{caption=false}{subfig}
\fi
\usepackage{subfig}
\makeatother

\usepackage{babel}
\providecommand{\propositionname}{Proposition}
\providecommand{\theoremname}{Theorem}

\begin{document}
\title{Designing Ad Auctions with Targeting Information}
\author{Srinivas Tunuguntla\thanks{Assistant Professor of Marketing at the Fuqua School of Business,
Duke University (email: srinivas.tunuguntla@duke.edu)}\hspace{4em}Carl F. Mela\thanks{The T. Austin Finch Foundation Professor Emeritus of Business Administration
and Scholar in Residence at the Fuqua School of Business, Duke University
(email: mela@duke.edu, phone: 919-660-7767).}\hspace{4em}Jason Pratt\thanks{General Manager and Senior Research Officer, Koddi (email:jason.pratt@koddi.com)}}
\maketitle
\begin{abstract}
\begin{singlespace}
Digital advertising publishers sell ad inventory that conveys targeting
information, such as demographic, contextual, or behavioral audience
segments, to advertisers. While revealing this information improves
ad relevance, it can reduce competition and lower auction revenues.
To address this trade-off, we develop the Information-Bundling Position
Auction (IBPA), a general auction mechanism for search and display
advertising that leverages targeting information while preserving
competition. The mechanism treats the realized audience segment as
the publisher's private information. For a given advertiser, IBPA
implements information bundling through mixed-bundle pricing over
segments; across advertisers, it applies the marginal-revenue framework
to allocate impressions and determine payments.

We show that IBPA is Bayesian incentive compatible and individually
rational: truthful bidding forms a Bayes--Nash equilibrium and advertisers
receive non-negative expected utility from participation. Moreover,
IBPA achieves at least 63\% of the revenue of the optimal feasible
mechanism and weakly dominates all scalar-bid disclosure mechanisms,
including generalized second-price (GSP) auctions. The mechanism also
resolves the trade-off between targeting precision and market thickness:
publisher revenue is increasing in information granularity and decreasing
in disclosure granularity.

Using auction-level data from a large retail media platform, we estimate
advertiser valuation distributions and simulate counterfactual outcomes.
Relative to GSP, IBPA increases publisher revenue by 91\%, allocation
rate by 26pp, advertiser welfare by 62\%, and total welfare by 79\%.
\end{singlespace}

\medskip{}

\begin{singlespace}
\vspace{0mm}\noindent \textbf{Keywords}: Display Advertising, Pricing,
Auctions, Reserve Prices, Optimization, Real-Time Bidding (RTB), Programmatic
Buying, Mechanism Design
\end{singlespace}
\end{abstract}

\thispagestyle{empty}
\setcounter{page}{0} 

\newpage{}

\section{Introduction}

Digital advertising is increasingly transacted through real-time auctions,
with global spend expected to reach nearly $\$1$ trillion by 2032,
with a 17\% CAGR.\footnote{\texttt{https://www.themarketintelligence.com/market-reports/programmatic-advertising-market-4734}}
Digital advertising contexts include search advertising on retail
media and search engines---typically sold using a generalized second
price auction or GSP \citep{Varian_2007}---and display ads---typically
sold in open exchanges via a first price auction and walled gardens
via a second price auction \citep{Choi2020}. Across these environments,
impression opportunities differ in terms of the consumer viewing the
ad---such as demographics, prior purchases, and browsing histories---and
the context in which the ad is delivered, including the publisher’s
page (e.g., surrounding content) and the ad format (e.g., banner,
tower). Advertisers’ valuations depend on such targeting information,
as both audience and context directly affect the expected value of
an impression.

This paper considers how a publisher should sell impressions when
it has private access to such targeting information. We model impressions
as belonging to different segments defined by their audience and contextual
attributes. For each arriving impression opportunity, the publisher
privately observes the realized segment, while advertisers do not.
Our central contribution is to design an auction mechanism for such
settings, where the value of an impression is segment-dependent and
the impression’s segment is privately observed by the publisher.

A key challenge is that advertiser valuations are multidimensional,
with each advertiser assigning different values to different segments.
Designing optimal auctions in such environments is well known to be
difficult \citep{kolesnikov2022beckmann}. In practice, publishers
address this challenge by restricting advertisers to submit a scalar
bid, thereby reducing the auction to a single dimension. The most
common implementation of such a scalar-bid auction is to disclose
the realized segment and elicit a bid conditional on that information.
This “full disclosure” regime improves match quality by allowing advertisers
to avoid irrelevant impressions. Conventional wisdom holds that such
disclosure should improve revenues by raising advertisers’ effective
bids. However, a growing body of work finds that disclosing granular
targeting information can reduce publisher revenue by thinning competition
within each auction \citep{Ada2022,Levin2010,Lu2015,Yao2011}.\footnote{Most empirical studies find that richer targeting information increases
ad prices and publisher revenue. The non-monotonic effects emphasized
here arise in settings where finer targeting substantially reduces
competition by thinning the set of bidders.} Publishers therefore face a trade-off between targeting precision
and market thickness---more granular targeting information improves
relevance and advertiser willingness-to-pay, but more granular disclosure
fragments demand and reduces auction competitiveness.

\citet{Rafieian2021,bergemann2022optimal} address this trade-off
by optimizing the disclosure policy while holding the scalar-bid auction
fixed. In particular, the publisher partitions segments into groups
and reveals only group membership, so that bids reflect expected values
over segments within each group. In contrast, we take a mechanism-design
perspective. Rather than restricting attention to scalar-bid auctions
and optimizing disclosure within that class, we redesign the auction
itself to directly accommodate multidimensional valuations.

We introduce a new auction mechanism---which we refer to as the Information-Bundling
Position Auction (IBPA)---that integrates insights from mixed bundling
in pricing with the marginal-revenue approach to mechanism design.
Rather than disclosing the realized segment and soliciting a scalar-bid,
the publisher keeps this information private and solicits a multidimensional
bid---one bid for each possible segment. In the single-advertiser
case, the optimal mechanism reduces to the publisher offering the
advertisers a mixed bundling menu of segments and prices, a form of
second-degree price discrimination. We extend this logic to multiple
advertisers using the marginal-revenue framework \citep{bulow1989simple,Alaei2013}.
For each advertiser, we construct a revenue curve, map the advertiser's
multidimensional bid to a position on that curve, and compute the
corresponding marginal revenue. The impression is then allocated to
the advertiser with the highest marginal revenue, and payments follow
a critical-quantile logic analogous to second price auctions.

We establish four theoretical properties of IBPA. First, we show that
the mechanism is Bayesian incentive compatible and individually rational:
truthful bidding forms a Bayes--Nash equilibrium and advertisers
receive non-negative expected utility from participation. Second,
while exact revenue maximization is generally intractable in multidimensional
environments, we show that IBPA achieves at least a $(1-1/e)\approx63\%$
fraction of the revenue of the optimal feasible mechanism. Third,
IBPA weakly dominates the class of scalar-bid disclosure mechanisms,
including generalized second-price auctions, irrespective of the distribution
of advertiser valuations or whether advertisers are symmetric. Finally,
IBPA resolves the targeting-precision versus market-thickness trade-off
identified in prior work: publisher revenue is always increasing in
the granularity of the information available to the publisher.

We empirically evaluate our mechanism using auction-level bidding
data from a major online retailer and its ad-tech partner. Using the
estimated distribution of advertiser valuations, segments, and click-through
rate parameters, we simulate counterfactual outcomes under alternative
mechanisms. We find that IBPA increases publisher revenue by 91\%,
allocation rate by 26pp, advertiser welfare by 62\%, and total welfare
by 79\%, relative to GSP.

In the remainder of the paper, we first review how our research builds
on the existing literature. We then present a numerical example that
illustrates the intuition behind the proposed mechanism and its relationship
to standard scalar-bid auctions. Next, we describe the auction environment,
formalize the publisher's optimization problem, and introduce the
proposed IBPA mechanism. We then establish the theoretical properties
of the proposed mechanism. Finally, we present an empirical application
using auction-level data from a large retail media platform to estimate
advertiser valuation distributions and evaluate the counterfactual
performance of IBPA relative to existing auction mechanisms. We conclude
by summarizing the contributions of the paper and discussing directions
for future research.

\section{Background and Related Literature}

\subsection{Digital Advertising}

A large and growing literature examines real-time bidding in display
advertising. Much of this work focuses on advertiser-side objectives,
including the response or return to advertising \citep{Gordon2023,Johnson2023,tunuguntla2026},
the design of advertising content \citep{Lee2018}, and optimal ad
targeting and bidding strategies \citep{Rafieian2021,Sayedi2018,Tunuguntla2020}.

A complementary stream examines the publisher side, focusing on the
design of auction mechanisms used to allocate advertising inventory.
This includes comparisons of first- and second-price auctions \citep{Despotakis_et_al_2021},
the optimization of reserve prices \citep{Choi2023a}, and the allocation
of inventory across channels such as direct sales and exchanges \citep{Ballocco_et_al_2025}.

We contribute to the publisher-side literature by studying auction
design when impressions are heterogeneous, advertiser valuations are
multidimensional, and impression characteristics are privately observed
by the publisher.

\subsection{Information Asymmetries in Auctions}

\label{subsec:lit-review-information-asymmetries}

A large literature studies how sellers should design auctions when
they possess private information about the objects being sold. Much
of this literature studies vertical quality information---where uncertainty
concerns the overall quality of the item and bidders share a common
ranking of outcomes---focusing on optimal information disclosure
and information acquisition \citep{Milgrom_Weber_1982,Bergemann2007,Li_et_al_2009,Shi2012,Kim2022,Chen2023}.
In contrast, digital advertising markets are characterized by horizontal
(match) information asymmetries, where the value of an impression
depends on how well it matches the preferences of different advertisers.
Audience and contextual information determine the relevance of an
impression, and different advertisers may value the same impression
differently.

Within this setting, the literature largely focuses on information
disclosure while holding the auction format fixed, typically as a
generalized second-price auction. Advertisers submit scalar bids,
and heterogeneity in valuations is accommodated through the information
disclosed prior to bidding.

A central insight from this literature is that disclosure can create
a trade-off between targeting precision and market thickness \citep{Levin2010}.
More granular information improves match quality and increases advertisers’
willingness to pay. While most empirical studies find that greater
targeting granularity increases publisher revenue, more granular disclosure
can also reduce competition by narrowing the set of relevant bidders.
In settings where this reduction in market thickness is sufficiently
severe, publisher revenue may become non-monotonic in the level of
information disclosed, a pattern documented in both theoretical and
empirical work \citep{Yao2011,Ada2022,Lu2015,Sousa_2025,wu2025platform,Rafieian2021}.

This paper takes a different perspective. We show that this trade-off
is not inherent to the environment, but arises from the restriction
to scalar-bid auction formats. By allowing advertisers to submit multidimensional
bids and by keeping impression-level information private to the publisher,
the auction can simultaneously exploit fine-grained targeting information
and preserve competition across advertisers.

\subsection{Mixed-Bundle Pricing}

A key insight underlying our mechanism is that the multi-advertiser
auction can be decomposed into a collection of single-advertiser problems.
Each such single-advertiser problem parallels the classic multi-product
pricing problem, where a seller offers multiple goods to a buyer with
heterogeneous preferences.

The multi-product pricing literature shows that revenue-maximizing
mechanisms in such environments take the form of second-degree price
discrimination implemented through mixed-bundle pricing \citep{Adams_Yellen_1976,rochet1985taxation,McAfee_et_al_1989,Armstrong_2016}.
In our setting, the ``goods'' correspond to impression segments,
and mixed bundling becomes a menu of segment bundles and prices. Rather
than screening advertisers using a single scalar bid, the publisher
screens them based on their full valuation vector across segments.
Thus, mixed bundling generalizes the role of reserve prices in standard
auctions by replacing a single screening threshold with a multidimensional
pricing menu.

This perspective highlights a key limitation of standard advertising
auctions. Restricting advertisers to scalar bids limits the publisher
to coarse forms of price discrimination mediated through information
disclosure. By eliciting multidimensional bids instead, the publisher
can screen across valuation vectors, expanding the set of feasible
mechanisms beyond those achievable through disclosure alone.

\subsection{The Marginal Revenue Framework}

\label{subsec:lit-marginal-rev}

We extend the bundling-based pricing logic from the single-advertiser
setting to the multi-advertiser environment through the marginal-revenue
approach. In the classical single-item auction with single-dimensional
private values, \citet{myerson1981optimal} shows that the revenue-maximizing
mechanism allocates the item to the bidder with the highest virtual
value. \citet{bulow1989simple} provide an equivalent interpretation
in terms of marginal revenue maximization: each bidder is associated
with a revenue curve, and the optimal allocation assigns the item
to the bidder with the highest marginal revenue, defined as the derivative
of this curve evaluated at the point implied by the bidder’s bid.
This equivalence provides both an economic interpretation and a tractable
method for implementation. Building on this insight, \citet{Alaei2013}
show that marginal-revenue maximization is revenue-optimal for any
setting with single-dimensional agent types and quasi-linear preferences.

Beyond single-dimensional settings, \citet{Alaei2013} extend the
marginal-revenue framework to certain multidimensional environments
in which the seller allocates a single resource subject to an ex ante
allocation constraint. Their examples include settings such as selling
an insurance contract with multiple deductible levels, where buyers
have different valuations for different contract terms. Although only
one contract is allocated, bidder types are multidimensional because
valuations vary with the choice of deductible. In such environments,
marginal-revenue mechanisms are generally not revenue-optimal unless
the type distribution satisfies ``revenue linearity'' \citep{Alaei2013},
but the framework continues to provide a tractable approach to mechanism
design and delivers strong approximation guarantees even when exact
optimality fails.

Our setting differs in a key aspect. An impression belongs to a realized
segment drawn by nature (as opposed to chosen by the seller), and
that segment is privately observed by the publisher (seller). Whether
an advertiser receives the impression therefore depends on the realized
segment. Thus, although our setting also involves allocating a single
resource subject to an ex ante feasibility constraint, an advertiser’s
allocation cannot be summarized by a single unconditional allocation
variable. Instead, the allocation is inherently state-contingent,
varying with the realized segment observed only by the publisher.
The mechanism must therefore specify a segment-contingent allocation
rule that depends on the publisher’s private information. We thus
extend the marginal-revenue framework to a distinct class of multidimensional
environments in which allocation is contingent on an \emph{exogenous}
state (segment).

As in the multidimensional environments studied by \citet{Alaei2013},
marginal-revenue mechanisms in our setting are not generally revenue-optimal
absent additional structure such as revenue linearity. Nevertheless,
the approach continues to yield tractable mechanisms with strong approximation
guarantees. Moreover, the mechanism we propose dominates the class
of scalar-bid, disclosure-based mechanisms.

\section{Numerical Example and Intuition}

\label{sec:Example}

Before introducing the formal model and mechanism, we begin with a
numerical example that illustrates the key economic forces and the
intuition underlying our approach. Throughout this section, we assume
that bids equal valuations; Section \ref{sec:Problem-Setup} shows
that this assumption can be supported by an incentive-compatible mechanism.

We proceed in two steps. We begin with a single advertiser, a single
ad slot, and two segments to illustrate price bundling. We then extend
to two advertisers to introduce the marginal-revenue approach.

\subsection{Single Advertiser: Price Bundling}

\label{subsec:example-single-adv}

Consider a publisher selling a single impression to one advertiser.
The impression can belong to one of two segments, $A$ or $B$, each
occurring with probability $1/2$. The publisher observes the realized
segment, while the advertiser knows only this prior distribution.
The advertiser's valuation depends on the segment, with $(v_{A},v_{B})$
independently distributed uniformly on $[0,1]$. We focus on a single
ad slot and normalize the click-through rate to one, so that valuations
can be interpreted as per-click values.

\paragraph{Scalar-bid mechanisms.}

In practice, most ad auctions restrict advertisers to submit a scalar
bid. Within this class, the publisher can vary the information disclosed
prior to bidding.

Under full disclosure, the publisher reveals the segment before the
auction. Conditional on segment $A$, the problem reduces to a single-item
auction with value $v_{A}\sim\text{Unif}[0,1]$. The optimal mechanism
is a second-price auction with reserve price $r_{A}=0.5$ yielding
expected revenue $0.25$. The same logic applies for segment $B$,
so total expected revenue is $0.25$.

Under null disclosure, the publisher does not reveal the segment and
the advertiser bids based on the expected value $v_{AB}=\tfrac{1}{2}(v_{A}+v_{B})$.
The optimal mechanism is again a second-price auction, but with a
reserve price $r_{AB}\approx0.41$, yielding expected revenue $\approx0.27$.

When there are more than two segments, the publisher may also adopt
partial disclosure by grouping segments into coarser categories and
revealing only the group containing the realized segment.

\paragraph{A pricing interpretation.}

Because there is a single advertiser, these scalar-bid mechanisms
can be equivalently viewed as posted-price schemes. Under full disclosure,
the publisher posts segment-specific prices $(r_{A}=0.5,r_{B}=0.5)$.
Under null disclosure, the publisher instead posts a single price
$r_{AB}\approx0.41$ for the bundled impression. In each case, the
advertiser purchases if its expected utility is non-negative. Thus,
scalar-bid mechanisms restrict the publisher to either segment-specific
pricing or pricing a pooled bundle of segments.

\paragraph{Mixed bundling via multidimensional bids.}

This observation suggests a natural extension. In pricing theory,
both separate pricing and pure bundling are special cases of mixed
bundling, which allows the seller to offer both individual segments
and bundles simultaneously. Mixed bundling weakly dominates either
extreme.

Consider the following mixed-bundle menu of options: (i) Buy segment
$A$ only: receive the impression and pay $\rho_{A}=0.66$ if the
realized segment is $A$, and receive nothing and pay zero otherwise,
(ii) Buy segment $B$ only: receive the impression and pay $\rho_{B}=0.66$
if the realized segment is $B$, and receive nothing and pay zero
otherwise, and (iii) Buy the bundle: receive the impression regardless
of segment, pay $\rho_{AB}=0.43$. Given $(v_{A},v_{B})$, the advertiser
selects the option that maximizes expected utility. This menu yields
expected revenue of approximately $0.275$, exceeding both full disclosure
and null disclosure.

Two features are worth emphasizing. First, this mechanism cannot be
implemented with a scalar bid: computing allocations and payments
requires conditioning on all of the advertiser’s segment-level valuations,
either directly through multidimensional bids or indirectly through
menu selection. Second, the menu generalizes reserve prices by replacing
a single screening threshold with a multidimensional pricing menu.

\begin{figure}[tbph]
\centering{}\includegraphics[scale=0.45]{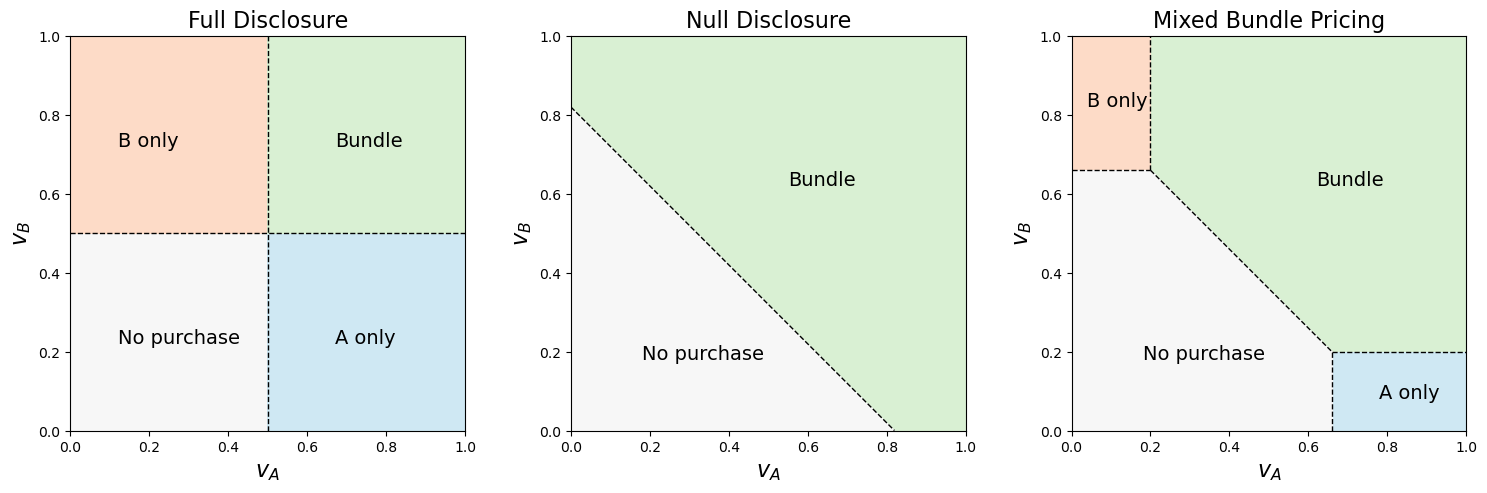}\caption{Allocation regions in advertiser valuation space under full disclosure,
null disclosure, and mixed bundling.}
\label{fig:single_adv_state_space}
\end{figure}

\paragraph{Key insight.}

Allowing multidimensional bids enables second-degree price discrimination
over valuation vectors rather than a single scalar statistic, expanding
the set of feasible mechanisms. More generally, with $T$ segments,
the publisher can price up to $2^{T}$ bundles.

Figure \ref{fig:single_adv_state_space} illustrates the partitions
of the valuation space $(v_{A},v_{B})$ under the three mechanisms.
Under full disclosure, allocation depends only on the realized segment,
producing vertical and horizontal reserve-price boundaries. Under
null disclosure, allocation depends on the advertiser's expected valuation
for the bundled impression, producing a diagonal cutoff. Mixed bundling
creates a richer partition based on the advertiser's full valuation
vector, allowing finer price discrimination than scalar-bid mechanisms.

\subsection{Single Advertiser Revenue Curve}

\label{subsec:example-rev-curve}

To extend the price-bundling logic to multiple advertisers, we summarize
each advertiser’s multidimensional valuation distribution by a revenue
curve, which maps the maximum revenue that can be extracted from an
advertiser as a function of the advertiser’s ex-ante probability of
receiving the impression. The revenue curve compresses the multidimensional
pricing problem into a one-dimensional object indexed only by allocation
probability, which allows us to compare advertisers using a scalar
marginal revenue.

We illustrate the revenue curve's construction using the two-segment
example from Section \ref{subsec:example-single-adv}. There, we identified
a mixed bundling menu with prices $\rho_{A}=0.66,\rho_{B}=0.66,\rho_{AB}=0.43$
that maximizes the publisher’s revenue. More generally, consider all
possible choices of $(\rho_{A},\rho_{B},\rho_{AB})$. Each such menu
induces (i) an ex-ante probability of allocation---the probability
that the advertiser receives the impression, averaged over the distribution
of advertiser valuations and segment realizations---and (ii) an ex-ante
expected revenue for the publisher. Plotting expected revenue against
allocation probability across all such menus, and taking the (weakly
increasing) concave hull, yields the advertiser’s revenue curve.

The revenue curve admits an equivalent and useful characterization.
For each $q\in[0,1]$, let $R(q)$ denote the maximum expected revenue
that the publisher can extract from the advertiser subject to the
constraint that the ex-ante probability of allocation is at most $q$.
The resulting function $R(q)$ coincides with the concave hull described
above. Because the revenue curve is concave, the marginal revenue
$R'(q)$ is decreasing in $q$.

\begin{figure}[tbph]
\centering{}\subfloat[Feasible revenue--allocation probability pairs and the induced revenue
curve under mixed-bundle pricing. Each point corresponds to a pricing
menu, and the red line traces the concave hull.]{\begin{centering}
\includegraphics[scale=0.5]{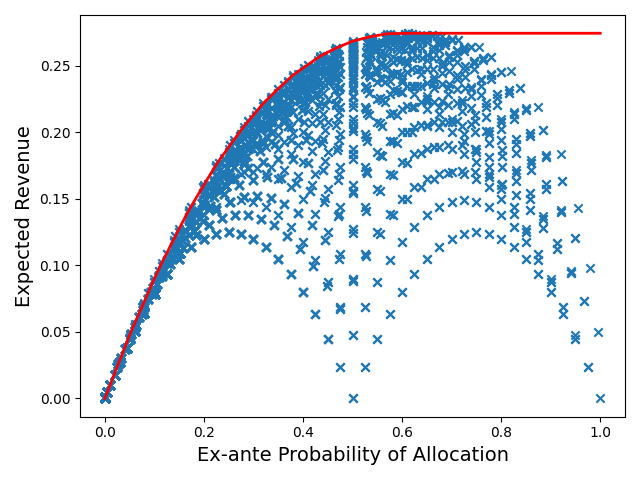}
\par\end{centering}
\label{fig:example-rev-curves-mixed-bundle}}\subfloat[Revenue curves across mechanism classes]{\begin{centering}
\includegraphics[scale=0.5]{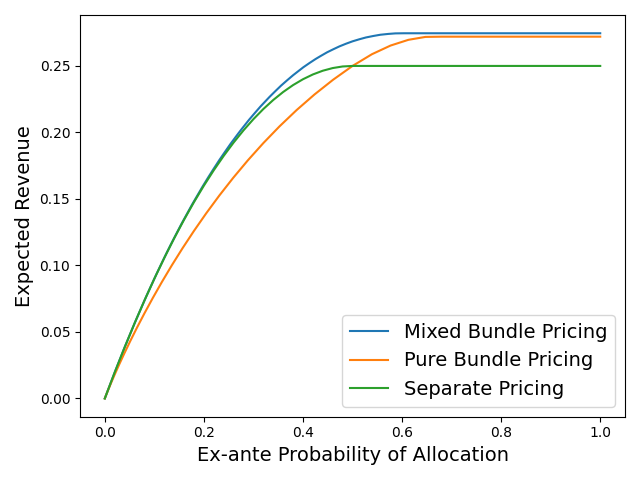}
\par\end{centering}
\label{fig:example-rev-curves-diff-pricing}}\caption{Revenue curves and their construction from single-advertiser mechanisms}
\label{fig:example-rev-curves}
\end{figure}

Figure \ref{fig:example-rev-curves-mixed-bundle} illustrates the
revenue curve for the two-segment example. Each point on the figure
corresponds to a different menu of mixed-bundle prices $(\rho_{A},\rho_{B},\rho_{AB})$.
Figure \ref{fig:example-rev-curves-diff-pricing} compares the mixed
bundling revenue curve with those obtained under separate pricing
(varying $\rho_{A},\rho_{B}$ with $\rho_{AB}=\infty$) and pure bundling
(varying $\rho_{AB}$ with $\rho_{A}=\rho_{B}=\infty$). Since both
are special cases of mixed bundling, their revenue curves lie weakly
below that of mixed bundling.

The revenue curve is advertiser-specific and provides a summary of
the advertiser’s multidimensional valuation distribution. Advertisers
with different valuation distributions therefore induce different
revenue curves. These curves are constructed ex ante from the advertisers'
priors and form the key inputs to the marginal-revenue approach described
next.

\subsection{Multiple Advertisers: Marginal Revenue Approach}

\label{subsec:example-marginal-rev}

We now turn to the multi-advertiser setting. Here, the publisher must
determine both how to allocate impressions across competing advertisers
and how to compute the corresponding payments for the winning advertiser.
The marginal revenue approach provides a tractable way to address
both problems.

When an auction occurs, advertisers submit their multidimensional
bids (equal to valuations), and the mechanism proceeds as follows.
First, for each advertiser, map its realized valuation vector to a
scalar position $q$---which we refer to as a quantile---on its
revenue curve and compute marginal revenue $R'(q)$. Second, allocate
the impression to the advertiser with the highest marginal revenue.
Third, compute the winning advertiser's payment. We now describe each
step in turn.

\paragraph{Mapping valuations to quantiles and computing marginal revenue.}

Fix an advertiser and its revenue curve. Each $q\in[0,1]$ on the
revenue curve corresponds to a single-advertiser mechanism $M(q)$,
which can be interpreted as a menu of mixed-bundle prices---e.g.,
$(\rho_{A},\rho_{B},\rho_{AB})$ in the two-segment case. By construction,
$M(q)$ induces an ex-ante allocation probability $q$ and expected
revenue $R(q)$. The collection of mechanisms $\{M(q)\}_{q\in[0,1]}$
that define the revenue curve provides the basis for mapping valuation
vectors to quantiles.

Consider our two-segment example with independent uniform valuations,
to illustrate this mapping from advertiser valuations to quantiles.
Fix a realized segment. For each $q$, the mechanism $M(q)$ determines
whether a given realized valuation vector would result in the impression
being allocated for that segment. We then map an advertiser’s realized
valuation vector to the smallest $q$ such that $M(q)$ would allocate
the impression conditional on the realized segment. This mapping is
segment-specific and is where the publisher’s private information
about the realized segment enters. The resulting $q$ determines the
advertiser’s marginal revenue $R'(q)$. Note that this quantile depends
only on the advertiser’s own valuation vector and revenue curve, and
is independent of other advertisers.

\begin{figure}[tbph]
\centering{}\subfloat[Segment A]{\begin{centering}
\includegraphics[scale=0.5]{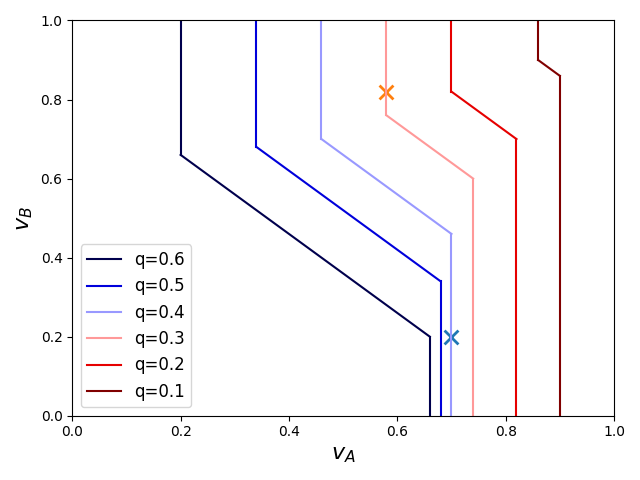}
\par\end{centering}
\label{fig:example-quantile-mapping-A}}\subfloat[Segment B]{\begin{centering}
\includegraphics[scale=0.5]{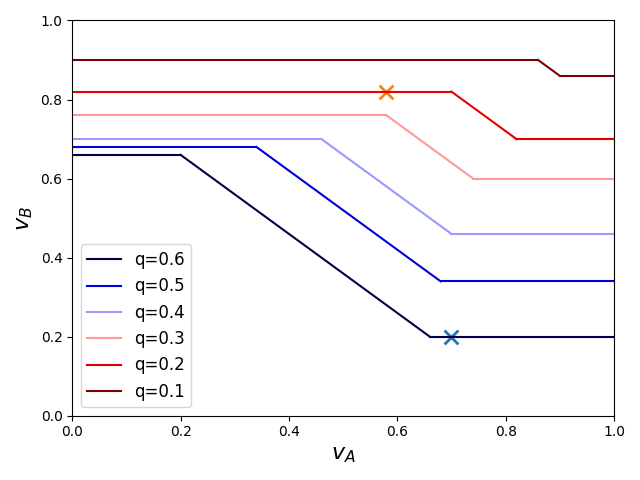}
\par\end{centering}
\label{fig:example-quantile-mapping-B}}\caption{Mapping valuation vectors to quantiles using single-advertiser mechanisms.
Advertiser 1 with realized valuation $(0.7,0.2)$ is shown as a blue
cross and advertiser 2 with realized valuation $(0.58,0.82)$ as an
orange cross; colored lines indicate allocation boundaries for different
values of $q$.}
\label{fig:example-quantile-mapping}
\end{figure}

Figure \ref{fig:example-quantile-mapping} illustrates the mapping
from realized valuation vectors to quantiles in the two-segment example.
The figure plots the allocation boundaries induced by the single-advertiser
mechanisms $M(q)$ for different values of $q$, for each realized
segment. We also plot two hypothetical advertisers: advertiser 1 with
realized valuations $(0.7,0.2)$ and advertiser 2 with realized valuations
$(0.58,0.82)$ for segments $A$ and $B$, respectively. The allocation
boundaries shown in the figure can also be interpreted as iso-quantile
curves, with each curve corresponding to valuation vectors that map
to the same quantile.

Consider segment $A$ (Figure \ref{fig:example-quantile-mapping-A}).
For each advertiser, we identify the smallest $q$ such that the corresponding
mechanism $M(q)$ would allocate the impression given the advertiser’s
realized valuation vector. For advertiser 1, this yields $q_{1}=0.4$,
while for advertiser 2, it yields $q_{2}=0.3$. The same procedure
applies for segment $B$ (Figure \ref{fig:example-quantile-mapping-B}),
yielding $q_{1}=0.6$ and $q_{2}=0.2$.

\paragraph{Allocation.}

Once each advertiser’s quantile is determined, its marginal revenue
$R'(q)$ is obtained from the revenue curve. The impression is then
allocated to the advertiser with the highest marginal revenue, with
ties broken arbitrarily.

In our example, both advertisers have the same valuation distribution
and therefore have identical revenue curves. Since marginal revenue
is decreasing in $q$, the advertiser with the lower quantile has
the higher marginal revenue and receives the impression. For segment
$A$, we have $q_{1}=0.4$ and $q_{2}=0.3$, so advertiser 2 wins.
The same conclusion holds for segment $B$. In particular, advertiser
2 wins segment $A$ even though $v_{2A}=0.58<v_{1A}=0.7$, because
the allocation depends on the advertisers’ full valuation vectors
through their induced marginal revenues, rather than only on valuations
for the realized segment.

\paragraph{Payment.}

For the winning advertiser, we compute the highest quantile $q^{\text{crit}}$
at which it would still win the impression given the other advertisers'
bids. The advertiser is then charged according to the mechanism $M(q^{\text{crit}})$
corresponding to this critical quantile.

This payment rule is analogous to a second-price auction: just as
the winner in a second-price auction pays the smallest bid at which
it would still win, here the winner pays according to the highest
quantile at which it would still win. In the case of identical revenue
curves, this corresponds to using the second-lowest quantile, directly
paralleling the second-price logic.

In our example, for segment $A$, advertiser 2 wins with quantile
$q_{2}=0.3$, and the critical quantile is $q^{\text{crit}}=q_{1}=0.4$,
the quantile of the competing advertiser. Thus, advertiser 2 is charged
according to the mechanism $M(0.4)$. For segment B, advertiser 2
again wins, with critical quantile $q^{\text{crit}}=q_{1}=0.6$, and
is therefore charged according to $M(0.6)$.

\paragraph{Summary.}

Together, these examples illustrate the main ingredients of IBPA:
price bundling, revenue curves, quantile mapping, marginal-revenue
allocation, and critical-quantile payments. We now formalize these
ideas in a general setting with arbitrary valuation distributions,
heterogeneous advertisers, and multiple ad slots.

\section{Problem Setup and Proposed Mechanism}

\label{sec:Problem-Setup}

\subsection{Model}

\label{subsec:Model}

Consider a publisher selling $S$ ad slots on a page, indexed by $s\in[S]$.\footnote{We use the notation $[K]=\{1,2,\dots,K\}$.}
Throughout the paper, we model both search and display advertising
as position auctions. Search advertising corresponds to the general
multi-slot case, whereas display advertising is a special case with
a single slot. Differences in pricing conventions can be accommodated
through the click-through-rate parameters and valuation definitions.
For example, CPM pricing is equivalent to setting click-through rates
equal to one and interpreting valuations as values per impression.

Each page belongs to a segment $t\in[T]$, defined by contextual page
attributes (e.g., content category) and/or consumer characteristics
(e.g., demographics, past purchases, and browsing history). Let 
\[
\mathbf{p}=(p_{1},p_{2},\ldots,p_{T}),\quad\text{with}\quad\sum_{t=1}^{T}p_{t}=1,
\]
denote the probability distribution over segments.

There are $A$ advertisers, indexed by $a\in[A]$. Advertiser $a$
has a per-click valuation $v_{at}\ge0$ for an impression of segment
$t$; impressions that do not receive a click generate no value. Differences
in $v_{at}$ across segments can be interpreted as arising from differences
in expected post-click outcomes, such as conversion probabilities
or purchase values, or other downstream advertiser objectives. Let
$\mathbf{v}_{a}=(v_{a1},\ldots,v_{aT})$ denote the advertiser's type
(valuation vector).

At the time of each impression, the publisher observes the realized
segment $t$, whereas the advertisers only know the distribution $\mathbf{p}$.
Each advertiser privately observes its own valuation vector $\mathbf{v}_{a}$,
while the publisher knows only a prior distribution $F_{a}$ over
$\mathbb{R}_{+}^{T}$ for each advertiser $a$. We allow for heterogeneous
priors $\{F_{a}\}_{a=1}^{A}$, and we assume that advertisers draw
their types---valuation vectors---independently of each other.

One interpretation of $F_{a}$ has the standard Bayesian interpretation
as uncertainty over advertiser valuations. In repeated-auction environments
such as digital advertising, it can be equivalently interpreted as
the distribution of advertiser valuations across impression opportunities.
Such variation may arise because advertisers possess information beyond
the targeting information observed by the publisher, including first-party
customer data, information derived from activity on the advertiser’s
website, or third-party data sources. Throughout the paper, we adopt
the standard Bayesian formulation while recognizing that repeated
digital advertising auctions provide a natural empirical interpretation
of the prior.

This formulation follows the standard Bayesian mechanism design framework
\citep{myerson1981optimal}. The timing is as follows: the publisher
first commits to a mechanism; then the publisher observes the realized
segment while each advertiser privately observes its valuation vector;
finally, the mechanism determines allocations of slots and payments.
Within this framework, ex-ante refers to the stage before the segment
and advertiser valuations are realized. Interim refers to the perspective
of a single player---advertiser or publisher---after observing its
private information but before knowing the realizations of others,
when expectations are taken over the types of the remaining players.
Ex-post refers to the stage after both the segment and all advertisers’
valuations are realized. We assume all agents are risk-neutral.

Without loss of generality, we focus on direct mechanisms \citep{myerson1981optimal},
in which each advertiser $a$ submits a bid vector $\mathbf{b}_{a}=(b_{a1},\ldots,b_{aT})$,
corresponding to bids for each segment. The mechanism then determines
outcomes as a function of all submitted bids, denoted by $\mathbf{B}=(\mathbf{b}_{1},\ldots,\mathbf{b}_{A})$.

A direct mechanism is characterized by allocation functions $x_{ast}(\mathbf{B})$
and an expected payment $m_{a}(\mathbf{B})$ for each advertiser.
$x_{ast}\in[0,1]$ denotes the probability that advertiser $a$ receives
slot $s$ for segment $t$, while $m_{a}$ denotes the expected payment
of advertiser $a$, where the expectation is taken over the mechanism's
randomization, segment, and click realizations.\footnote{Only expected payments are needed to compute the publisher's revenue.
The mechanism does not depend on how payments are disaggregated at
the slot or segment level. Thus, specific per-slot payments $m_{ast}$
can be implemented in any convenient form as long as they aggregate
to the expected payment $m_{a}$.} 

Feasibility requires that each advertiser is assigned at most one
slot for each segment, i.e., 
\[
\sum_{s}x_{ast}\leq1\quad\forall a,t
\]
and that each slot is assigned to at most one advertiser, i.e., 
\[
\sum_{a}x_{ast}\leq1\quad\forall s,t
\]

\subsection{Advertiser Incentive Constraints}

\label{subsec:Advertiser-Incentive-Constraints}

Given a mechanism---allocation and payment functions---advertisers
submit bids that maximize their expected utility. Because valuations
are specified per-click, advertiser utility depends on the click-through
rate (CTR). It is standard practice in the literature to model the
CTR of an advertiser $a$ in slot $s$ as a product of the advertiser's
quality $\gamma_{a}$ and the slot effect $\alpha_{s}$ \citep{athey2010structural}.
We further allow the CTR to vary with segment $t$ through segment
effects $\beta_{t}$. Thus, the CTR of advertiser $a$ in slot $s$
for segment $t$ is given by 
\[
\bar{c}_{ast}=\alpha_{s}\,\beta_{t}\,\gamma_{a}
\]
We assume that slots are numbered so that $\alpha_{1}\geq\alpha_{2}\geq\cdots\geq\alpha_{S}$.
Throughout the paper, we refer to slot $1$ as the highest slot and
slot $S$ as the lowest slot. Furthermore, to normalize the scale,
we set the effect of the top slot to $\alpha_{1}=1$ and the effect
of the first segment to $\beta_{1}=1$ without loss of generality.

The expected utility of advertiser $a$, when bidding $\mathbf{b}_{a}$
with true valuation vector $\mathbf{v}_{a}$, is then 
\[
u_{a}(\mathbf{b}_{a},\mathbf{v}_{a})=\mathbb{E}\!\left[\sum_{t,s}p_{t}\,\alpha_{s}\beta_{t}\gamma_{a}v_{at}\,x_{ast}(\mathbf{b}_{a},\mathbf{b}_{-a})\;-\;m_{a}(\mathbf{b}_{a},\mathbf{b}_{-a})\right],
\]
where the expectation is taken over the bids of all other advertisers
$\mathbf{b}_{-a}$. Each advertiser chooses bids to maximize this
expected utility.

By the revelation principle \citep{myerson1981optimal}, we can restrict
our attention to mechanisms that are incentive compatible without
loss of generality. A mechanism is Bayesian incentive compatible (BIC)
if truthful bidding constitutes a Bayes-Nash equilibrium, that is,
\[
u_{a}(\mathbf{v}_{a},\mathbf{v}_{a})\;\geq\;u_{a}(\mathbf{b}_{a},\mathbf{v}_{a})\quad\forall\,\mathbf{b}_{a},\;\forall\,\mathbf{v}_{a},\;\forall a.
\]
We impose this constraint in our publisher optimization problem to
ensure that bids equal true valuations, i.e., $\mathbf{V}=(\mathbf{v}_{1},\ldots,\mathbf{v}_{A})=\mathbf{B}=(\mathbf{b}_{1},\ldots,\mathbf{b}_{A})$.

In addition, mechanisms must satisfy a participation, or Bayesian
individual rationality (BIR), constraint: each advertiser's expected
utility---in equilibrium---must be non-negative, 
\[
u_{a}(\mathbf{v}_{a},\mathbf{v}_{a})\;\geq\;0\quad\forall\,\mathbf{v}_{a},\;\forall a.
\]

\subsection{Publisher Optimization Problem}

Given these incentive constraints, the publisher chooses a mechanism
to maximize expected revenue, defined as the sum of expected payments
across all advertisers\footnote{More generally, the publisher's objective could involve a combination
of revenue and advertiser utilities (e.g., to reflect competitive
pressures across publishers), but the structure of the analysis remains
unchanged.}. The publisher's problem can thus be summarized as choosing allocation
and payment rules to maximize expected revenue subject to feasibility,
incentive compatibility, and individual rationality constraints: 
\begin{equation}
\begin{aligned}\max_{\{x_{ast}(\cdot),\,m_{a}(\cdot)\}} & \quad\mathbb{E}_{\mathbf{V}}\!\left[\sum_{a}m_{a}(\mathbf{V})\right]\\[1em]
\text{s.t.} & \quad\sum_{s}x_{ast}(\mathbf{V})\;\leq\;1,\quad\forall a,t,\;\forall\mathbf{V}\quad\text{(Feasibility: one slot per advertiser)}\\[0.5em]
 & \quad\sum_{a}x_{ast}(\mathbf{V})\;\leq\;1,\quad\forall s,t,\;\forall\mathbf{V}\quad\text{(Feasibility: one advertiser per slot)}\\[0.5em]
 & \quad u_{a}(\mathbf{v}_{a},\mathbf{v}_{a})\;\geq\;u_{a}(\mathbf{b}_{a},\mathbf{v}_{a}),\quad\forall\mathbf{b}_{a},\;\forall\mathbf{v}_{a},\;\forall a\quad\text{(BIC)}\\[0.5em]
 & \quad u_{a}(\mathbf{v}_{a},\mathbf{v}_{a})\;\geq\;0,\quad\forall\mathbf{v}_{a},\;\forall a\quad\text{(BIR)}.
\end{aligned}
\label{eq:publisher_problem}
\end{equation}

\subsection{Information-Bundling Position Auction (IBPA)}

\label{subsec:Proposed-Mechanism}

To address the publisher’s problem, we develop the Information-Bundling
Position Auction (IBPA), which applies the marginal-revenue approach
to this environment with multidimensional advertiser valuations, multiple
slots, and publisher-private segment information. The mechanism reduces
the joint advertiser-slot-segment problem to a tractable collection
of single-advertiser, single-slot components.

Implementing the marginal-revenue approach in our setting raises two
challenges. First, unlike standard single-resource environments, the
publisher allocates multiple slots for each possible segment. At first
glance, this suggests that the single-advertiser problem is multidimensional
across both slots and segments, so that no one-dimensional revenue
curve exists. We show, however, that the single-advertiser problem
depends only on the highest available slot. Since revenue curves are
constructed ex ante, before the realization of segments or advertiser
valuations, we can integrate over segments and obtain a one-dimensional
revenue curve for each advertiser.

Second, the marginal-revenue approach requires mapping each advertiser’s
valuation vector to a scalar quantile on its revenue curve. In standard
single-dimensional settings, this mapping follows from the natural
ordering of scalar values. Here, advertiser types are multidimensional
valuation vectors and therefore have no natural ordering. We resolve
this by using the publisher’s private information: for any slot and
realized segment, we order advertiser types according to their probability
of receiving the slot under the prior distribution of valuations,
yielding a segment-specific quantile mapping.

Using these revenue curves and quantile mappings, IBPA computes advertisers’
marginal revenues, assigns slots in decreasing order of slot effect,
and determines payments consistent with incentive compatibility. The
remainder of this section formalizes the single-advertiser mechanisms,
constructs their revenue curves, and develops the resulting marginal-revenue
allocation and payment rules.

\subsubsection{Single Advertiser Mechanisms}

\label{subsec:Ex-ante-Constrained-Single}

We begin with the single-advertiser problem, which forms the building
block of IBPA. As illustrated in Section \ref{subsec:example-single-adv},
the mechanism reduces to a menu of (possibly randomized) bundles of
segments and associated prices. The marginal-revenue approach extends
this single-advertiser mechanism to multiple advertisers.

Fix advertiser $a_{0}$ with quality effect $\gamma_{a_{0}}$ on the
click-through rate (CTR). Let $\tilde{S}\subseteq[S]$ denote the
set of available slots, and let $s_{0}$ be the slot with the highest
slot effect $\alpha_{s_{0}}$.

With a single advertiser, the publisher’s problem in Equation \ref{eq:publisher_problem}
reduces to choosing allocation functions $x_{a_{0}st}(\mathbf{v}_{a_{0}})\;\forall s\in\tilde{S},\;\forall t\;$
and an expected payment function $m_{a_{0}}(\mathbf{v}_{a_{0}})$,
to maximize expected revenue, subject to incentive compatibility (IC),
individual rationality (IR), and the feasibility constraint that the
advertiser cannot be assigned more than one slot.

By the taxation principle \citep{rochet1985taxation}, any single-agent
IC and IR mechanism can be represented as a menu of options from which
the advertiser self-selects. Each menu option $k$ specifies probabilistic
allocations across slots and segments, $\chi_{st}^{(k)}\;\forall s\in\tilde{S},\forall t$,
together with a payment $\alpha_{s_{0}}\gamma_{a_{0}}\mu^{(k)}$.\footnote{We express the payment in this scaled form for convenience; multiplying
by $\alpha_{s_{0}}\gamma_{a_{0}}$ is without loss of generality.}

Given its realized valuation vector $\mathbf{v}_{a_{0}}$, the advertiser
chooses the menu option that maximizes its utility. If the advertiser
chooses option $k$, then $x_{a_{0}st}=\chi_{st}^{(k)}\;\forall s,t$,
and $m_{a_{0}}=\alpha_{s_{0}}\gamma_{a_{0}}\mu^{(k)}$. The resulting
utility is
\begin{equation}
u_{a_{0}}(\mathbf{v}_{a_{0}})=\sum_{t,s}p_{t}\,\alpha_{s}\beta_{t}\gamma_{a_{0}}v_{a_{0}t}\,\chi_{st}^{(k)}\;-\;\alpha_{s_{0}}\gamma_{a_{0}}\mu^{(k)}.\label{eq:single-adv-utility}
\end{equation}

This representation of the single-advertiser mechanism as a menu of
options allows for a further simplification: any allocation that assigns
positive probability to lower slots can be replaced by an equivalent
allocation that assigns probability only to the highest available
slot $s_{0}$, with probabilities appropriately rescaled. This transformation
preserves the advertiser’s expected utility for every valuation vector,
and therefore leaves publisher revenue unchanged. The following proposition
formalizes this result.
\begin{prop}
\label{prop:only-higher-slot}For any mechanism $M$ that includes
a menu option $k$ with $\chi_{\tilde{s}\tilde{t}}^{(k)}>0$ for some
$\tilde{s}\neq s_{0},\;\tilde{t}$, there exists another mechanism
$M'$ in which $\chi_{st}^{(k)}=0\;\forall s\neq s_{0}\;\forall t$
such that the publisher's expected revenue under $M'$ is equal to
that under $M$.
\end{prop}

\begin{proof}
See Web Appendix \ref{appendix:Omitted-Proofs-Model}.
\end{proof}
Hence, without loss of generality, we can restrict attention to menus
in which each option specifies allocations only for slot $s_{0}$
across segments $\chi_{t}\;\forall t$ and a corresponding payment
$\alpha_{s_{0}}\gamma_{a_{0}}\mu^{(k)}$. This representation reduces
the single-advertiser mechanism to a menu of (possibly randomized)
bundles of segments and associated prices, which generalizes the mixed-bundle
pricing in Section \ref{subsec:example-single-adv}. Thus, the single-advertiser
mechanism is precisely a multidimensional pricing problem over segments

\subsubsection{Single Advertiser Revenue Curve}

\label{subsec:Revenue-Curve}

We now formalize the revenue curve introduced in Section \ref{subsec:example-rev-curve}.
For a given slot, each single-advertiser mechanism---a menu of randomized
bundles of segments and associated prices---induces (i) an ex-ante
probability that the advertiser receives the slot, averaged over the
distribution of advertiser valuations and segment realizations, and
(ii) an ex-ante expected revenue for the publisher. The revenue curve
is obtained by solving for the single-advertiser mechanism that maximizes
expected revenue subject to the constraint that the ex-ante probability
of allocation is at most $q$, for each $q\in[0,1]$.

Fix advertiser $a$ and let $s$ denote the highest available slot.
Consider a menu $\{\chi^{(k)},\mu^{(k)}\}_{k\in[K]}$, where $\chi^{(k)}$
specifies a randomized bundle of segments and $\mu^{(k)}$ the associated
payment. Given such a menu, the advertiser draws a valuation vector
$\mathbf{v}_{a}\sim F_{a}$ and selects the option that maximizes
its utility as defined in Equation \ref{eq:single-adv-utility}. Let
$\xi_{k}$ denote the probability that the advertiser selects option
$k$.

Conditional on selecting option $k$, the advertiser receives slot
$s$ with probability $\sum_{t}p_{t}\chi_{t}^{(k)}$ and the publisher
receives revenue $\alpha_{s}\gamma_{a}\mu^{(k)}$. The ex-ante constrained
single-advertiser problem is therefore
\begin{equation}
\max_{K,\{\chi^{(k)},\mu^{(k)}\}}\;\alpha_{s}\gamma_{a}\sum_{k=1}^{K}\xi_{k}\mu^{(k)}\quad\text{subject to}\quad\sum_{k=1}^{K}\xi_{k}\Bigg(\sum_{t}p_{t}\chi_{t}^{(k)}\Bigg)\le q.\label{eq:single-adv-problem}
\end{equation}

We denote the optimal value of this problem as a function of $q$
by $R_{as}(q)$, which defines the advertiser’s revenue curve for
slot $s$.\footnote{\label{fn:finite-menu} For expositional convenience, we represent
the mechanism as a menu of length $K$. More generally, the taxation
principle permits menu representations with infinitely many or even
a continuum of menu options. Our use of finite menus should therefore
be interpreted as a notational simplification rather than a restriction
on the underlying class of incentive-compatible and individually rational
mechanisms.} Intuitively, the revenue curve describes how the advertiser’s contribution
to publisher revenue changes as the advertiser’s allocation probability
for the slot increases. When constructed for all advertisers, these
curves capture the trade-off inherent in increasing one advertiser’s
allocation probability relative to others.\footnote{Because revenue curves are constructed ex-ante, the allocation constraint
averages not only over the advertiser’s valuation distribution but
also over segment realizations.}

The following proposition establishes that the revenue curve is non-decreasing
and concave. This result is consistent with standard economic intuition:
as the allocation probability increases, the mechanism exhibits diminishing
returns, so marginal revenues decline with $q$. To establish the
result, let $\mathcal{M}$ denote the class of mechanisms over which
the constrained single-advertiser problem is optimized, and assume
that $\mathcal{M}$ is closed under convex combinations.\footnote{A convex combination of mechanisms $M_{1},\ldots,M_{n}$ is defined
as selecting mechanism $M_{i}$ with probability $\lambda_{i}$, where
$\lambda_{i}\geq0$ for all $i$ and $\sum_{i=1}^{n}\lambda_{i}=1$.}
\begin{prop}
\label{prop:rev-curve-concave}When the class of mechanisms $\mathcal{M}$
is closed under convex combinations, the revenue curve $R_{as}(q)$
is non-decreasing and concave in $q$. Consequently, the marginal
revenue $R_{as}'(q)$ is non-negative and weakly decreasing in $q$.
\end{prop}

\begin{proof}
See Web Appendix \ref{appendix:Omitted-Proofs-Model}.
\end{proof}
The revenue curve also exhibits additional structure that further
simplifies the marginal-revenue approach.
\begin{prop}
\label{prop:rev-curve-factors}For any mechanism class $\mathcal{M}$,
the revenue curve factorizes as 
\[
R_{as}(q)=\alpha_{s}\gamma_{a}\,\Phi_{a}(q),
\]
where $\Phi_{a}(q)$ depends only on advertiser $a$'s valuation distribution
and not on slot $s$.
\end{prop}

\begin{proof}
See Web Appendix \ref{appendix:Omitted-Proofs-Model}.
\end{proof}
The function $\Phi_{a}(q$) is obtained by solving the slot-normalized
version of the single advertiser problem in Equation \ref{eq:single-adv-problem},
given by

\begin{equation}
\Phi_{a}(q)\;=\;\max_{K,\{\chi^{(k)},\mu^{(k)}\}}\;\sum_{k=1}^{K}\xi_{k}\,\mu^{(k)}\quad\text{s.t.}\quad\sum_{k=1}^{K}\xi_{k}\!\left(\sum_{t}p_{t}\,\chi_{t}^{(k)}\right)\le q.\label{eq:single-adv-slot-normalized}
\end{equation}
We therefore refer to $\Phi_{a}(q)$ as the advertiser's normalized
revenue curve henceforth.

The factorization of the revenue curve into slot and advertiser components
has two important implications. First, revenue curves for a given
advertiser across slots are simply scaled versions of one another,
with slot effects entering only multiplicatively. As a consequence,
when computing marginal revenues from advertiser valuations, the mapping
to a quantile $q$ on the revenue curve depends only on $\Phi_{a}(q)$,
which is independent of the slot. Hence the quantile mapping is itself
slot-independent.

Second, marginal revenues take the form $R'_{as}(q)=\alpha_{s}\gamma_{a}\Phi'_{a}(q)$.
For a given advertiser, this expression implies that marginal revenue
is higher for higher slots. Thus, to maximize marginal revenue, it
is natural to allocate in order of slots from highest to lowest. At
the same time, for a fixed slot, advertisers' marginal revenues are
ranked by $\gamma_{a}\Phi'_{a}(q)$. This parallels the generalized
second-price auction, where advertisers are ordered by the product
of a quality score and a bid. In our formulation, $\gamma_{a}$ plays
the role of the quality score, while $\Phi'_{a}(q)$ can be interpreted
as a virtual value---the derivative of the scaled revenue curve---similar
to Myerson's optimal auction for a single item \citep{myerson1981optimal,bulow1989simple}.
We refer to $\Phi_{a}'(q)$ as the advertiser's normalized marginal
revenue or virtual value henceforth.

\subsubsection{Marginal Revenue Approach}

\label{subsec:Marginal-Revenue-Approach}

The slot-normalized revenue curves are computed offline for each advertiser,
prior to any auction. During an auction, advertisers draw valuation
vectors from their respective priors and submit corresponding bid
vectors. For now, we assume truthful bidding, so that bids equal valuations;
later, we show that the resulting mechanism is Bayesian incentive
compatible and individually rational (see Section \ref{subsec:Incentive-Compatibility-proof}),
thereby justifying this assumption.

Given the submitted valuation vectors and the realized segment observed
by the publisher, the mechanism proceeds in three steps, as illustrated
in \ref{subsec:example-marginal-rev}. First, for each advertiser,
the publisher maps the valuation vector to a position on the advertiser’s
revenue curve, referred to as a quantile. Second, the publisher computes
each advertiser’s marginal revenue as the derivative of the revenue
curve at the corresponding quantile and allocates slots using the
resulting marginal revenues. Third, the publisher computes payments
for the winning advertisers.

To build intuition, we begin with a special case in which the collection
of mechanisms $\{M_{a}(q)\}_{q\in[0,1]}$---where $M_{a}(q)$ denotes
the mechanism for advertiser $a$ that solves the slot-normalized
single-advertiser problem in Equation \ref{eq:single-adv-slot-normalized}
under allocation constraint $q$, achieving the normalized revenue
$\Phi_{a}(q)$---exhibits three additional structural properties.
First, allocations are deterministic: for every menu option $k$,
the allocation vector $\chi^{(k)}\in\{0,1\}^{T}$ is binary across
segments. Second, for each segment $t$, let $\Omega_{at}(q)$ denote
the set of valuation vectors allocated under $M_{a}(q)$ when the
realized segment is $t$. These allocation regions satisfy nestedness:
if $q_{1}\le q_{2}$, then $\Omega_{at}(q_{1})\subseteq\Omega_{at}(q_{2})$
for every segment $t$. Intuitively, as the allocation constraint
is relaxed, the allocation region expands to include lower valuations
while continuing to allocate higher valuations. Third, the allocation
regions are probability-matched under the prior, so that $F_{a}(\Omega_{at}(q))=q$
for every segment $t$ and every $q\in[0,1]$.

These properties are restrictive: in general, the mechanisms $\{M_{a}(q)\}_{q\in[0,1]}$
may involve randomization and need not generate nested allocation
regions. Nevertheless, this special case captures the core mechanics
of the marginal-revenue approach, allows the quantile mapping to be
described geometrically, and mirrors the numerical example in Section
\ref{sec:Example}. The more general construction is presented in
Web Appendix \ref{appendix:Quantile-Mapping}. We now describe the
three steps of the marginal-revenue approach in turn.

\paragraph{Mapping valuations to quantiles.}

We begin by constructing a mapping from an advertiser’s valuation
vector to a position on its revenue curve. When the mechanisms $\{M_{a}(q)\}_{q\in[0,1]}$
satisfy the nestedness property above, the allocation regions $\Omega_{at}(q)$
expand monotonically as $q$ increases. Moreover, because these regions
satisfy $F_{a}(\Omega_{at}(q))=q$, each value of $q$ indexes the
probability mass of the corresponding allocation region. Relaxing
the allocation constraint from $q$ to $q+dq$ therefore adds the
valuation vectors $\Omega_{at}(q+dq)\backslash\Omega_{at}(q)$, whose
contribution to publisher revenue is the marginal revenue $\Phi_{a}(q+dq)-\Phi_{a}(q)=\Phi_{a}'(q)dq$.
These newly admitted valuation vectors can thus be interpreted as
the marginal types at quantile $q$.

This yields a natural quantile mapping. For realized segment $t$,
define the advertiser's quantile as the smallest allocation constraint
under which its valuation vector is allocated:
\[
q_{at}=Q_{at}(\mathbf{v}_{a})=\inf\{q\in[0,1]:\mathbf{v}_{a}\in\Omega_{at}(q)\}.
\]
with $q_{at}=1$ if the advertiser is never allocated for any $q<1$.
Lower quantiles correspond to valuation vectors admitted under tighter
allocation constraints and therefore have higher marginal revenues,
exactly as illustrated in Figure \ref{fig:example-quantile-mapping}.

Finally, Proposition \ref{prop:rev-curve-factors} showed that $R_{as}(q)=\alpha_{s}\gamma_{a}\Phi_{a}(q)$.
Since slot effects enter only multiplicatively, the underlying mechanisms
$M_{a}(q)$ are identical across slots, implying that the quantile
mapping is slot-independent.

\paragraph{Slot Allocations.}

With the valuation-to-quantile mappings in place, the publisher computes
each advertiser’s marginal revenue for the realized segment. For advertiser
$a$, let $q_{at}=Q_{at}(\mathbf{v}_{a})$ denote the advertiser’s
quantile for segment $t$. The marginal revenue of advertiser $a$
for slot $s$ is then $R_{as}'(q_{at})=\alpha_{s}\gamma_{a}\Phi_{a}'(q_{at})$.

Proposition \ref{prop:rev-curve-factors} established that marginal
revenues across slots differ only through the slot multiplier $\alpha_{s}$.
Consequently, maximizing total marginal revenue reduces to processing
slots in decreasing order of $\alpha_{s}$ and assigning each slot
to the advertiser with the highest marginal revenue.

Within each slot $s$, advertisers are ranked by $\gamma_{a}\Phi_{a}'(q_{at})$.
Slots are assigned in this order until either all slots or all advertisers
are exhausted, excluding advertisers with non-positive marginal revenues.

This allocation rule parallels standard scalar-bid position auctions,
in which advertisers are ranked by the product of a quality score
and a scalar bid. In our mechanism, however, the scalar bid is replaced
by the advertiser’s normalized marginal revenue $\Phi_{a}'(q_{at})$,
which depends on the advertiser’s full valuation vector and the realized
segment. Consequently, allocations depend on richer information about
advertiser preferences across segments, rather than on a single scalar
statistic.

\paragraph{Payments.}

We now generalize the critical-quantile payment logic introduced in
Section \ref{subsec:example-marginal-rev}. In the single-slot setting
considered there, an advertiser faces a single critical quantile below
which it receives the slot. The single-advertiser mechanism associated
with this threshold determines its payment. With multiple slots, each
slot has a corresponding critical quantile, and an advertiser's payment
depends on the critical quantiles for its assigned slot and all lower
slots.

To formalize this idea, set $\alpha_{S+1}=0$ and decompose each slot
effect into incremental click-through-rate layers $\Delta_{j}=\alpha_{j}-\alpha_{j+1},\;j\in[S]$.
Receiving slot $s$ is equivalent to receiving layers $j=s,\ldots,S$,
since $\alpha_{s}=\sum_{j=s}^{S}\Delta_{j}$.

Consider advertiser $a$, with quantile $q_{at}$, assigned slot $s$.
For each layer $j\geq s$, holding the other advertisers' quantiles
fixed, define the layer-$j$ critical quantile as
\[
\widehat{q}_{at}^{\,j}=\sup\left\{ q\in[0,1]:a\text{ would continue to receive layer }j\text{ if its quantile were }q\right\} .
\]
Because $\gamma_{a}\Phi_{a}'(q)$ is weakly decreasing in $q$, advertiser
$a$ receives layer $j$ for all quantiles below this cutoff. Moreover,
the cutoff depends only on the other advertisers' quantiles.

Let $\widetilde{a}_{j}$ denote the advertiser ranked $j+1$, when
one exists. Then $\widehat{q}_{at}^{\,j}$ is the largest quantile
satisfying $\gamma_{a}\Phi_{a}'\!\left(\widehat{q}_{at}^{\,j}\right)\geq\gamma_{\widetilde{a}_{j}}\Phi_{\widetilde{a}_{j}}'\!\left(q_{\widetilde{a}_{j}t}\right)$.
If no advertiser below rank $j$ has a positive score, the critical
quantile is instead determined by $\gamma_{a}\Phi_{a}'\!\left(\widehat{q}_{at}^{\,j}\right)>0$.
Thus, $\widehat{q}_{at}^{\,j}$ identifies the boundary at which advertiser
$a$ loses layer $j$ and falls below slot $j$.

The single-advertiser mechanism $M_{a}(\widehat{q}_{at}^{\,j})$ determines
the payment for layer $j$. Given advertiser $a$'s valuation vector,
let $\chi_{a\tau}^{\,j}$ and $\mu_{a}^{\,j}$ denote the allocation
vector and normalized expected payment of the selected menu option
under $M_{a}(\widehat{q}_{at}^{\,j})$, and define $\lambda_{a}^{\,j}=\sum_{\tau=1}^{T}p_{\tau}\chi_{a\tau}^{\,j}.$

By Proposition \ref{prop:rev-curve-factors}, layer $j$ scales the
normalized single-advertiser mechanism by $\Delta_{j}\gamma_{a}$.
Its expected incremental payment is therefore $\Delta_{j}\gamma_{a}\mu_{a}^{\,j}$.
To collect this payment only when the advertiser receives the layer,
one convenient implementation charges $\Delta_{j}\gamma_{a}\frac{\mu_{a}^{\,j}}{\lambda_{a}^{\,j}}$
conditional on allocation. Averaging over segment realizations and
the mechanism's randomization then yields the required expected incremental
payment $\Delta_{j}\gamma_{a}\mu_{a}^{\,j}$.

Summing across the layers, the advertiser's per-impression payment
is 
\[
m_{ast}^{\mathrm{imp}}=\gamma_{a}\sum_{j=s}^{S}(\alpha_{j}-\alpha_{j+1})\frac{\mu_{a}^{\,j}}{\lambda_{a}^{\,j}}.
\]
If payments are collected per click, as is common in practice, the
per-click payment is obtained by dividing by the realized click-through
rate: 
\[
m_{ast}^{\mathrm{click}}=\frac{m_{ast}^{\mathrm{imp}}}{\alpha_{s}\beta_{t}\gamma_{a}}=\frac{1}{\alpha_{s}\beta_{t}}\sum_{j=s}^{S}(\alpha_{j}-\alpha_{j+1})\frac{\mu_{a}^{\,j}}{\lambda_{a}^{\,j}}.
\]

This construction is the position-auction analogue of the Myerson
payment identity \citep{myerson1981optimal}. Allocation of each incremental
click-through-rate layer is monotone in the advertiser's quantile,
and the corresponding critical-quantile payment implements the incentive-compatible
payment associated with that allocation rule. Equivalently, just as
a winner in a second-price auction pays according to the weakest bid
at which it would continue to win, an advertiser here pays for each
layer according to the weakest quantile at which it would continue
to receive that layer. With a single slot, there is only one layer,
and the formula reduces to the critical-quantile payment rule in Section
\ref{subsec:example-marginal-rev}.

\subsubsection{IBPA Summary}

\paragraph*{Economic Logic of Pricing and Allocation.}

In sum, IBPA combines several linked components. First, it converts
each advertiser’s multidimensional valuation problem into a one-dimensional
revenue-ranking problem, making the publisher’s allocation problem
tractable. For advertiser $a$, the normalized revenue curve $\Phi_{a}(q)$
gives the maximum expected revenue that the publisher can obtain when
the advertiser’s ex-ante probability of receiving an impression is
at most $q$. Each point $q$ on this curve corresponds to a single-advertiser
pricing menu that implements the associated allocation probability.

Second, the revenue curve summarizes the trade-off underlying the
choice of pricing menu. Under a tight allocation constraint, the mechanism
allocates to only valuation vectors that support relatively selective
pricing menus and generate high revenue as a result. Relaxing the
allocation constraint increases the probability that the impression
is allocated, but does so by extending allocation toward weaker valuation
vectors with lower incremental revenue. Thus, $\Phi_{a}(q)$ captures
the trade-off between allocating more frequently and extracting revenue
from the marginal allocation.

Third, at auction time, the publisher uses the realized impression
segment to map each advertiser’s valuation vector to a quantile $q_{at}$
on the advertiser's revenue curve. In the nested deterministic case,
$q_{at}$ is the smallest allocation probability at which the corresponding
single-advertiser mechanism would allocate the impression to that
valuation vector. A lower quantile therefore identifies a valuation
vector that is profitable to serve even under a tighter allocation
constraint and consequently carries a higher marginal revenue. Importantly,
this is a revenue-based index of the advertiser’s full valuation vector,
rather than simply its valuation for the realized segment.

Fourth, IBPA allocates slots according to the resulting marginal-revenue
scores, $\gamma_{a}\Phi'_{a}(q)$, assigning higher slots to advertisers
with larger scores. In this way, the mechanism balances the increased
likelihood of allocating the impression against the declining revenue
contribution of expanding allocation. Payments are determined by the
critical-quantile rule in the nested deterministic case and by the
corresponding interim payment rule in the general construction.

\paragraph{Offline and Online Implementation.}

IBPA naturally separates into an offline stage and an online auction
stage. In the offline stage, the publisher solves the constrained
single-advertiser problems to construct the advertiser-specific normalized
revenue curves and segment-specific valuation-to-quantile mappings.
These objects depend only on advertisers’ valuation distributions
and can therefore be computed before auctions occur.

In the online stage, after observing the realized segment and receiving
advertisers’ valuation vectors, the publisher maps each valuation
vector to a quantile, computes its marginal-revenue score, allocates
slots greedily in decreasing order of these scores, and computes the
associated payments. Algorithm \ref{alg:ibpa} in Web Appendix \ref{appendix:algorithm}
summarizes these steps, and Table \ref{tab:notation} provides a summary
of the notation.

This separation is important for scalability. The computationally
intensive construction of revenue curves and quantile mappings is
performed offline, while the online auction requires only quantile
evaluation, marginal-revenue computation, allocation, and payment
calculation.

\section{Theoretical Analysis}

\label{sec:Theoretical-Results}

This section establishes four theoretical properties of IBPA. First,
the mechanism is a feasible Bayesian incentive compatible and individually
rational mechanism. Second, although multidimensional mechanism design
environments generally do not admit tractable exact solutions, IBPA
inherits strong approximation guarantees from the marginal-revenue
framework. Third, IBPA weakly dominates the class of scalar-bid disclosure
mechanisms, including generalized second-price auctions. Finally,
IBPA resolves the targeting-precision versus market-thickness trade-off
identified in prior work.

\subsection{Incentive Compatibility and Feasibility of IBPA}

\label{subsec:Incentive-Compatibility-proof}

We first establish that IBPA satisfies the feasibility, Bayesian incentive
compatibility (BIC), and Bayesian individual rationality (BIR) constraints
in Equation \ref{eq:publisher_problem}. The result follows from the
fact that the interim mechanism induced by the marginal-revenue allocation
rule can be represented as a convex combination of single-advertiser
mechanisms that are themselves incentive compatible and individually
rational. Here, the interim mechanism refers to the mechanism faced
by a given advertiser after integrating over the valuations of the
other advertisers and the mechanism’s randomization. Consequently,
truthful reporting constitutes a Bayes--Nash equilibrium, allowing
the subsequent analysis to treat submitted bid vectors as valuation
vectors. Feasibility follows directly from the slot allocation rule,
which assigns each slot to at most one advertiser and each advertiser
to at most one slot.
\begin{prop}
\label{prop:IC-IR-Feasibility} The IBPA mechanism satisfies feasibility,
BIC, and BIR.
\end{prop}

\begin{proof}
See Web Appendix \ref{appendix:Omitted-Proofs-Results}.
\end{proof}

\subsection{Revenue Approximation}

Having established that IBPA is a feasible BIC and BIR mechanism,
we next compare its revenue to that of the optimal feasible BIC and
BIR mechanism. Because our setting is multidimensional---advertisers
possess valuation vectors over segments, and the publisher’s allocation
rule is segment-contingent---exact characterization of the revenue-maximizing
BIC and BIR mechanism is intractable absent additional structural
assumptions on the valuation distributions \citep{kolesnikov2022beckmann,Alaei2013}.
Nevertheless, the marginal-revenue structure of IBPA yields a strong
approximation guarantee relative to the optimal mechanism.

Let $\mathrm{Rev}^{\mathrm{OPT}}$ denote the maximum expected revenue
attainable by any feasible BIC and BIR mechanism for the publisher's
problem in Equation \ref{eq:publisher_problem}, and let $\mathrm{Rev}^{\mathrm{IBPA}}$
denote the expected revenue generated by IBPA.
\begin{thm}
\label{thm:ibpa-approximation} Suppose advertisers' valuation vectors
are drawn independently across advertisers, while allowing arbitrary
dependence across segments within an advertiser. Then
\[
\mathrm{Rev}^{\mathrm{IBPA}}\ge\left(1-\frac{1}{e}\right)\mathrm{Rev}^{\mathrm{OPT}}
\]
\end{thm}

\begin{proof}
See Web Appendix \ref{appendix:Omitted-Proofs-Results}.
\end{proof}
The proof proceeds in two steps. First, we construct an ex-ante relaxation
of the publisher's problem in Equation \ref{eq:publisher_problem}.
Rather than requiring the feasibility constraints on allocation to
hold for every realization of advertiser valuations and segments,
we require them to hold only ex-ante, i.e., in expectation. This relaxation
enlarges the feasible set of mechanisms, since every ex-post feasible
BIC and BIR mechanism is also feasible for the ex-ante relaxed problem.
The optimal value of the relaxed problem therefore provides an upper
bound on the revenue attainable by any feasible mechanism. Second,
we compare IBPA to this ex-ante benchmark. Results from the marginal-revenue
literature imply that the ex-post feasible marginal-revenue mechanism
achieves at least a $(1-1/e)$ fraction of the corresponding ex-ante
relaxation. Applying this result to IBPA yields the approximation
guarantee.

Theorem \ref{thm:ibpa-approximation} implies that IBPA obtains at
least $(1-1/e)\approx0.632$ fraction of the revenue of the optimal
feasible BIC and BIR mechanism, regardless of the valuation distributions
or the complexity of the environment. Thus, although exact multidimensional
revenue maximization is generally intractable, IBPA provides a tractable
mechanism with a strong worst-case performance guarantee.

The next subsection provides a complementary benchmark by comparing
IBPA with the class of scalar-bid disclosure mechanisms, including
generalized second-price auctions.

\subsection{Dominance over Scalar-Bid Disclosure Mechanisms}

\label{subsec:Dominance-over-Scalar-Bid}

Standard advertising auctions address multidimensional valuations
by restricting advertisers to submit scalar bids while varying the
information disclosed before bidding. A disclosure policy partitions
the segment space into groups and reveals only the group containing
the realized segment, so advertisers bid based on their expected value
for the disclosed group.

Following \citet{Rafieian2021}, we formalize disclosure through partitions
of the segment space. Let $\mathcal{P}_{\mathrm{disc}}$ be a partition
of the set of segments $[T]$ into $D$ disjoint disclosure blocks,
and let $g_{\mathrm{disc}}:[T]\rightarrow[D]$ denote the associated
block membership function. When an impression of segment $t$ arrives,
the publisher observes $t$ but discloses only the block $g_{\mathrm{disc}}(t)$
to advertisers. Because advertisers submit scalar bids, advertiser
$a$'s bid for a disclosed block $d\in[D]$ reflects its expected
valuation over the segments contained in that block.

We refer to any mechanism that operates in this way as a scalar-bid
disclosure mechanism. Formally, for a disclosure partition $\mathcal{P}_{\mathrm{disc}}$,
let $M_{\mathrm{SB}}(\mathcal{P}_{\mathrm{disc}})$ denote any feasible
scalar-bid (SB) mechanism in which, after disclosure block $d$ is
revealed, advertisers submit a scalar bid for that block, and allocations
and payments depend only on the disclosed block and the submitted
scalar bids, not on distinctions among segments within the block.
This class includes generalized second-price auctions, first-price
auctions, and other disclosure-based auction formats that restrict
advertisers to scalar bids conditional on the disclosed information.
Let $\mathrm{Rev}^{M_{\mathrm{SB}}}(\mathcal{P}_{\mathrm{disc}})$
denote the expected publisher revenue generated by such a mechanism.

The following theorem shows that IBPA weakly dominates the class of
scalar-bid disclosure mechanisms, for any disclosure partition.
\begin{thm}
\label{thm:dominance-scalar-bid} For any disclosure partition $\mathcal{P}_{\mathrm{disc}}$
and any scalar-bid disclosure mechanism $M_{\mathrm{SB}}(\mathcal{P}_{\mathrm{disc}})$,
\[
\mathrm{Rev}^{\mathrm{IBPA}}\ge\mathrm{Rev}^{M_{\mathrm{SB}}}(\mathcal{P}_{\mathrm{disc}}).
\]
\end{thm}

\begin{proof}
See Web Appendix \ref{appendix:Omitted-Proofs-Results}.
\end{proof}
The proof proceeds in two steps. First, for any disclosure partition,
a scalar-bid disclosure mechanism reduces the environment to a single-parameter
position auction, where each advertiser is characterized by its expected
value for the disclosed block. By the marginal-revenue characterization
of optimal auctions in single-parameter environments, the optimal
scalar-bid mechanism is therefore a marginal-revenue mechanism \citep{bulow1989simple}.
Second, every single-advertiser mechanism feasible under a scalar-bid
disclosure policy can also be implemented within IBPA by restricting
allocations and payments to depend only on disclosure blocks. IBPA
therefore optimizes over a strictly larger set of mechanisms and therefore
weakly dominates every scalar-bid disclosure mechanism.

An immediate implication of Theorem \ref{thm:dominance-scalar-bid}
is that IBPA weakly dominates generalized second-price (GSP) auctions
under any disclosure regime, since GSP is a scalar-bid disclosure
mechanism. More broadly, the dominance result highlights the fundamental
limitation of scalar-bid mechanisms. By requiring advertisers to compress
segment-level valuations into a single bid before allocation and pricing
occur, scalar-bid mechanisms discard information about heterogeneity
in advertiser preferences across segments. In contrast, IBPA elicits
the advertiser's full valuation vector and applies the marginal-revenue
framework directly to this richer information while preserving incentive
compatibility. As a result, IBPA can implement a strictly larger set
of allocation and pricing rules, allowing the publisher to better
exploit differences in advertiser valuations and thereby generate
weakly higher revenue.

\subsection{Monotonicity in Information and Disclosure Granularity}

The previous subsections compared IBPA against two revenue benchmarks:
the optimal feasible BIC and BIR mechanism, and the class of scalar-bid
disclosure mechanisms. We now turn to a complementary question: how
does publisher revenue under IBPA vary with the granularity of the
information available to the publisher and the granularity of the
information disclosed to advertisers?

The question has two dimensions. The first is information granularity:
how finely should the publisher distinguish segments internally? The
second is disclosure granularity: how much of this information should
be revealed to advertisers before bidding? Under IBPA, these decisions
are distinct because advertisers submit multidimensional bids.

We formalize the publisher’s internal information using a partition
$\mathcal{P}_{\mathrm{info}}$ of the segment space $[T]$ into $I$
blocks, with block-membership function $g_{\mathrm{info}}:[T]\rightarrow[I]$.
Under information partition $\mathcal{P}_{\mathrm{info}}$, the publisher
observes only the information block $g_{\mathrm{info}}(t)$ containing
the realized segment $t$, and therefore cannot condition allocations
or payments on distinctions finer than the blocks in $\mathcal{P}_{\mathrm{info}}$.
Disclosure is represented by a partition $\mathcal{P}_{\mathrm{disc}}$,
with block-membership function $g_{\mathrm{disc}}$.

We compare the granularity of two partitions using set refinement
\citep{Rafieian2021}. Let $\mathcal{P}_{1}$ and $\mathcal{P}_{2}$
be two partitions of $[T]$. We say that $\mathcal{P}_{1}$ is at
least as granular as $\mathcal{P}_{2}$---denoted $\mathcal{P}_{1}\succeq\mathcal{P}_{2}$---if
for every block $B\in\mathcal{P}_{1}$, there exists a block $C\in\mathcal{P}_{2}$
such that $B\subseteq C$. That is, $\mathcal{P}_{1}$ is more granular
than $\mathcal{P}_{2}$ whenever it splits the segment space more
finely.

A pair $(\mathcal{P}_{\mathrm{info}},\mathcal{P}_{\mathrm{disc}})$
is feasible if $\mathcal{P}_{\mathrm{info}}\succeq\mathcal{P}_{\mathrm{disc}}$,
so every disclosure block is a union of information blocks. Advertisers
then submit bids for the information blocks contained within the disclosed
block. The following theorem shows that the publisher revenue under
IBPA is monotonically increasing in the information granularity and
monotonically decreasing in the disclosure granularity.
\begin{thm}
\label{thm:granularity-monotonicity}For any $\mathcal{P}_{\mathrm{info}}^{(1)}\succeq\mathcal{P}_{\mathrm{info}}^{(2)}$
and for all $\mathcal{P}_{\mathrm{disc}}$ such that both pairs $(\mathcal{P}_{\mathrm{info}}^{(1)},\mathcal{P}_{\mathrm{disc}})$
and $(\mathcal{P}_{\mathrm{info}}^{(2)},\mathcal{P}_{\mathrm{disc}})$
are feasible,
\[
\mathrm{Rev}^{\mathrm{IBPA}}(\mathcal{P}_{\mathrm{info}}^{(1)},\mathcal{P}_{\mathrm{disc}})\;\ge\;\mathrm{Rev}^{\mathrm{IBPA}}(\mathcal{P}_{\mathrm{info}}^{(2)},\mathcal{P}_{\mathrm{disc}}).
\]
Similarly, for any $\mathcal{P}_{\mathrm{disc}}^{(1)}\succeq\mathcal{P}_{\mathrm{disc}}^{(2)}$
and for all $\mathcal{P}_{\mathrm{info}}$ such that both pairs $(\mathcal{P}_{\mathrm{info}},\mathcal{P}_{\mathrm{disc}}^{(1)})$
and $(\mathcal{P}_{\mathrm{info}},\mathcal{P}_{\mathrm{disc}}^{(2)})$
are feasible,
\[
\mathrm{Rev}^{\mathrm{IBPA}}(\mathcal{P}_{\mathrm{info}},\mathcal{P}_{\mathrm{disc}}^{(1)})\;\le\;\mathrm{Rev}^{\mathrm{IBPA}}(\mathcal{P}_{\mathrm{info}},\mathcal{P}_{\mathrm{disc}}^{(2)}).
\]
\end{thm}

\begin{proof}
See Web Appendix \ref{appendix:Omitted-Proofs-Results}.
\end{proof}
Taken together, the two monotonicity results imply that publisher
revenue is maximized under the full-information/null-disclosure regime.
That is, the publisher benefits from observing the most refined segment
information available while revealing no information about the realized
segment prior to bidding. The mechanism described in Section \ref{sec:Problem-Setup}
corresponds exactly to this regime.

The key insight is that finer information and coarser disclosure enlarge
the set of feasible single-advertiser mechanisms. A mechanism feasible
under a coarser information partition can always be replicated under
a finer one, while a mechanism feasible under finer disclosure can
always be replicated under coarser disclosure. The monotonicity results
follow immediately from these nested feasible sets.

Theorem \ref{thm:granularity-monotonicity} stands in contrast to
prior work \citep{Levin2010,Yao2011,Ada2022,Rafieian2021} showing
that publisher revenue under scalar-bid mechanisms such as GSP need
not be monotone in the granularity of targeting information. Under
such mechanisms, advertisers must compress their segment-level valuations
into a single bid, causing information granularity and disclosure
granularity to become coupled. As a result, increasing the amount
of information available to advertisers improves targeting precision
but may simultaneously reduce competition, generating the non-monotonic
trade-offs documented in the literature.

Under IBPA, information granularity and disclosure granularity can
be varied independently because advertisers submit multidimensional
bids. Finer information expands the set of feasible mechanisms, while
greater disclosure restricts it. Revenue therefore moves monotonically
in the direction predicted by second-degree price discrimination:
it increases with information granularity and decreases with disclosure
granularity.

\subsection{Computational Complexity and Approximation}

IBPA is defined through a collection of constrained single-advertiser
problems that determine advertisers’ normalized revenue curves. Solving
these problems exactly may require optimization over a rich class
of multidimensional pricing mechanisms and can become computationally
challenging as the number of segments increases. 

To address this issue, in Web Appendix \ref{appendix:Approximation},
we develop two computationally tractable approximation classes. The
first restricts attention to deterministic segment bundles, yielding
a finite menu representation closely related to mixed-bundling mechanisms.
The second imposes an additive pricing structure across segments that
can be optimized in time polynomial in the number of segments. Building
on approximation results from the multidimensional pricing literature,
we show that these approximation classes retain strong worst-case
revenue guarantees relative to the unrestricted single-advertiser
problem.

\section{Empirical Application}

The theoretical results establish that IBPA outperforms scalar-bid
disclosure mechanisms, including GSP. This section quantifies the
magnitude of those gains in a realistic empirical setting. Using auction-level
data from a large retail media platform, we estimate the primitives
required to implement the mechanism, including the prior distribution
of advertiser valuations, the distribution of segments, and slot effects
on click-through rates. With these primitives in hand, we compute
publisher revenue under IBPA and compare it against optimal scalar-bid
mechanisms under alternative disclosure regimes, as well as GSP benchmarks.

\subsection{Data}

\label{subsec:Data}

Our analysis draws on auction-level data from a large retail media
platform collected during the period May 17, 2025 to July 14, 2025.
The dataset contains all position auctions conducted on the platform
during this interval for one product category. For each auction, we
observe the realized segment, the number of ad slots available, and
participating advertisers' bids and quality scores ($\gamma_{a}$).

The platform classifies impressions into segments that combine audience
and device information. Segments combine consumer behavior (three
categories), demographics (eight categories), and device type (three
categories), yielding $3\times8\times3=72$ distinct segments.\footnote{Although additional audience attributes are available, we focus on
these dimensions because the platform considers them the most relevant
for advertiser targeting.}

The platform employs the generalized second-price (GSP) auction format
under the full disclosure. Advertisers therefore observe the realized
segment before bidding and submit a bid specific to that segment.
Auction outcomes follow the standard GSP rule: advertisers are ranked
by the product of their bid and quality score, slots are assigned
in that order, and each advertiser pays the minimum bid needed to
retain its assigned slot.

To estimate click-through rate (CTR) parameters, we use a separate
dataset covering the same time period as the auction data. This dataset
contains daily-level observations for each advertiser, reporting the
number of impressions and clicks received at each slot position. These
daily aggregates identify the slot multipliers $\alpha_{s}$ using
variation in advertisers’ realized positions over time, following
\citet{athey2010structural}.

The auctions are sealed bid. Advertisers combine the targeting information
disclosed by the publisher with private information---such as first-party
customer data, conversion histories, website activity, or third-party
data---when determining bids. Consequently, valuations may vary across
impression opportunities even within the same segment, motivating
the private-information framework used in the paper.

\subsubsection*{Data Summary}

Table \ref{tab:summary-stats} summarizes the data. The sample contains
9.6 million auctions, 72 segments, and 1,207 advertisers. Most auctions
($\approx66\%$) contain either 8 or 20 slots.

\begin{table}
\centering
\begin{tabular}{cc}
\hline 
Statistic & Value\tabularnewline
\hline 
\hline 
Auctions & 9.6M\tabularnewline
Advertisers & 1,207\tabularnewline
Segments & 72\tabularnewline
Slot distribution & \begin{cellvarwidth}[t]
\centering
1 slot (14\%), 3 slots (11\%),

8 slots (42\%), 20 slots (24\%),

other (9\%)
\end{cellvarwidth}\tabularnewline
Submitted bids & 44.4M\tabularnewline
Average bid & \$5.31\tabularnewline
Average quality score & 0.0096\tabularnewline
\hline 
\end{tabular}

\caption{Data summary statistics}

\label{tab:summary-stats}
\end{table}

\begin{figure}[tbh]
\centering{}\includegraphics[scale=0.7]{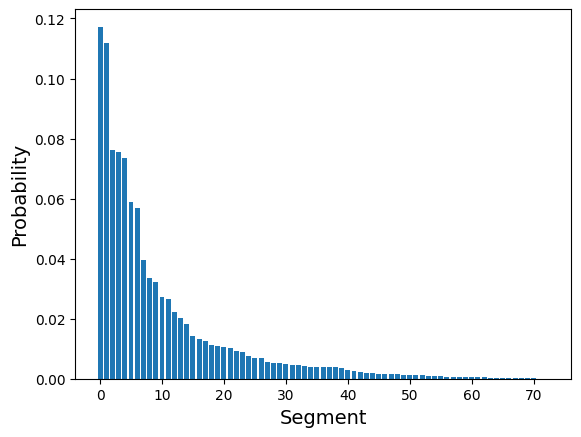}\caption{Distribution of segments}
\label{fig:inventory-type-distribution}
\end{figure}

\begin{figure}[tbh]
\centering{}\subfloat[Advertiser-level]{\centering{}\includegraphics[scale=0.5]{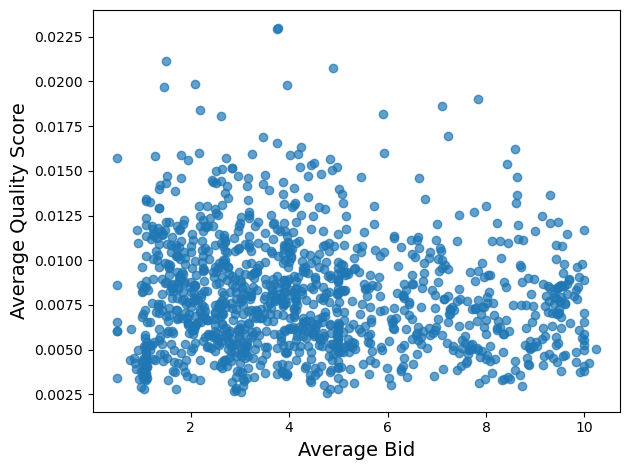}}\subfloat[Segment-level]{\centering{}\includegraphics[scale=0.5]{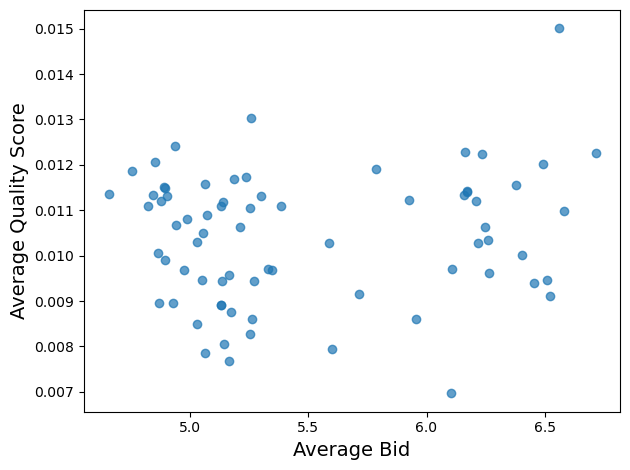}}\caption{Distribution of average bid vs. quality score}
\label{fig:bid-qs-scatter}
\end{figure}

Figure \ref{fig:inventory-type-distribution} shows substantial variation
in segment frequency; the most common segment accounts for approximately
12\% of auctions. These empirical frequencies define the segment probabilities
$p_{t}$. Advertiser bids and quality scores also exhibit substantial
heterogeneity. The average bid is $\$5.31$ per-click, and the average
quality score ($\gamma$) is 0.0096. Figure \ref{fig:bid-qs-scatter}
illustrates this heterogeneity by plotting average bids and quality
scores by advertiser and by segment.

\begin{figure}[tbh]
\centering{}\includegraphics[scale=0.7]{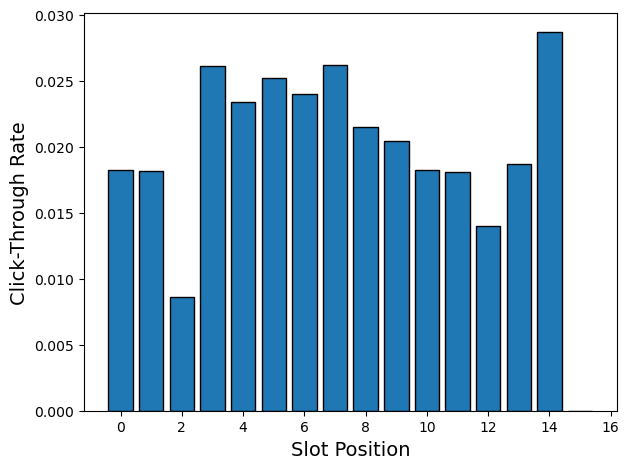}\caption{Variation of click-through rate by slot}
\label{fig:slot-ctr}
\end{figure}

The CTR dataset contains 11.4M impressions and 198,609 clicks during
the sample period, yielding an average click-through rate of 0.0174.
Figure \ref{fig:slot-ctr} plots the empirical CTR by slot position.
Observed CTR need not decline monotonically with slot position because
it reflects both slot effects and the quality of advertisers assigned
to each slot; we estimate the underlying slot effects separately.

\subsection{Estimating Primitives}

\label{subsec:Estimating-Primitives}

To evaluate the performance of the marginal-revenue mechanism and
the benchmark mechanisms, we require several primitives: (i) the distribution
of advertiser valuations $\{F_{a}\}_{a\in[A]}$, (ii) the distribution
of advertiser quality scores $\{\gamma_{a}\}$, (iii) slot effects
on click-through rates $\{\alpha_{s}\}$, and (iv) the distribution
of segments $\{p_{t}\}$. The distribution of segments follows directly
from the empirical frequencies computed in Section \ref{subsec:Data}.
Similarly, the empirical distribution of advertiser quality scores
is used directly. The remaining primitives---slot effects and valuation
distributions---are identified from the data using the framework
of \citet{athey2010structural} and the equilibrium structure of generalized
second-price auctions. We describe their estimation in Web Appendix
\ref{appendix:Primitives}.

\subsection{Comparison of Mechanisms}

\label{subsec:Comparison-of-Mechanisms}

Using the primitives estimated in Section \ref{subsec:Estimating-Primitives},
we compute publisher revenue under IBPA and compare it against two
classes of benchmarks. First, we consider optimal scalar-bid mechanisms
under alternative disclosure regimes, which correspond to the theoretical
benchmark in Theorem \ref{thm:dominance-scalar-bid}. Second, we compare
IBPA to generalized second-price (GSP) auctions under the same disclosure
regimes, as GSP remains the dominant auction format in search and
retail media advertising. We first describe the mechanisms considered
and then outline our counterfactual simulation procedure.

\subsubsection{IBPA}

Our primary mechanism is IBPA under the full-information/null-disclosure
regime $\left(\mathcal{P}_{\mathrm{info}}^{\mathrm{full}},\mathcal{P}_{\mathrm{disc}}^{\mathrm{null}}\right)$,
which Theorem \ref{thm:granularity-monotonicity} identifies as the
revenue-maximizing information and disclosure regime.\footnote{We implement the additive-pricing approximation $\mathrm{IBPA}_{\mathrm{add}}$
described in Web Appendix \ref{appendix:Approximation}. Since $\mathrm{IBPA}_{\mathrm{add}}$
restricts the class of feasible single-advertiser mechanisms, the
resulting revenue estimates provide a conservative lower bound on
the performance of the unrestricted IBPA mechanism.} The constrained single-advertiser problems (Equation \ref{eq:single-adv-slot-normalized})
are solved numerically using a genetic algorithm.\footnote{We use a genetic algorithm to account for the possibility of multiple
local optima. Results are robust to algorithmic tuning parameters;
we use: population size 100, 50 parents per generation (tournament
selection), 5 elites, crossover probability 0.8 (blending), and mutation
probability 0.2 (random mutation).}

For each advertiser, we solve the constrained problem on a grid of
allocation probabilities $q\in[0,1]$. The normalized revenue curve
$\Phi_{a}(q)$ is then approximated by a piecewise-linear interpolation
across this grid. This yields a conservative approximation, as the
true revenue curve is concave and therefore weakly above the piecewise-linear
fit. Quantile mappings and auction outcomes---allocations and payments---are
computed as described in Section \ref{subsec:Marginal-Revenue-Approach}.

\subsubsection{Optimal Scalar-Bid Mechanisms}

Theorem \ref{thm:dominance-scalar-bid} establishes that IBPA weakly
dominates the class of scalar-bid disclosure mechanisms. To quantify
the magnitude of this advantage, we first construct optimal scalar-bid
benchmarks under alternative disclosure regimes.

As discussed in Section \ref{subsec:Dominance-over-Scalar-Bid}, for
a given disclosure partition $\mathcal{P}$, the corresponding revenue-maximizing
scalar-bid mechanism is the marginal-revenue mechanism in the induced
single-parameter environment. These mechanisms are Bayesian incentive
compatible, so advertisers bid truthfully and allocations and payments
follow the corresponding marginal-revenue rules.

Ideally, one might evaluate all possible partitions of the 72 segments,
but this is computationally infeasible. Instead, we exploit the five
underlying segment attributes: consumer behavior (1 attribute), demographics
(3 attributes), and device type (1 attribute). By toggling each attribute
on or off, we generate $2^{5}=32$ meaningful disclosure regimes.

In our empirical setting, the null-disclosure partition\textit{\emph{
$\mathcal{P}=\mathcal{P}_{\mathrm{disc}}^{\mathrm{null}}$ yields
the highest revenue. We therefore report results for both the null-disclosure
and full-disclosure regimes, denoted OPT-SB-ND and OPT-SB-FD, respectively.}}

\subsubsection{GSP Benchmarks}

\label{subsec:GSP-Benchmarks}

We additionally compare IBPA to generalized second-price (GSP) auctions,
which remain the dominant auction format in search and retail media
advertising. We report GSP under the same full-disclosure and null-disclosure
regimes considered above, denoted GSP-FD and GSP-ND, respectively.

Unlike IBPA and the optimal scalar-bid mechanisms, GSP is not incentive
compatible. Consequently, advertiser behavior depends on the equilibrium
concept employed. In principle, the appropriate comparison would use
a Bayes-Nash equilibrium under private information. However, Bayes-Nash
equilibria for GSP position auctions with advertiser quality scores
are generally not tractable and may fail to exist \citep{gomes2009bayes}.\footnote{Bayes-Nash equilibria have been characterized for certain GSP environments
without quality scores. To our knowledge, no analogous characterization
exists for the quality-score environment considered here.} We therefore follow \citet{Edelman_et_al_2007,Varian_2007} and compute
an envy-free equilibrium. Because such equilibria are not unique,
we report the envy-free upper bound---the equilibrium yielding the
highest publisher revenue---thereby providing a conservative benchmark
for comparison with IBPA.

For each GSP benchmark, reserve prices are optimized to maximize publisher
revenue.. Specifically, we evaluate publisher revenue over a grid
of reserve prices, where the grid granularity matches the minimum
bid increment observed in the data, and select the revenue-maximizing
reserve. Following the standard treatment in the sponsored-search
literature (e.g., \citealp{athey2010structural}), the reserve price
is implemented as an additional bidder with quality score equal to
one. All reported GSP results therefore correspond to the revenue-maximizing
reserve price within the considered grid.

\subsubsection{Counterfactual Performance Comparison}

\label{subsec:Counterfactuals}

To compare mechanisms in a common environment, we simulate 10 million
auctions using the primitives from Section \ref{subsec:Estimating-Primitives}.
For each simulated auction, we (1) set the number of slots to $8$
and use the estimated slot effects $\{\alpha_{s}\}$, (2) draw the
number of advertisers and their quality scores $\gamma_{a}$ from
the empirical distributions, (3) draw a valuation vector $v_{a}$
for each advertiser from the estimated distributions $\{F_{at}\}$,
and (4) draw the realized segment $t$ from the empirical distribution
$\{p_{t}\}$. 

\begin{figure}[tbph]
\centering{}\includegraphics[scale=0.7]{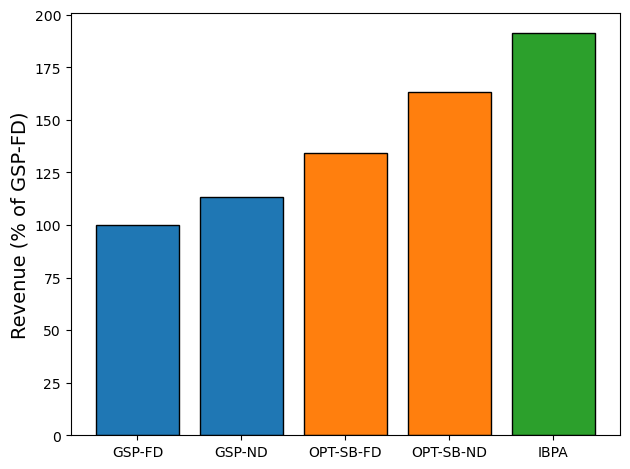}\caption{Revenue comparison}
\label{fig:rev-comparison}
\end{figure}

Figure \ref{fig:rev-comparison} reports publisher revenue across
all mechanisms. Consistent with Theorem \ref{thm:dominance-scalar-bid},
IBPA outperforms the optimal scalar-bid benchmarks under both disclosure
regimes. Moreover, the optimal scalar-bid mechanisms outperform the
corresponding GSP benchmarks. The ordering of revenues matches our
theoretical predictions:
\[
\begin{aligned} & \mathrm{IBPA}\;>\;\mathrm{OPT-SB-FD}\;>\;\mathrm{GSP-FD}\\
\text{{and}\ensuremath{\quad}} & \mathrm{IBPA}\;>\;\mathrm{OPT-SB-ND}\;>\;\mathrm{GSP-ND}.
\end{aligned}
\]

Relative to GSP-FD, IBPA increases publisher revenue by 91\%. This
improvement can be decomposed into two components. First, the corresponding
optimal scalar-bid mechanism (OPT-SB-FD) increases revenue by 34\%
relative to GSP-FD, reflecting the value of revenue-optimal allocation
and pricing within the class of scalar-bid mechanisms. Second, IBPA
increases revenue by an additional 42\% relative to OPT-SB-FD, reflecting
the value of multidimensional bidding and the richer price-discrimination
opportunities enabled by IBPA.

A similar pattern emerges under null disclosure. Relative to GSP-ND,
IBPA increases revenue by 69\%. Of this gain, 44\% is attributable
to moving from GSP-ND to OPT-SB-ND, while the remaining 17\% arises
from moving from OPT-SB-ND to IBPA.

While our theoretical guarantees concern revenue, we also measure
advertiser welfare (expected sum of utilities) and total welfare (publisher
revenue plus advertiser welfare). Table \ref{tab:metrics} reports
percentage changes in revenue, advertiser welfare, and total welfare,
along with percentage-point changes in allocation rate, relative to
the GSP--FD benchmark across alternative mechanisms. While welfare
improvements are not theoretically guaranteed for arbitrary primitives,
in our empirical setting, both metrics improve under the proposed
mechanism relative to the benchmark---advertiser welfare by 62\%
and total welfare by 79\%. The improvements appear to be driven by
allocation rate---in our simulations, the probability of an impression
being sold under the proposed mechanism is 47\% compared to 21\% under
the benchmark. Higher allocation rate indicates that impressions are
more frequently matched to advertisers who value them, improving outcomes
for all participants.

\begin{table}
\centering
\begin{tabular}{cccccc}
\hline 
Metric & GSP-FD & GSP-ND & OPT-SB-FD & OPT-SB-ND & IBPA\tabularnewline
\hline 
\hline 
Revenue & +0\% & +14\% & +34\% & +63\% & +91\%\tabularnewline
Advertiser Welfare & +0\% & -28\% & +49\% & +69\% & +62\%\tabularnewline
Total Welfare & +0\% & -4\% & +41\% & +66\% & +79\%\tabularnewline
Allocation Rate & +0pp & +19pp & +2pp & +25pp & +26pp\tabularnewline
\hline 
\end{tabular}

\caption{Performance of mechanisms relative to GSP-FD}

\label{tab:metrics}
\end{table}

The results also provide some insight into the sources of the revenue
gains. While allocation and pricing decisions are jointly determined
by the mechanism and therefore do not admit a clean additive decomposition,
the comparison across mechanisms is informative. Moving from GSP-FD
to the optimal scalar-bid mechanism increases revenue by 34\% while
increasing allocation rate by only 2 percentage points. In contrast,
moving from the optimal scalar-bid mechanism to IBPA increases revenue
by an additional 42\% and allocation rate by an additional 24 percentage
points. These patterns suggest that a substantial portion of the gains
from IBPA arises from improved matching between advertisers and impressions,
which increases the probability that impressions are allocated. At
the same time, the additional revenue gain relative to the optimal
scalar-bid benchmark indicates that multidimensional bidding and the
richer screening opportunities available under IBPA also contribute
materially to performance.

\subsubsection{Revenue and Information Granularity}

\label{subsec:Revenue-and-Information}

The previous subsection compared mechanisms under full publisher information,
using each mechanism's revenue-maximizing disclosure regime. We now
examine how publisher revenue changes as publisher information becomes
more granular.\footnote{This exercise is relevant for understanding the potential revenue
implications of privacy policies that restrict the information publishers
may use for advertising. Such policies can be viewed as coarsening
the information available to the publisher.} For IBPA, we maintain the revenue-maximizing null-disclosure regime
while varying the information available to the publisher. For scalar-bid
mechanisms, information granularity and disclosure granularity are
equivalent, since advertisers can condition their bids only on disclosed
information. We therefore report the scalar-bid mechanisms under full
disclosure throughout this analysis.\footnote{For scalar-bid mechanisms, information and disclosure granularity
cannot be varied independently. A fully informed publisher disclosing
a coarse partition is equivalent to a less-informed publisher observing
only that partition and fully disclosing it. Thus, when varying publisher
information, it is without loss of generality to impose full disclosure.}

\begin{figure}[tbph]
\centering{}\includegraphics[scale=0.7]{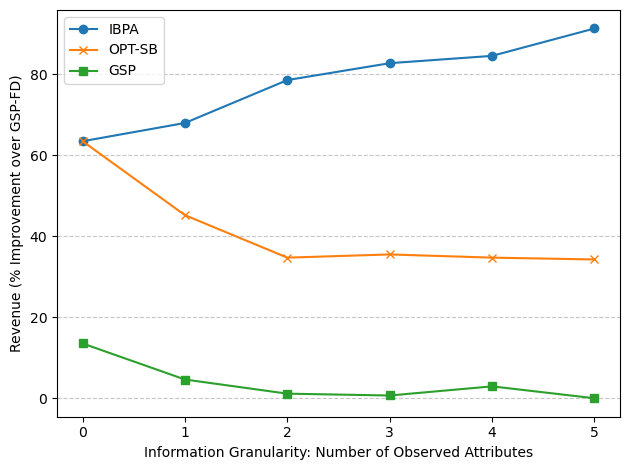}\caption{Revenue and information granularity}
\label{fig:revenue-info-granularity}
\end{figure}

Starting from no publisher information, we sequentially reveal the
five segment attributes (consumer behavior, three demographic attributes,
and device type) in a random order. Figure \ref{fig:revenue-info-granularity}
reports publisher revenue as additional attributes become available.

Consistent with Theorem \ref{thm:granularity-monotonicity}, publisher
revenue under IBPA increases monotonically with information granularity.
In contrast, neither the optimal scalar-bid mechanism nor GSP exhibits
monotonicity. Additional information can reduce revenue because finer
segmentation thins competition while scalar bidding prevents the mechanism
from fully exploiting the richer information.

At null publisher information, IBPA and the optimal scalar-bid mechanism
coincide because both reduce to the same single-parameter environment.
As publisher information becomes more granular, the revenue gap between
the two mechanisms widens in our empirical setting, illustrating the
increasing value of multidimensional bidding. Consistent with the
earlier revenue comparisons, GSP underperforms the corresponding optimal
scalar-bid mechanism across all information regimes.

\section{Discussion \& Conclusion}

\label{sec:Conclusion}

This paper develops the Information-Bundling Position Auction (IBPA),
a mechanism for selling digital advertising inventory when advertiser
valuations depend on audience and contextual information privately
observed by the publisher. Unlike standard scalar-bid auctions, IBPA
separates targeting precision from market thickness by allowing advertisers
to submit multidimensional bids while treating the realized impression
segment as the publisher's private information.

For each advertiser, IBPA constructs a menu of bundle prices that
implements second-degree price discrimination over the advertiser's
valuation vector. Across advertisers, impressions are allocated using
the marginal-revenue framework.

The resulting mechanism is feasible, Bayesian incentive compatible,
and individually rational. Although exact revenue maximization is
generally intractable, IBPA achieves a $(1-1/e)$-approximation to
the optimal feasible mechanism. It also weakly dominates scalar-bid
disclosure mechanisms, including GSP, and resolves the targeting-precision
versus market-thickness trade-off. Consequently, publisher revenue
is increasing in the granularity of the information available to the
publisher.

Using data from a retail media platform, we estimate advertiser valuations
and evaluate the performance of IBPA in counterfactual simulations.
Relative to GSP, IBPA increases publisher revenue by approximately
91\%, advertiser welfare by 62\%, and total welfare by 79\%. The welfare
gains arise primarily from improved allocation rates, as impressions
are more frequently matched to advertisers who value them.

Several extensions and considerations merit discussion. First, our
analysis abstracts from advertiser budget constraints, multi-homing,
and competition among publishers. Although IBPA satisfies individual
rationality, a budget-constrained advertiser may reallocate spending
across platforms if its marginal return changes. Publishers can accommodate
such concerns by imposing minimum advertiser-utility constraints or
incorporating advertiser welfare into their objectives. Because IBPA
is constructed from single-advertiser revenue problems, these modifications
can be incorporated directly into the mechanism.

Relatedly, the counterfactual revenue gains estimated for a single
publisher need not scale proportionally under industry-wide adoption.
Aggregate advertiser budgets may be fixed or slow-moving, and higher
allocation rates may increase ad load, potentially affecting consumer
engagement. Publishers therefore choose auction design jointly with
ad supply, pacing, and other policies that balance short-run revenue
against advertiser retention and consumer experience. A full equilibrium
analysis of publisher competition, advertiser budget allocation, and
consumer engagement is an important direction for future work.

Second, practical implementation requires addressing both bid complexity
and computational scalability. Although IBPA requires multidimensional
bids, these bids can be represented compactly. Advertising platforms
already allow advertisers to combine base bids with modifiers for
device, demographics, behavior, and other targeting attributes. These
inputs naturally generate segment-contingent valuation vectors without
requiring advertisers to submit an independent bid for every segment.
Modern machine-learning bidding systems can also generate segment-contingent
valuation vectors in a single model evaluation, mitigating the computational
burden on advertisers.

Most of the publisher’s computational burden occurs offline, when
constructing revenue curves and quantile mappings. The online mechanism
then uses these mappings to compute marginal revenues, allocations,
and payments.\footnote{In our implementation, an online auction required approximately 400
milliseconds on a standard laptop. Constructing a revenue curve required
several minutes with a 10-point quantile grid and approximately one
hour with a 100-point grid.} Scalability nevertheless remains important because the number of
segments can grow rapidly with the number of targeting attributes.
Practical implementations may therefore require structured bid representations,
dimension reduction, or restricted pricing menus, such as the additive-pricing
approximation used in our empirical analysis.

Finally, practical adoption requires commitment and transparency.
Incentive compatibility ensures that truthful reporting is optimal
conditional on the announced mechanism, but advertisers must also
trust the publisher to follow its allocation and payment rules \citep{akbarpour2020credible}.
This concern is not unique to IBPA: GSP similarly requires confidence
in platform-determined quality scores, rankings, and pricing rules.
An important direction for future work therefore concerns the role
of commitment, transparency, and verifiability in ad auction design.

Overall, this paper shows that the tension between targeting precision
and competition is not inherent to digital advertising markets. By
combining information bundling with marginal-revenue allocation, IBPA
provides a tractable alternative to scalar-bid disclosure mechanisms,
provides strong performance guarantees, and opens several directions
for research on implementation, platform competition, commitment,
and dynamic advertiser participation.

\clearpage{}

\begin{singlespace}
\bibliographystyle{jmr}
\bibliography{InfoPricingReferences}

\end{singlespace}

\clearpage{}

\appendix
\setcounter{page}{1} 
\begin{center}
{\Large Web Appendix}{\Large\par}
\par\end{center}

\section{Proofs from Section \ref{sec:Problem-Setup}}

\label{appendix:Omitted-Proofs-Model}
\begin{proof}[\textbf{Proof of Proposition \ref{prop:only-higher-slot}}]
Consider any single-advertiser IC/IR mechanism $M$, represented
(by the taxation principle) as a menu of lottery pricings $\{(\chi^{(k)},\mu^{(k)})\}_{k}$,
where each lottery $k$ specifies allocation probabilities $\chi_{st}^{(k)}\in[0,1]$
for slots $s\in\tilde{S}$ and segments $t\in[T]$, and an expected
payment $m^{(k)}=\alpha_{s_{0}}\gamma_{a_{0}}\mu^{(k)}$. Fix a lottery
pricing $k$.

Define, for each segment $t$, the CTR-weighted (slot-collapsed) allocation
probability 
\[
\bar{\chi}_{t}^{(k)}\;\equiv\;\sum_{s\in\tilde{S}}\frac{\alpha_{s}}{\alpha_{s_{0}}}\chi_{st}^{(k)}.
\]
Now construct a modified lottery pricing $(\chi^{(k)\prime},\mu^{(k)\prime})$
that allocates only the highest available slot $s_{0}$ as follows:
\[
\chi_{s_{0}t}^{(k)\prime}=\bar{\chi}_{t}^{(k)}\quad\forall t,\qquad\chi_{st}^{(k)\prime}=0\quad\forall s\neq s_{0},\forall t,\qquad\mu^{(k)\prime}=\mu^{(k)}.
\]

For each $t$, since $\alpha_{s}\le\alpha_{s_{0}}$ for all $s\in\tilde{S}$,
\[
\bar{\chi}_{t}^{(k)}=\sum_{s\in\tilde{S}}\frac{\alpha_{s}}{\alpha_{s_{0}}}\chi_{st}^{(k)}\;\le\;\sum_{s\in\tilde{S}}\chi_{st}^{(k)}\;\le\;1,
\]
where the last inequality follows from the feasibility constraint
that the advertiser receives at most one slot for any given segment.
Hence the modified lottery pricing is also feasible.

For any valuation vector $\mathbf{v}_{a_{0}}=(v_{a_{0}1},\ldots,v_{a_{0}T})$,
the expected utility from lottery $k$ under $M$ is 
\[
U^{(k)}(\mathbf{v}_{a_{0}})=\sum_{t}\sum_{s}p_{t}\,\alpha_{s}\beta_{t}\gamma_{a_{0}}v_{a_{0}t}\chi_{st}^{(k)}-\alpha_{s_{0}}\gamma_{a_{0}}\mu^{(k)}.
\]
Under the modified lottery pricing, 
\[
U^{(k)\prime}(\mathbf{v}_{a_{0}})=\sum_{t}p_{t}\,\alpha_{s_{0}}\beta_{t}\gamma_{a_{0}}v_{a_{0}t}\chi_{s_{0}t}^{(k)\prime}-\alpha_{s_{0}}\gamma_{a_{0}}\mu^{(k)\prime}.
\]
By construction, $\alpha_{s_{0}}\chi_{s_{0}t}^{(k)\prime}=\alpha_{s_{0}}\bar{\chi}_{t}^{(k)}=\sum_{s}\alpha_{s}\chi_{st}^{(k)}$
for every $t$, and $\mu^{(k)\prime}=\mu^{(k)}$. Therefore $U^{(k)\prime}(\mathbf{v}_{a_{0}})=U^{(k)}(\mathbf{v}_{a_{0}})$
for all valuation vectors $\mathbf{v}_{a_{0}}$.

Thus replacing lottery pricing $k$ by $(\chi^{(k)\prime},\mu^{(k)\prime})$
leaves the utility of that menu item unchanged pointwise in $v_{a_{0}}$,
so the advertiser's menu choice behavior is unchanged. Because the
payment parameter is unchanged, publisher revenue from this modified
lottery pricing is also unchanged. Repeating this argument for every
lottery pricing in the mechanism establishes the result.
\end{proof}
$\ $
\begin{proof}[\textbf{Proof of Proposition \ref{prop:rev-curve-concave}}]
For notational convenience, we drop the advertiser and slot subscripts
and write $R(q)$ for the optimal expected revenue of the ex-ante
single-advertiser problem with allocation constraint $q$. Let $M(q)$
denote an optimal mechanism that achieves $R(q)$, and let $\Lambda(q)$
be its ex-ante allocation probability. By feasibility, $\Lambda(q)\leq q$.

To prove monotonicity, take any $q_{1},q_{2}\in[0,1]$. Any mechanism
feasible under constraint $q_{1}$---in particular $M(q_{1})$---satisfies
$\Lambda(q_{1})\le q_{1}\le q_{2}$ and is therefore feasible at constraint
$q_{2}$. Since $R(q_{2})$ is the maximum revenue over all mechanisms
feasible at constraint $q_{2}$, we obtain $R(q_{2})\ge R(q_{1})$.
Thus, $R(q)$ is non-decreasing in $q$, and consequently, $R'(q)\ge0\:\forall q$.

Take any $q_{1},q_{2}\in[0,1]$ with $q_{1}<q_{2}$ and any $\nu\in[0,1]$.
Let $q=\nu q_{1}+(1-\nu)q_{2}$. Consider the convex combination of
mechanisms 
\[
\widetilde{M}\;\equiv\;\nu\,M(q_{1})+(1-\nu)\,M(q_{2}).
\]
The ex-ante allocation probability under $\widetilde{M}$ is 
\[
\widetilde{\Lambda}\;=\;\nu\,\Lambda(q_{1})+(1-\nu)\,\Lambda(q_{2})\;\le\;\nu q_{1}+(1-\nu)q_{2}\;\equiv\;q,
\]
so $\widetilde{M}$ is feasible for the problem with allocation constraint
$q$.

Moreover, the expected revenue of $\widetilde{M}$ is 
\[
\text{Revenue}(\widetilde{M})\;=\;\nu\,R(q_{1})+(1-\nu)\,R(q_{2}).
\]

Since $R(q)$ is defined as the maximum expected revenue among all
mechanisms feasible at constraint $q$, it follows that 
\[
R(q)\;\ge\;\text{Revenue}(\widetilde{M})\;=\;\nu\,R(q_{1})+(1-\nu)\,R(q_{2}).
\]

This inequality holds for every $q_{1},q_{2}\in[0,1]$ and every $\nu\in[0,1]$
with $q=\nu q_{1}+(1-\nu)q_{2}$. Hence $R(\cdot)$ is concave on
$[0,1]$. Concavity immediately implies that the (right) derivative
$R'(q)$ exists almost everywhere and is weakly decreasing wherever
it exists. 
\end{proof}
$\ $
\begin{proof}[\textbf{Proof of Proposition \ref{prop:rev-curve-factors}}]
Fix an advertiser $a$ and a slot $s$. Recall that the ex-ante single-advertiser
problem with allocation constraint $q$ chooses a menu of lotteries
$\{(\chi^{(k)},\mu^{(k)})\}_{k}$ to maximize 
\[
R_{as}(q)\;=\;\max_{K,\{\chi^{(k)},\mu^{(k)}\}}\;\alpha_{s}\gamma_{a}\sum_{k=1}^{K}\xi_{k}\,\mu^{(k)}\quad\text{s.t.}\quad\sum_{k=1}^{K}\xi_{k}\!\left(\sum_{t}p_{t}\,\chi_{t}^{(k)}\right)\le q,
\]
where $\xi_{k}$ is the probability that lottery $k$ is chosen ex-ante
under the prior.

Define $\Phi_{a}(q)$ to be the optimal value of the \emph{slot-normalized}
problem that keeps the same feasible menus and allocation constraint
but omits the multiplicative factor $\alpha_{s}\gamma_{a}$ from the
objective: 
\[
\Phi_{a}(q)\;=\;\max_{K,\{\chi^{(k)},\mu^{(k)}\}}\;\sum_{k=1}^{K}\xi_{k}\,\mu^{(k)}\quad\text{s.t.}\quad\sum_{k=1}^{K}\xi_{k}\!\left(\sum_{t}p_{t}\,\chi_{t}^{(k)}\right)\le q,
\]

To prove the claim, it is sufficient to show that $\Phi_{a}(q)$ is
independent of the slot $s$.

Consider any fixed menu $\{(\chi^{(k)},\mu^{(k)})\}_{k}$. Take two
options $k_{1},k_{2}$ from this menu. The advertiser's expected utility
difference between $k_{1}$ and $k_{2}$ when the highest remaining
slot is $s$ is 
\[
\sum_{t}p_{t}\,\alpha_{s}\beta_{t}\gamma_{a}\,v_{at}\,\Big(\chi_{t}^{(k_{1})}-\chi_{t}^{(k_{2})}\Big)\;-\;\alpha_{s}\gamma_{a}\Big(\mu^{(k_{1})}-\mu^{(k_{2})}\Big).
\]

Factoring out $\alpha_{s}\gamma_{a}>0$, the sign of this difference---and
thus the choice between $k_{1}$ and $k_{2}$---is determined by
\[
\sum_{t}p_{t}\,\beta_{t}\,v_{at}\,\Big(\chi_{t}^{(k_{1})}-\chi_{t}^{(k_{2})}\Big)\;-\;\Big(\mu^{(k_{1})}-\mu^{(k_{2})}\Big),
\]
which is independent of $s$. Therefore, for any menu, the induced
choice probabilities do not depend on $s$. The allocation constraint
$\sum_{k}\xi_{k}\big(\sum_{t}p_{t}\chi_{t}^{(k)}\big)\le q$ is also
independent of $s$.

Consequently, the feasible set of menus for the normalized problem
is the same for every $s$, and the only appearance of $(\alpha_{s},\gamma_{a})$
in the original objective is as a multiplicative factor. It follows
that the optimal value satisfies 
\[
R_{as}(q)\;=\;\alpha_{s}\gamma_{a}\,\Phi_{a}(q),
\]
where $\Phi_{a}(q)$ is independent of the slot $s$. This proves
the claim. 
\end{proof}

\clearpage{}

\section{General Quantile Mapping and Payments}

\label{appendix:Quantile-Mapping}

In Section \ref{subsec:Marginal-Revenue-Approach}, we described the
valuation-to-quantile mapping and payment rule under a transparent
special case in which the single-advertiser mechanisms $\{M_{a}(q)\}_{q\in[0,1]}$
have nested deterministic allocations. In that case, a valuation vector
can be mapped to the smallest allocation constraint $q$ under which
the corresponding single-advertiser mechanism allocates the impression,
and payments admit a critical-quantile interpretation.

This appendix describes the general construction, which does not require
nestedness or deterministic allocations. The key difference from the
special case in the main text is that the quantile mapping may be
randomized, and payments are not generally computed using critical
quantiles. Instead, we first construct the interim mechanism induced
by marginal-revenue maximization and then implement the corresponding
allocation and payment rules through a randomized quantile mapping.

Because Proposition \ref{prop:rev-curve-factors} implies that an
advertiser's normalized marginal-revenue score $\gamma_{a}\Phi_{a}'(q)$
is common across slots, we construct a single quantile for each advertiser
and realized segment. The construction therefore does not treat exact
slot assignments separately. Instead, it summarizes the advertiser's
position allocation in top-slot-equivalent units.

\subsection{Preliminaries}

The marginal revenue approach requires a mapping from advertiser valuation
vectors to quantiles. We begin with the observation that the induced
quantiles must be uniformly distributed under the advertiser's prior
\citep{Alaei2013}. That is, for each advertiser $a$ and realized
segment $t$, the final mapping $Q_{at}$ should satisfy 
\[
Q_{at}(\mathbf{v}_{a})\sim\mathrm{Unif}[0,1]\qquad\text{when }\mathbf{v}_{a}\sim F_{a}.
\]

Suppose first that such a quantile mapping has been constructed. Given
a vector of quantiles $\mathbf{q}=(q_{1},\ldots,q_{A})$, the marginal-revenue
allocation rule assigns slots in decreasing order of the scores $\gamma_{a}\Phi_{a}'(q_{a})$,
excluding advertisers with non-positive marginal revenue and breaking
ties arbitrarily.

Let $I_{as}^{MR}(q,\mathbf{Q}_{-a})\in\{0,1\}$ indicate that advertiser
$a$ is assigned slot $s$ by the marginal-revenue allocation rule
when its own quantile is $q$. Define advertiser $a$'s quantile-space,
top-slot-equivalent allocation curve as 
\[
\omega_{a}^{MR}(q)=\mathbb{E}_{\mathbf{Q}_{-a}}\left[\sum_{s=1}^{S}\alpha_{s}I_{as}^{MR}(q,\mathbf{Q}_{-a})\right],
\]
where the expectation is taken over the independently drawn quantiles
of the other advertisers and over any tie-breaking randomization.

Equivalently, setting $\alpha_{S+1}=0$ and $\Delta_{j}=\alpha_{j}-\alpha_{j+1}$,
\[
\omega_{a}^{MR}(q)=\sum_{j=1}^{S}\Delta_{j}\Pr\left\{ a\text{ is among the top }j\text{ eligible advertisers}\,\middle|\,q_{a}=q\right\} .
\]
Since $\Phi_{a}$ is concave, $\Phi_{a}'(q)$ is weakly decreasing
in $q$. Consequently, each probability in the equation above, and
therefore $\omega_{a}^{MR}(q)$, is weakly decreasing in $q$.

The use of the top-slot-equivalent allocation curve is important in
the multi-slot environment. The probability that an advertiser receives
a particular slot need not be monotone in its quantile, because weakening
its score may move it from a higher slot to a lower slot. In contrast,
its click-through-rate-weighted allocation is weakly decreasing in
its quantile.

\subsection{Interim Mechanism as a Convex Combination of $\{M_{a}(q)\}$}

The function $\omega_{a}^{MR}(q)$ determines the interim mechanism
faced by advertiser $a$. In particular, the interim mechanism is
a convex combination of the optimal single-advertiser mechanisms $\{M_{a}(q)\}_{q\in[0,1]}$,
with weights determined by the decrease in $\omega_{a}^{MR}(q)$ \citep{Alaei2013}.
Thus, the interim mechanism $\overline{M}_{a}$ can be expressed as\footnote{Formally, the integral is interpreted with respect to the Lebesgue-{}-Stieltjes
measure induced by $\omega_{a}^{MR}$, including the appropriate boundary
mass at $q=1$. Any residual probability mass $1-\omega_{a}^{MR}(0)$
is assigned to the null mechanism $M_{a}(0)$.} 
\[
\overline{M}_{a}=\int_{0}^{1}M_{a}(\widetilde{q})\left(-d\omega_{a}^{MR}(\widetilde{q})\right).
\]

Operationally, $\overline{M}_{a}$ draws an allocation constraint
$\tilde{q}$ according to the distribution induced by $\omega_{a}^{MR}$
above and then offers advertiser $a$ the single-advertiser mechanism
$M_{a}(\tilde{q})$.

Let $\overline{\chi}_{at}(\mathbf{v}_{a})$ denote the probability
that $\overline{M}_{a}$ allocates the top-slot-equivalent impression
in segment $t$, when its valuation vector is $\mathbf{v}_{a}$ and
$\overline{\mu}_{a}(\mathbf{v}_{a})$ denote the corresponding normalized
expected payment. The objective of the quantile mapping and payment
construction is to implement this target interim allocation and payment
rule in the multi-advertiser auction.

Since every $M_{a}(q)$ is incentive compatible and individually rational
and their convex combination induced by $\omega_{a}^{MR}$ to construct
$\overline{M}_{a}$ does not depend on advertiser $a$'s valuation,
$\overline{M}_{a}$ is also incentive compatible and individually
rational.

\subsection{Constructing the Quantile Mapping}

For a fixed advertiser $a$ and segment $t$, the target allocation
rule $\overline{\chi}_{at}(\mathbf{v}_{a})$ induces an ordering over
valuation vectors. Higher values of $\overline{\chi}_{at}$ correspond
to stronger types: under the target interim mechanism, they receive
a greater top-slot-equivalent allocation.

The rank of a valuation vector $\mathbf{v}_{a}$ induced by such an
ordering is given by 
\[
\rho_{at}(\mathbf{v}_{a})=\Pr_{\mathbf{v}_{a}'\sim F_{a}}\left\{ \overline{\chi}_{at}(\mathbf{v}_{a}')>\overline{\chi}_{at}(\mathbf{v}_{a})\right\} .
\]

This ranking in turn induces a ranked interim allocation curve 
\[
g_{at}(r)=\sup\left\{ y\in[0,1]:\Pr_{\mathbf{v}_{a}\sim F_{a}}\left(\overline{\chi}_{at}(\mathbf{v}_{a})\ge y\right)\ge r\right\} .
\]

Thus, $g_{at}(r)$ is the allocation probability of a valuation vector
at rank $r$ when valuation vectors are sorted in decreasing order
of their target allocation probabilities. This curve is weakly decreasing
in $r$ by construction.

The ranking $\rho_{at}$ satisfies the property $\rho_{at}(\mathbf{v}_{a})\sim\mathrm{Unif}[0,1]$
when $\mathbf{v}_{a}\sim F_{a}$\footnote{$\rho_{at}(\mathbf{v}_{a})\sim\mathrm{Unif}[0,1]$ assuming that the
distribution of $x_{at}(\mathbf{v}_{a})$ when $\mathbf{v}_{a}\sim F_{a}$
is atomless. When the target allocation rule has atoms, ties can be
broken by an independent uniform random variable. In particular, one
may replace $\rho_{at}(\mathbf{v}_{a})$ by $\Pr\{x_{at}(\mathbf{V}_{a}')>x_{at}(\mathbf{v}_{a})\}+U_{at}^{0}\Pr\{x_{at}(\mathbf{V}_{a}')=x_{at}(\mathbf{v}_{a})\},$where
$U_{at}^{0}\sim\mathrm{Unif}[0,1]$. This randomized rank is uniform
without the atomlessness assumption.}. However, $\rho_{at}(\mathbf{v}_{a})$ cannot generally be used directly
as the final quantile. If the mechanism used $q_{at}=\rho_{at}(\mathbf{v}_{a})$,
then the advertiser's resulting ranked allocation curve would be $\omega_{a}^{MR}(\rho_{at}(\mathbf{v}_{a}))$,
which implements the quantile-space winning curve $\omega_{a}^{MR}$,
not the ranked interim allocation curve $g_{at}$. In general, $\omega_{a}^{MR}(r)\neq g_{at}(r)$.
Thus, the rank mapping has the correct distributional property, but
it need not implement the correct ranked allocation curve.

To obtain the correct implementation, we modify the rank mapping $\rho_{at}(\mathbf{v}_{a})$
using the interval-resampling construction of \citet{Alaei2013}.
The goal is to construct a randomized transformation 
\[
\Psi_{at}:[0,1]\to[0,1]
\]
such that, for $R\sim\mathrm{Unif}[0,1]$, 
\[
\Psi_{at}(R)\sim\mathrm{Unif}[0,1],
\]
and, for each rank $r$, 
\[
\mathbb{E}\left[\omega_{a}^{MR}\bigl(\Psi_{at}(r)\bigr)\right]=g_{at}(r),
\]
where the expectation is taken over the randomization in $\Psi_{at}$.

The first condition preserves the uniform distribution of quantiles.
The second condition ensures that the marginal-revenue allocation
rule implements the target allocation probability prescribed by the
interim mechanism $\overline{M}_{a}$.

The transformation $\Psi_{at}$ is constructed by interval resampling.
In a discrete approximation, order valuation vectors by the rank $\rho_{at}$.
Whenever the quantile-space allocation curve $\omega_{a}^{MR}$ differs
from the ranked target allocation curve $g_{at}$, identify an interval
of ranks over which averaging $\omega_{a}^{MR}$ produces the desired
level of $g_{at}$. Types whose preliminary ranks fall in this interval
are then resampled uniformly within the interval. Repeating this procedure
over the ordered type space yields a sequence of resampling intervals.
This procedure preserves the uniform distribution of final quantiles
while ensuring that the allocation probabilities induced by the marginal-revenue
allocation rule coincide with the target interim allocation probabilities.

The final segment-specific quantile mapping is 
\[
Q_{at}(\mathbf{v}_{a})=\Psi_{at}\bigl(\rho_{at}(\mathbf{v}_{a})\bigr).
\]
By construction, $Q_{at}(\mathbf{v}_{a})\sim\mathrm{Unif}[0,1]$ when
$\mathbf{v}_{a}\sim F_{a}$, and the marginal-revenue allocation rule
implements the target segment-contingent allocation probability $\overline{\chi}_{at}(\mathbf{v}_{a})$.
This construction nests the deterministic mapping in Section \ref{subsec:Marginal-Revenue-Approach}.

\subsection{Payments}

In the nested deterministic case described in the main text, payments
can be represented using critical quantiles. In the general case,
such a representation need not exist. Payments are instead inherited
from the target interim mechanism $\overline{M}_{a}$.

Let $\overline{\mu}_{a}(\mathbf{v}_{a})$ denote the normalized expected
payment under $\overline{M}_{a}$, and define 
\[
\overline{\lambda}_{a}(\mathbf{v}_{a})=\sum_{t=1}^{T}p_{t}\overline{\chi}_{at}(\mathbf{v}_{a}).
\]
Thus, $\overline{\lambda}_{a}(\mathbf{v}_{a})$ is advertiser $a$'s
ex-ante allocation in top-slot-equivalent units under the target interim
mechanism. Since $\alpha_{1}=1$, the corresponding expected payment
in the position auction is $\gamma_{a}\overline{\mu}_{a}(\mathbf{v}_{a})$.
To collect this payment only when the advertiser receives a slot,
one convenient implementation charges 
\[
m_{ast}^{\mathrm{imp}}(\mathbf{v}_{a})=\alpha_{s}\gamma_{a}\frac{\overline{\mu}_{a}(\mathbf{v}_{a})}{\overline{\lambda}_{a}(\mathbf{v}_{a})}.
\]
If payments are collected per click, the impression payment is divided
by the realized click-through rate $\alpha_{s}\beta_{t}\gamma_{a}$.
The resulting per-click payment is

\[
m_{ast}^{\mathrm{click}}(\mathbf{v}_{a})=\frac{\overline{\mu}_{a}(\mathbf{v}_{a})}{\beta_{t}\overline{\lambda}_{a}(\mathbf{v}_{a})}.
\]

In the nested deterministic case, the layer-specific critical-quantile
payments described in Section \ref{subsec:Marginal-Revenue-Approach}
provide an alternative disaggregation of the same expected payment.

\clearpage{}

\section{Notation and Algorithm Summary}

\label{appendix:algorithm}

\begin{table}[H]
\centering
\begin{tabular}{c>{\centering}p{8cm}}
\hline 
Notation & Definition\tabularnewline
\hline 
\noalign{\vskip0.5cm}
$a\in[A],\;s\in[S],\;t\in[T]$ & Advertiser, slot, and segment indices. \tabularnewline
\noalign{\vskip0.5cm}
$p_{t}$ & Probability of segment $t$. \tabularnewline
\noalign{\vskip0.5cm}
$\mathbf{v}_{a}=(v_{a1},\ldots,v_{aT})$  & Advertiser $a$'s valuation vector, where $v_{at}$ is the per-click
valuation for segment $t$. \tabularnewline
\noalign{\vskip0.5cm}
$F_{a}$  & Distribution of advertiser $a$'s valuation vector. \tabularnewline
\noalign{\vskip0.5cm}
$\mathbf{b}_{a}=(b_{a1},\ldots,b_{aT})$ & Advertiser $a$'s bid vector, where $b_{at}$ is the bid for segment
$t$. \tabularnewline
\noalign{\vskip0.5cm}
$\alpha_{s},\;\beta_{t},\;\gamma_{a}$  & Slot, segment, and advertiser effects in the click-through rate $\bar{c}_{ast}=\alpha_{s}\beta_{t}\gamma_{a}$. \tabularnewline
\noalign{\vskip0.5cm}
$x_{ast}(\mathbf{V}),\;m_{a}(\mathbf{V})$  & Allocation rule and expected payment rule. \tabularnewline
\noalign{\vskip0.5cm}
$R_{as}(q)=\alpha_{s}\gamma_{a}\Phi_{a}(q),\;\Phi_{a}(q),\;\Phi_{a}'(q)$  & Revenue curve, slot-normalized revenue curve, and normalized marginal
revenue. \tabularnewline
\noalign{\vskip0.5cm}
$M_{a}(q)$  & Optimal single-advertiser mechanism attaining $\Phi_{a}(q)$ under
allocation constraint $q$. \tabularnewline
\noalign{\vskip0.5cm}
$\Omega_{at}(q)$  & Allocation region induced by a deterministic mechanism $M_{a}(q)$
for segment $t$. \tabularnewline
\noalign{\vskip0.5cm}
$q_{at}=Q_{at}(v_{a})$  & Quantile associated with advertiser $a$'s valuation vector for segment
$t$. \tabularnewline
\noalign{\vskip0.5cm}
$\chi_{t}^{(k)},\;\mu^{(k)}$  & Allocation probability and payment parameter associated with menu
option $k$. \tabularnewline
\hline 
\noalign{\vskip0.5cm}
\end{tabular}

\caption{Summary of notation}

\label{tab:notation}
\end{table}

\begin{algorithm}
\caption{Information-Bundling Position Auction (IBPA)}

\begin{algorithmic}[1]

\STATE \textbf{Inputs:}
\STATE \quad Segment probabilities $p_t$
\STATE \quad Advertiser priors $F_a$
\STATE \quad Quality scores $\gamma_a$
\STATE \quad Slot effects $\alpha_s$

\STATE
\STATE \textbf{Outputs:}
\STATE \quad Allocations $x_{ast}$
\STATE \quad Payments $m_a$

\STATE
\item[] \textbf{Offline Stage: Construct Revenue Curves}
\FOR{each advertiser $a$}
    \STATE Solve the slot-normalized constrained single-advertiser problem for all $q\in[0,1]$ (Equation \ref{eq:single-adv-slot-normalized})
    \STATE Construct the normalized revenue curve $\Phi_a(q)$ (Section \ref{subsec:Revenue-Curve})
    \STATE Construct the segment-specific valuation-to-quantile mappings $Q_{at}(\mathbf{v}_a)$ for all segments $t$ (Section \ref{subsec:Marginal-Revenue-Approach} and Web Appendix \ref{appendix:Quantile-Mapping})
\ENDFOR

\vspace{0.5em}

\item[] \textbf{Online Stage: Marginal-Revenue Auction}
\FOR{each arriving impression}
    \STATE Publisher observes realized segment $t$
    \STATE Advertisers submit bid vectors $\mathbf{v}_a$ (truthful bidding in equilibrium)
    
    \FOR{each advertiser $a$}
        \STATE Compute quantile from valuation vector:
        \[
        q_{at}=Q_{at}(\mathbf{v}_a)
        \]
        \STATE Compute marginal revenue from quantile for each slot (Proposition \ref{prop:rev-curve-factors}):
        \[
        R'_{as}(q_{at})
        =
        \alpha_s\gamma_a\Phi'_a(q_{at})
        \]
    \ENDFOR

    \STATE Allocate slots greedily in decreasing order of $\alpha_s$, assigning each slot to the unassigned advertiser with the highest positive marginal revenue (Section \ref{subsec:Marginal-Revenue-Approach})
	\STATE Compute payment for each winning advertiser (Section \ref{subsec:Marginal-Revenue-Approach} and Web Appendix \ref{appendix:Quantile-Mapping})

\ENDFOR

\end{algorithmic}

\label{alg:ibpa}
\end{algorithm}

\clearpage{}

\section{Proofs from Section \ref{sec:Theoretical-Results}}

\label{appendix:Omitted-Proofs-Results}
\begin{proof}[\textbf{Proof of Proposition \ref{prop:IC-IR-Feasibility}}]
 We verify feasibility, Bayesian incentive compatibility, and Bayesian
individual rationality in turn.

Feasibility follows directly from the allocation rule. For each realized
segment, IBPA processes slots in decreasing order of slot effect and
assigns each slot to at most one advertiser. Once an advertiser is
assigned a slot, that advertiser is removed from consideration for
lower slots. Hence, for every valuation profile $\mathbf{V}$, segment
$t$, and slot $s$, 
\[
\sum_{a}x_{ast}(\mathbf{V})\leq1\quad\text{and}\quad\sum_{s}x_{ast}(\mathbf{V})\leq1,
\]
so the feasibility constraints in Equation~\eqref{eq:publisher_problem}
are satisfied.

We next establish incentive compatibility. Fix an advertiser $a$.
Conditional on advertiser $a$'s reported valuation vector, the marginal-revenue
allocation rule and the quantile construction induce an interim mechanism
for advertiser $a$, where the expectation is taken over the valuations
of the remaining advertisers and over any randomization in the mechanism.
By construction (see Web Appendix \ref{appendix:Quantile-Mapping}),
this interim mechanism can be represented as a convex combination
of the single-advertiser mechanisms $\{M_{a}(q)\}_{q\in[0,1]}$, with
weights induced by the marginal-revenue allocation rule. Each single-advertiser
mechanism $M_{a}(q)$ is incentive compatible by construction, since
it solves the single-agent constrained problem over incentive-compatible
mechanisms. A convex combination of incentive-compatible mechanisms
is incentive compatible: for any true valuation vector $\mathbf{v}_{a}$
and any report $\mathbf{b}_{a}$, truthful reporting yields weakly
higher expected utility in each component mechanism, and therefore
also in their expectation. Hence truthful reporting maximizes advertiser
$a$'s interim expected utility. Since this argument applies to every
advertiser $a$, truthful multidimensional reporting constitutes a
Bayes--Nash equilibrium, and the IBPA mechanism is BIC.

Finally, individual rationality follows from the same convex-combination
argument. Each single-advertiser mechanism $M_{a}(q)$ is individually
rational by construction, so truthful participation gives advertiser
$a$ weakly non-negative expected utility in every component mechanism.
Any convex combination of these mechanisms therefore also gives weakly
non-negative interim expected utility. Thus the IBPA mechanism satisfies
BIR.

Combining feasibility, BIC, and BIR establishes the result. 
\end{proof}
$\ $
\begin{proof}[\textbf{Proof of Theorem \ref{thm:ibpa-approximation}}]
 We prove the result in two steps. First, we upper-bound the revenue
of the optimal feasible BIC and BIR mechanism by an ex-ante relaxation.
Second, we show that IBPA obtains at least a $(1-1/e)$ fraction of
this ex-ante relaxation. 

Throughout the proof, we use $\alpha_{1}=1$ and let $\alpha_{S+1}=0$.
For notational simplicity, we write $\Phi_{a}'(q)$ for the marginal
revenue of advertiser $a$. At points where $\Phi_{a}$ is not differentiable,
take any subgradient; the argument is unchanged by the usual ironing
or tie-breaking conventions.

Consider any feasible BIC and BIR mechanism $M$ for the publisher's
problem in Equation \ref{eq:publisher_problem}, with allocation rule
$x_{ast}^{M}(\mathbf{V})$ and payment rule $m_{a}^{M}(\mathbf{V})$.
For advertiser $a$, define the interim allocation and payment rules
induced by $M$, obtained by integrating over the valuation vectors
of the other advertisers: 
\[
X_{ast}^{M}(\mathbf{v}_{a})=\mathbb{E}_{\mathbf{V}_{-a}}\left[x_{ast}^{M}(\mathbf{v}_{a},\mathbf{V}_{-a})\right],\qquad\bar{m}_{a}^{M}(\mathbf{v}_{a})=\mathbb{E}_{\mathbf{V}_{-a}}\left[m_{a}^{M}(\mathbf{v}_{a},\mathbf{V}_{-a})\right].
\]
Because $M$ is BIC and BIR, this interim single-advertiser mechanism
is IC and IR for advertiser $a$.

Now collapse advertiser $a$'s interim slot allocation into top-slot-equivalent
impression units. Define 
\[
\tilde{x}_{at}^{M}(\mathbf{v}_{a})=\sum_{s=1}^{S}\alpha_{s}X_{ast}^{M}(\mathbf{v}_{a}).
\]
as the probability of receiving the top slot that would generate the
same click-through-rate-weighted allocation as the original slot assignment.
Thus $\tilde{x}_{at}^{M}(\mathbf{v}_{a})$ converts the original slot
allocation into an equivalent top-slot allocation measured in click-through-rate
units. Since the original mechanism assigns advertiser $a$ to at
most one slot for each realized segment and $\alpha_{s}\le1$, we
have 
\[
0\le\tilde{x}_{at}^{M}(\mathbf{v}_{a})\le1\qquad\text{for all }a,t,\mathbf{v}_{a}.
\]
Moreover, the collapsed allocation preserves advertiser $a$'s interim
utility: 
\[
\sum_{t=1}^{T}p_{t}\beta_{t}\gamma_{a}v_{at}\tilde{x}_{at}^{M}(\mathbf{v}_{a})-\bar{m}_{a}^{M}(\mathbf{v}_{a})=\mathbb{E}_{\mathbf{V}_{-a}}\left[\sum_{t=1}^{T}\sum_{s=1}^{S}p_{t}\alpha_{s}\beta_{t}\gamma_{a}v_{at}x_{ast}^{M}(\mathbf{v}_{a},\mathbf{V}_{-a})-m_{a}^{M}(\mathbf{v}_{a},\mathbf{V}_{-a})\right].
\]
Thus the collapsed rule $(\tilde{x}_{a}^{M},\bar{m}_{a}^{M})$ is
a feasible IC and IR single-advertiser mechanism for the top slot.

Let 
\begin{equation}
z_{a}^{M}=\mathbb{E}_{\mathbf{v}_{a}}\left[\sum_{t=1}^{T}p_{t}\tilde{x}_{at}^{M}(\mathbf{v}_{a})\right]=\mathbb{E}_{\mathbf{V},t}\left[\sum_{s=1}^{S}\alpha_{s}x_{ast}^{M}(\mathbf{V})\right]\label{eq:top-slote-equivalent}
\end{equation}
denote advertiser $a$'s ex-ante top-slot-equivalent allocation probability
under $M$. By the definition of the normalized single-advertiser
revenue curve $\Phi_{a}$, no IC and IR single-advertiser mechanism
that allocates top-slot-equivalent impression with ex-ante probability
at most $z_{a}^{M}$ can extract more than $\gamma_{a}\Phi_{a}(z_{a}^{M})$
from advertiser $a$. Therefore, 
\[
\mathbb{E}_{V}\left[m_{a}^{M}(\mathbf{V})\right]\le\gamma_{a}\Phi_{a}(z_{a}^{M}).
\]
Summing over advertisers gives 
\begin{equation}
\mathrm{Rev}^{M}=\sum_{a=1}^{A}\mathbb{E}_{\mathbf{V}}\left[m_{a}^{M}(\mathbf{V})\right]\le\sum_{a=1}^{A}\gamma_{a}\Phi_{a}(z_{a}^{M}),\label{eq:Rev}
\end{equation}
where $\mathrm{Rev}^{M}$ is the expected revenue of the mechanism
$M$.

We next characterize the ex-ante feasibility of the vector $z^{M}=(z_{1}^{M},\ldots,z_{A}^{M})$.
For any subset of advertisers $B\subseteq[A]$, ex-post feasibility
of $M$ implies that, for every valuation profile $\mathbf{V}$ and
segment $t$, 
\[
\sum_{a\in B}\sum_{s=1}^{S}\alpha_{s}x_{ast}^{M}(\mathbf{V})\le\sum_{j=1}^{\min\{|B|,S\}}\alpha_{j}.
\]
The right-hand side is the largest total slot effect that can be assigned
to advertisers in $B$: assign up to $|B|$ advertisers to the highest
available slots. Taking expectations over $\mathbf{V}$ and $t$ and
using \ref{eq:top-slote-equivalent}, we obtain 
\[
\sum_{a\in B}z_{a}^{M}\le\sum_{j=1}^{\min\{|B|,S\}}\alpha_{j}\qquad\forall B\subseteq[A].
\]
Define the ex-ante impression-allocation polytope 
\[
\mathcal{P}_{\alpha}=\left\{ z\in\mathbb{R}_{+}^{A}:\sum_{a\in B}z_{a}\le\sum_{j=1}^{\min\{|B|,S\}}\alpha_{j}\quad\forall B\subseteq[A]\right\} .
\]
The preceding argument shows that every ex-post feasible mechanism
induces an ex-ante vector $z^{M}\in\mathcal{P}_{\alpha}$. Hence every
feasible BIC and BIR mechanism $M$ satisfies 
\[
\mathrm{Rev}^{M}\le\sum_{a=1}^{A}\gamma_{a}\Phi_{a}(z_{a}^{M})\le\max_{z\in\mathcal{P}_{\alpha}}\sum_{a=1}^{A}\gamma_{a}\Phi_{a}(z_{a}),
\]
where the first inequality follows from equation \ref{eq:Rev}. Define
the ex-ante relaxation value as 
\[
\mathrm{Rev}^{\mathrm{EA}}\equiv\max_{z\in\mathcal{P}_{\alpha}}\sum_{a=1}^{A}\gamma_{a}\Phi_{a}(z_{a}).
\]
Taking the supremum over all feasible BIC and BIR mechanisms $M$
yields a bound on the optimal revenue 
\[
\mathrm{Rev}^{\mathrm{OPT}}\le\mathrm{Rev}^{\mathrm{EA}}.
\]

It remains to compare IBPA to the ex-ante relaxation. Consider the
following single-dimensional analog of the publisher's multidimensional
problem, where each advertiser $a$'s type is represented by a scalar
quantile $q_{a}\sim\mathrm{Unif}[0,1]$, advertiser $a$'s revenue
curve is $G_{a}(q)=\gamma_{a}\Phi_{a}(q)$, and the feasibility constraint
is the same ex-ante service polytope $\mathcal{P}_{\alpha}$. The
ex-ante relaxation of this analog is exactly $\mathrm{Rev}^{\mathrm{EA}}$.

By construction of the segment-specific quantile mappings in IBPA,
for every advertiser $a$ and segment $t$, $Q_{at}(\mathbf{v}_{a})\sim\mathrm{Unif}[0,1]$,
when $\mathbf{v}_{a}\sim F_{a}$. Since advertisers' valuation vectors
are independent across advertisers, the quantiles $\left(Q_{1t}(\mathbf{v}_{1}),\ldots,Q_{At}(\mathbf{v}_{A})\right)$
are independent across advertisers for every realized segment $t$.
This argument does not require independence across segments within
an advertiser's valuation vector; arbitrary within-advertiser dependence
is already incorporated into the single-advertiser revenue curve $\Phi_{a}$
and the quantile mappings $Q_{at}$.

Conditional on the realized segment $t$, IBPA assigns slots to maximize
total marginal revenue: 
\[
\max_{x}\sum_{a=1}^{A}\sum_{s=1}^{S}\alpha_{s}\gamma_{a}\Phi_{a}'\!\left(Q_{at}(\mathbf{v}_{a})\right)x_{as}
\]
subject to the assignment constraints that each advertiser receives
at most one slot and each slot is assigned to at most one advertiser.
This is precisely the marginal-revenue mechanism for the single-dimensional
analog, with revenue curves $G_{a}(q)=\gamma_{a}\Phi_{a}(q)$ and
feasibility constraint $\mathcal{P}_{\alpha}$. The payment rule of
IBPA implements the interim mechanism corresponding to this marginal-revenue
allocation rule, so the revenue of IBPA in the original environment
equals the revenue of the marginal-revenue mechanism in the analog.

We now apply the feasibility-based approximation result. To do so,
it suffices to verify that the weighted rank functions induced by
the position-auction feasibility constraint are monotone and submodular.
Fix any nonnegative weight vector $w=(w_{1},\ldots,w_{A})$. For a
set of available advertisers $D\subseteq[A]$, define the weighted
rank function induced by the position auction as 
\[
f_{w}(D)=\max_{\substack{x_{as}\in\{0,1\}\\
\sum_{s}x_{as}\le1,\ \sum_{a}x_{as}\le1
}
}\sum_{a\in D}\sum_{s=1}^{S}\alpha_{s}w_{a}x_{as}.
\]
Equivalently, $f_{w}(D)$ assigns the largest weight in $D$ to the
highest slot, the second-largest weight to the second-highest slot,
and so on. Let $r_{k}^{w}(D)$ denote the weighted rank function of
the $k$-unit uniform matroid, i.e., the sum of the $k$ largest weights
in $D$. Then
\[
f_{w}(D)=\sum_{k=1}^{S}(\alpha_{k}-\alpha_{k+1})\,r_{k}^{w}(D).
\]
Each $r_{k}^{w}(\cdot)$ is monotone and submodular, and the coefficients
$\alpha_{k}-\alpha_{k+1}$ are nonnegative. Therefore $f_{w}(\cdot)$
is monotone and submodular.

By the correlation-gap bound of \citet{yan2011mechanism}, as used
in the feasibility-based approximation framework of \citet{Alaei2013},
the ex-post feasible marginal-revenue mechanism for any environment
whose weighted rank functions are monotone submodular obtains at least
a $(1-1/e)$ fraction of the ex-ante relaxation. Applying this result
to the single-dimensional analog gives 
\[
\mathrm{Rev}^{\mathrm{IBPA}}\ge\left(1-\frac{1}{e}\right)\mathrm{Rev}^{\mathrm{EA}}.
\]
Combining this lower bound with $\mathrm{Rev}^{\mathrm{OPT}}\le\mathrm{Rev}^{\mathrm{EA}}$,
we obtain 
\[
\mathrm{Rev}^{\mathrm{IBPA}}\ge\left(1-\frac{1}{e}\right)\mathrm{Rev}^{\mathrm{OPT}}.
\]
This proves the theorem. 
\end{proof}
$\ $
\begin{proof}[\textbf{Proof of Theorem \ref{thm:dominance-scalar-bid}}]
 Fix an arbitrary disclosure partition $\mathcal{P}_{\mathrm{disc}}=\{B_{1},\ldots,B_{D}\}$.
Let 
\[
p_{d}=\sum_{t\in B_{d}}p_{t}
\]
denote the probability that the realized segment belongs to disclosure
block $d$. For each advertiser $a$, define the block-level value
\[
\bar{v}_{ad}=\sum_{t\in B_{d}}\frac{p_{t}}{p_{d}}\beta_{t}v_{at}.
\]
Equivalently, if the segment effect $\beta_{t}$ is absorbed into
the valuation, then $\bar{v}_{ad}$ is simply advertiser $a$'s expected
valuation over the segments in block $d$. Conditional on disclosure
block $d$, receiving slot $s$ therefore gives advertiser $a$ expected
gross value $\alpha_{s}\gamma_{a}\bar{v}_{ad}.$

Consider any scalar-bid disclosure mechanism $M_{\mathrm{SB}}(\mathcal{P}_{\mathrm{disc}})$.
Because the mechanism elicits only a scalar bid for the disclosed
block, and allocations and payments depend only on this scalar bid
and the disclosed block, the mechanism treats all segments in $B_{d}$
as a single object from the advertiser's perspective. Thus, conditional
on disclosure block $d$, $M_{\mathrm{SB}}(\mathcal{P}_{\mathrm{disc}})$
induces a single-parameter position-auction environment in which advertiser
$a$'s type is $\bar{v}_{ad}$.

If $M_{\mathrm{SB}}(\mathcal{P}_{\mathrm{disc}})$ is not itself a
direct truthful mechanism, fix any Bayes--Nash equilibrium of the
mechanism. By the revelation principle, the allocation and payment
rules induced by this equilibrium can be implemented by a direct BIC
and BIR mechanism in the corresponding single-parameter environment,
with the same expected publisher revenue. Hence the revenue of any
scalar-bid disclosure mechanism is bounded above by the revenue of
the optimal BIC and BIR mechanism in the block-level single-parameter
environment induced by $\mathcal{P}_{\mathrm{disc}}$.

Let $\mathrm{Rev}^{\mathrm{IBPA}}(\mathcal{P}_{\mathrm{disc}})$ denote
the revenue of the marginal-revenue mechanism in this auxiliary block-level
environment, where the objects being sold are the disclosure blocks
$B_{1},\ldots,B_{D}$ and the realized block is fully disclosed. Since
this is a standard single-parameter environment with quasi-linear
preferences and position-auction feasibility constraints, the marginal-revenue
mechanism is revenue-maximizing among all feasible BIC and BIR mechanisms.
Therefore, 
\[
\mathrm{Rev}^{M_{\mathrm{SB}}}(\mathcal{P}_{\mathrm{disc}})\le\mathrm{Rev}^{\mathrm{IBPA}}(\mathcal{P}_{\mathrm{disc}}).
\]

It remains to compare this auxiliary block-level IBPA benchmark with
the proposed IBPA mechanism, which operates on the full segment space
and does not disclose the realized segment. We do this by comparing
the underlying single-advertiser revenue curves.

Fix advertiser $a$, slot $s$, and allocation constraint $q\in[0,1]$.
Let $\bar{\Phi}_{a}^{\mathcal{P}_{\mathrm{disc}}}(q)$ denote the
normalized single-advertiser revenue curve in the auxiliary block-level
environment, and let $\Phi_{a}(q)$ denote the normalized single-advertiser
revenue curve under the full segment space with no disclosure.

We claim that 
\[
\Phi_{a}(q)\ge\bar{\Phi}_{a}^{\mathcal{P}_{\mathrm{disc}}}(q)\quad\text{for all }a,q.
\]
To see this, take any single-advertiser mechanism feasible in the
auxiliary block-level environment under allocation constraint $q$.
Under that mechanism, after block $d$ is disclosed, the advertiser
faces a menu of lotteries over that block. Let a generic option in
the menu for block $d$ be denoted by $\ell_{d}$, with allocation
probability $\bar{\chi}_{d}(\ell_{d})$ and normalized payment $\bar{\mu}_{d}(\ell_{d})$,
both conditional on block $d$.

We now construct a no-disclosure single-advertiser mechanism on the
original segment space that replicates this block-level mechanism.
The no-disclosure menu contains one option for every tuple 
\[
\ell=(\ell_{1},\ldots,\ell_{D}),
\]
where $\ell_{d}$ is an option from the block-$d$ menu. For such
a tuple, define the segment-level allocation rule and normalized payment
by 
\[
\chi_{t}(\ell)=\bar{\chi}_{d}(\ell_{d})\quad\text{for every }t\in B_{d},
\]
and 
\[
\mu(\ell)=\sum_{d=1}^{D}p_{d}\bar{\mu}_{d}(\ell_{d}).
\]

For any valuation vector $v_{a}$, the advertiser's utility from tuple
$\ell$ in the constructed no-disclosure mechanism is 
\[
\sum_{d=1}^{D}\sum_{t\in B_{d}}p_{t}\beta_{t}v_{at}\bar{\chi}_{d}(\ell_{d})-\sum_{d=1}^{D}p_{d}\bar{\mu}_{d}(\ell_{d})=\sum_{d=1}^{D}p_{d}\left[\bar{v}_{ad}\bar{\chi}_{d}(\ell_{d})-\bar{\mu}_{d}(\ell_{d})\right].
\]
Because the menu contains all tuples of block-level options, maximizing
this expression over $\ell$ is equivalent to choosing, for each block
$d$, the option that the advertiser would have chosen after observing
$d$ in the auxiliary block-level mechanism. Thus, the constructed
no-disclosure mechanism replicates the same ex-ante allocation probability,
the same expected payment, and the same utility for every valuation
vector. In particular, it satisfies the same ex-ante allocation constraint
$q$ and generates the same expected revenue.

Therefore, every single-advertiser mechanism feasible in the auxiliary
block-level environment is also feasible under the full segment space
with no disclosure. This proves that $\Phi_{a}(q)\ge\bar{\Phi}_{a}^{\mathcal{P}_{\mathrm{disc}}}(q)\quad\text{for all }a,q$.
Since the slot-specific revenue curve factorizes as 
\[
R_{as}(q)=\alpha_{s}\gamma_{a}\Phi_{a}(q),
\]
this implies that the full-segment/no-disclosure revenue curves weakly
dominate the auxiliary block-level revenue curves for every advertiser
and slot: 
\[
R_{as}(q)\ge\bar{R}_{as}^{\mathcal{P}_{\mathrm{disc}}}(q)\quad\text{for all }a,s,q.
\]

The marginal-revenue mechanism preserves this pointwise ordering of
single-advertiser revenue curves: if one environment yields weakly
higher single-advertiser revenue curves for every advertiser, slot,
and allocation probability, then the corresponding multi-advertiser
marginal-revenue mechanism yields weakly higher expected revenue.
Applying this monotonicity property gives 
\[
\mathrm{Rev}^{\mathrm{IBPA}}\ge\mathrm{Rev}^{\mathrm{IBPA}}(\mathcal{P}_{\mathrm{disc}}).
\]

Combining the above inequality with $\mathrm{Rev}^{M_{\mathrm{SB}}}(\mathcal{P}_{\mathrm{disc}})\le\mathrm{Rev}^{\mathrm{IBPA}}(\mathcal{P}_{\mathrm{disc}})$,
we obtain 
\[
\mathrm{Rev}^{\mathrm{IBPA}}\ge\mathrm{Rev}^{\mathrm{IBPA}}(\mathcal{P}_{\mathrm{disc}})\ge\mathrm{Rev}^{M_{\mathrm{SB}}}(\mathcal{P}_{\mathrm{disc}}).
\]
Since the disclosure partition $\mathcal{P}_{\mathrm{disc}}$ and
the scalar-bid disclosure mechanism $M_{\mathrm{SB}}(\mathcal{P}_{\mathrm{disc}})$
were arbitrary, the result follows. 
\end{proof}
$\ $
\begin{proof}[\textbf{Proof of Theorem \ref{thm:granularity-monotonicity}}]
 We prove the two monotonicity statements by comparing the single-advertiser
revenue curves induced by the relevant information and disclosure
regimes. The argument parallels the proof of Theorem \ref{thm:dominance-scalar-bid}.

For any information partition $\mathcal{P}_{\mathrm{info}}$ and disclosure
partition $\mathcal{P}_{\mathrm{disc}}$ such that $\mathcal{P}_{\mathrm{info}}\succeq\mathcal{P}_{\mathrm{disc}}$,
let 
\[
R_{as}^{\mathcal{P}_{\mathrm{info}},\mathcal{P}_{\mathrm{disc}}}(q)
\]
denote advertiser $a$'s slot-$s$ revenue curve under the pair $(\mathcal{P}_{\mathrm{info}},\mathcal{P}_{\mathrm{disc}})$.
By Proposition \ref{prop:rev-curve-factors}, this curve factorizes
as 
\[
R_{as}^{\mathcal{P}_{\mathrm{info}},\mathcal{P}_{\mathrm{disc}}}(q)=\alpha_{s}\gamma_{a}\Phi_{a}^{\mathcal{P}_{\mathrm{info}},\mathcal{P}_{\mathrm{disc}}}(q),
\]
where $\Phi_{a}^{\mathcal{P}_{\mathrm{info}},\mathcal{P}_{\mathrm{disc}}}(q)$
is the corresponding normalized single-advertiser revenue curve.

We first establish monotonicity in information granularity. Fix a
disclosure partition $\mathcal{P}_{\mathrm{disc}}$, and let $\mathcal{P}_{\mathrm{info}}^{(1)}\succeq\mathcal{P}_{\mathrm{info}}^{(2)}$,
with both pairs $(\mathcal{P}_{\mathrm{info}}^{(1)},\mathcal{P}_{\mathrm{disc}})$
and $(\mathcal{P}_{\mathrm{info}}^{(2)},\mathcal{P}_{\mathrm{disc}})$
feasible. Fix an advertiser $a$, a slot $s$, and an allocation constraint
$q\in[0,1]$.

Consider any single-advertiser mechanism feasible under the coarser
information partition $\mathcal{P}_{\mathrm{info}}^{(2)}$. Under
this mechanism, after disclosure block $d\in\mathcal{P}_{\mathrm{disc}}$
is revealed, the advertiser faces a menu whose options specify allocation
probabilities over the information blocks contained in $d$. Let a
generic menu option be denoted by $\ell_{d}$, and let $\bar{\chi}_{C}(\ell_{d})$
be the allocation probability assigned to coarse information block
$C\in\mathcal{P}_{\mathrm{info}}^{(2)}$ with $C\subseteq d$. Let
$\bar{\mu}(\ell_{d})$ denote the associated normalized payment.

We now construct a mechanism under the finer information partition
$\mathcal{P}_{\mathrm{info}}^{(1)}$ that replicates this coarser
mechanism. For every fine information block $i\in\mathcal{P}_{\mathrm{info}}^{(1)}$,
let $C(i)$ be the unique block in $\mathcal{P}_{\mathrm{info}}^{(2)}$
such that $i\subseteq C(i)$. Define the allocation rule under the
finer partition by 
\[
\chi_{i}(\ell_{d})=\bar{\chi}_{C(i)}(\ell_{d})\quad\text{for every }i\subseteq d,
\]
and keep the same normalized payment, 
\[
\mu(\ell_{d})=\bar{\mu}(\ell_{d}).
\]

This construction preserves utilities pointwise. To see this, let
$p_{i}=\sum_{t\in i}p_{t}$ and define advertiser $a$'s expected
value for information block $i$ as 
\[
\bar{v}_{ai}=\frac{1}{p_{i}}\sum_{t\in i}p_{t}\beta_{t}v_{at}.
\]
For any coarse information block $C$, the corresponding coarse-block
value is 
\[
\bar{v}_{aC}=\frac{1}{p_{C}}\sum_{i\subseteq C}p_{i}\bar{v}_{ai}.
\]
Therefore, the utility from option $\ell_{d}$ under the finer-information
mechanism is 
\[
\sum_{i\subseteq d}\frac{p_{i}}{p_{d}}\bar{v}_{ai}\chi_{i}(\ell_{d})-\mu(\ell_{d})=\sum_{C\subseteq d}\frac{p_{C}}{p_{d}}\bar{v}_{aC}\bar{\chi}_{C}(\ell_{d})-\bar{\mu}(\ell_{d}),
\]
which is exactly the utility from the corresponding option under the
coarser-information mechanism. Hence the advertiser makes the same
menu choice, the publisher obtains the same expected payment, and
the ex-ante allocation probability is unchanged. In particular, the
constructed mechanism satisfies the same allocation constraint $q$
and generates the same expected revenue.

Thus every single-advertiser mechanism feasible under $(\mathcal{P}_{\mathrm{info}}^{(2)},\mathcal{P}_{\mathrm{disc}})$
is also feasible under $(\mathcal{P}_{\mathrm{info}}^{(1)},\mathcal{P}_{\mathrm{disc}})$.
It follows that, for every advertiser $a$, slot $s$, and allocation
constraint $q$, 
\[
R_{as}^{\mathcal{P}_{\mathrm{info}}^{(1)},\mathcal{P}_{\mathrm{disc}}}(q)\ge R_{as}^{\mathcal{P}_{\mathrm{info}}^{(2)},\mathcal{P}_{\mathrm{disc}}}(q).
\]
That is, finer information weakly raises the single-advertiser revenue
curves.

We next establish monotonicity in disclosure granularity. Fix an information
partition $\mathcal{P}_{\mathrm{info}}$, and let $\mathcal{P}_{\mathrm{disc}}^{(1)}\succeq\mathcal{P}_{\mathrm{disc}}^{(2)}$,
with both pairs $(\mathcal{P}_{\mathrm{info}},\mathcal{P}_{\mathrm{disc}}^{(1)})$
and $(\mathcal{P}_{\mathrm{info}},\mathcal{P}_{\mathrm{disc}}^{(2)})$
feasible. Thus, each block $D\in\mathcal{P}_{\mathrm{disc}}^{(2)}$
is a union of blocks $d\in\mathcal{P}_{\mathrm{disc}}^{(1)}$.

Fix advertiser $a$, slot $s$, and allocation constraint $q$. Consider
any single-advertiser mechanism feasible under the finer disclosure
partition $\mathcal{P}_{\mathrm{disc}}^{(1)}$. Under this mechanism,
after a fine disclosure block $d\in\mathcal{P}_{\mathrm{disc}}^{(1)}$
is revealed, the advertiser faces a menu of options. Let $\ell_{d}$
denote an option from the menu associated with fine disclosure block
$d$, with allocation probabilities $\bar{\chi}_{i}(\ell_{d})$ for
information blocks $i\in\mathcal{P}_{\mathrm{info}}$ satisfying $i\subseteq d$,
and normalized payment $\bar{\mu}(\ell_{d})$.

We construct a mechanism under the coarser disclosure partition $\mathcal{P}_{\mathrm{disc}}^{(2)}$
that replicates the finer-disclosure mechanism. Fix a coarse disclosure
block $D\in\mathcal{P}_{\mathrm{disc}}^{(2)}$, and let 
\[
\mathcal{D}(D)=\{d\in\mathcal{P}_{\mathrm{disc}}^{(1)}:d\subseteq D\}
\]
be the set of fine disclosure blocks contained in $D$. The menu offered
after $D$ is disclosed contains one option for every tuple 
\[
\ell_{D}=(\ell_{d})_{d\in\mathcal{D}(D)},
\]
where $\ell_{d}$ is an option from the fine-disclosure menu associated
with block $d$.

For such a tuple, define the allocation probability for each information
block $i\subseteq d$ by 
\[
\chi_{i}(\ell_{D})=\bar{\chi}_{i}(\ell_{d}),
\]
and define the normalized payment by 
\[
\mu(\ell_{D})=\sum_{d\in\mathcal{D}(D)}\frac{p_{d}}{p_{D}}\bar{\mu}(\ell_{d}),
\]
where $p_{d}=\sum_{t\in d}p_{t}$ and $p_{D}=\sum_{d\in\mathcal{D}(D)}p_{d}$.

For any valuation vector $v_{a}$, the advertiser's utility from tuple
$\ell_{D}$ conditional on coarse disclosure block $D$ is 
\[
\sum_{i\subseteq D}\frac{p_{i}}{p_{D}}\bar{v}_{ai}\chi_{i}(\ell_{D})-\mu(\ell_{D}).
\]
Using the construction above, this can be written as 
\[
\sum_{d\in\mathcal{D}(D)}\frac{p_{d}}{p_{D}}\left[\sum_{i\subseteq d}\frac{p_{i}}{p_{d}}\bar{v}_{ai}\bar{\chi}_{i}(\ell_{d})-\bar{\mu}(\ell_{d})\right].
\]
Because the menu under the coarse disclosure block $D$ contains all
tuples of fine-block menu options, maximizing this expression over
$\ell_{D}$ is equivalent to choosing, separately for each fine disclosure
block $d\subseteq D$, the option that the advertiser would have chosen
when $d$ was directly disclosed under the finer-disclosure mechanism.
Therefore, the coarser-disclosure mechanism replicates the same ex-ante
allocation probabilities and the same expected payments as the finer-disclosure
mechanism. In particular, it satisfies the same allocation constraint
$q$ and generates the same expected revenue.

Thus every single-advertiser mechanism feasible under $(\mathcal{P}_{\mathrm{info}},\mathcal{P}_{\mathrm{disc}}^{(1)})$
is also feasible under $(\mathcal{P}_{\mathrm{info}},\mathcal{P}_{\mathrm{disc}}^{(2)})$.
Hence, for every advertiser $a$, slot $s$, and allocation constraint
$q$, 
\[
R_{as}^{\mathcal{P}_{\mathrm{info}},\mathcal{P}_{\mathrm{disc}}^{(1)}}(q)\le R_{as}^{\mathcal{P}_{\mathrm{info}},\mathcal{P}_{\mathrm{disc}}^{(2)}}(q).
\]
That is, coarser disclosure weakly raises the single-advertiser revenue
curves.

Finally, as in the proof of Theorem \ref{thm:dominance-scalar-bid},
the marginal-revenue mechanism preserves pointwise dominance of the
single-advertiser revenue curves: if one regime yields weakly higher
revenue curves $R_{as}(q)$ for every advertiser $a$, slot $s$,
and allocation probability $q$, then the corresponding multi-advertiser
IBPA mechanism yields weakly higher expected publisher revenue. Applying
this monotonicity property to the revenue-curve dominance derived
above proves the theorem. 
\end{proof}

\clearpage{}

\section{Computational Complexity and Approximation}

\label{appendix:Approximation}

We now discuss the computational challenges of solving the ex-ante
constrained single-advertiser problem. Solving the ex-ante constrained
single-advertiser problem in Section \ref{sec:Problem-Setup} requires
optimizing over the full class $\mathcal{M}$ of arbitrary menu of
randomized bundle pricings, where each menu option specifies a allocation-probability
vector $\chi\in[0,1]^{T}$ and an associated expected payment $\mu$.
While the problem is conceptually straightforward, the optimal menu
can have arbitrary length; consequently, computation can become expensive
when the number of segments $T$ is large \citep{manelli2007multidimensional,hart2013menu}.

To address this complexity, we consider two structured families of
mechanisms that approximate $\mathcal{M}$ while retaining strong
revenue guarantees. The first restricts allocation-probability vectors
to be binary, reducing the menu to a collection of deterministic bundles.
The second further restricts payments to follow a simple additive
form. Both classes retain strong performance guarantees by preserving
the core price-discrimination structure of the full mechanism while
also simplifying computation. We also discuss how the main theoretical
results in Section \ref{sec:Theoretical-Results} extend to these
approximations.

\subsection{Binary-Allocation Menus}

Our first approximation solves the single-advertiser problem by restricting
attention to menus of bundle pricings with binary allocation-probability
vectors. Formally, define
\[
\mathcal{M}_{\mathrm{bin}}^{0}=\{(\chi,\mu)\in\mathcal{M}:\chi_{t}\in\{0,1\}\;\forall t\}.
\]
A menu option in $\mathcal{M}_{\mathrm{bin}}^{0}$ therefore corresponds
to offering the advertiser a subset of segments together with an associated
payment: the advertiser receives exactly those segments for which
$\chi_{t}=1$. To ensure individual rationality, we normalize the
payment for the zero-allocation vector to be $\mu(\boldsymbol{0})=0$.

Since there are at most $2^{T}$ binary allocation-probability vectors
and the zero vector is normalized, solving the optimal single-advertiser
mechanism for any ex-ante allocation constraint $q$ requires computing
at most $2^{T}-1$ payments, one for each non-zero allocation vector.
Although this growth is exponential in $T$, it is far more tractable
than optimizing over the full mechanism class $\mathcal{M}$, whose
optimal menus may require infinitely many (non-binary) randomized
bundles.

Mechanisms in $\mathcal{M}_{\mathrm{bin}}^{0}$ are closely related
to the classic mixed bundling approach in multi-product monopoly pricing.
Each binary allocation vector corresponds to a bundle of segments,
and the mechanism offers a menu of such bundles with associated prices
to implement second-degree price discrimination.

Let $\mathcal{M}_{\mathrm{bin}}$ be the convex closure of $\mathcal{M}_{\mathrm{bin}}^{0}$.
That is, $\mathcal{M}_{\mathrm{bin}}$ includes all convex combinations
of mechanisms in $\mathcal{M}_{\mathrm{bin}}^{0}$. We denote the
marginal-revenue extension of this restricted single-advertiser class
by $\mathrm{IBPA}_{\mathrm{bin}}$.

Despite its simplicity, $\mathcal{M}_{\mathrm{bin}}$ enjoys strong
approximation guarantees for the single-advertiser problem: when valuations
across segments are drawn independently (i.e., from a product distribution),
the optimal mechanism in $\mathcal{M}_{\mathrm{bin}}$ achieves a
constant-factor approximation to the optimal revenue in $\mathcal{\mathcal{M}}$,
and under arbitrary correlations, it achieves an $O(\log T)$ approximation
(see Proposition \ref{prop:deteministic-approx}).

The following propositions show that the theoretical results from
Section \ref{sec:Theoretical-Results} continue to hold for $\mathrm{IBPA}_{\mathrm{bin}}$.
\begin{prop}
For every disclosure partition $\mathcal{P}_{\mathrm{disc}}$ and
any scalar-bid disclosure mechanism $M_{\mathrm{SB}}(\mathcal{P}_{\mathrm{disc}})$,
\[
\mathrm{Rev}^{\mathrm{IBPA}_{\mathrm{bin}}}\ge\mathrm{Rev}^{M_{\mathrm{SB}}}(\mathcal{P}_{\mathrm{disc}}).
\]
\end{prop}

The proof follows Theorem \ref{thm:dominance-scalar-bid} with $\mathcal{M}_{\mathrm{bin}}$
replacing $\mathcal{M}$. For any disclosure partition $\mathcal{P}_{\mathrm{disc}}$,
a scalar-bid disclosure mechanism induces a block-level single-parameter
position auction, whose revenue is bounded by the corresponding marginal-revenue
mechanism. The single-agent mechanisms in this auxiliary problem are
threshold rules, or convex combinations of threshold rules after ironing,
and hence belong to $\mathcal{M}_{\mathrm{bin}}$. The tuple construction
from Theorem \ref{thm:dominance-scalar-bid} embeds these block-level
binary mechanisms into the full-information/null-disclosure environment
without changing utilities, allocation probabilities, or payments.
Thus the $\mathcal{M}_{\mathrm{bin}}$ revenue curves weakly dominate
the scalar-bid block-level curves, implying the result.
\begin{prop}
For any $\mathcal{P}_{\mathrm{info}}^{(1)}\succeq\mathcal{P}_{\mathrm{info}}^{(2)}$
and for all $\mathcal{P}_{\mathrm{disc}}$ such that both pairs $(\mathcal{P}_{\mathrm{info}}^{(1)},\mathcal{P}_{\mathrm{disc}})$
and $(\mathcal{P}_{\mathrm{info}}^{(2)},\mathcal{P}_{\mathrm{disc}})$
are feasible,
\[
\mathrm{Rev}^{\mathrm{IBPA}_{\mathrm{bin}}}(\mathcal{P}_{\mathrm{info}}^{(1)},\mathcal{P}_{\mathrm{disc}})\;\ge\;\mathrm{Rev}^{\mathrm{IBPA}_{\mathrm{bin}}}(\mathcal{P}_{\mathrm{info}}^{(2)},\mathcal{P}_{\mathrm{disc}}).
\]
Similarly, for any $\mathcal{P}_{\mathrm{disc}}^{(1)}\succeq\mathcal{P}_{\mathrm{disc}}^{(2)}$
and for all $\mathcal{P}_{\mathrm{info}}$ such that both pairs $(\mathcal{P}_{\mathrm{info}},\mathcal{P}_{\mathrm{disc}}^{(1)})$
and $(\mathcal{P}_{\mathrm{info}},\mathcal{P}_{\mathrm{disc}}^{(2)})$
are feasible,
\[
\mathrm{Rev}^{\mathrm{IBPA}_{\mathrm{bin}}}(\mathcal{P}_{\mathrm{info}},\mathcal{P}_{\mathrm{disc}}^{(1)})\;\le\;\mathrm{Rev}^{\mathrm{IBPA}_{\mathrm{bin}}}(\mathcal{P}_{\mathrm{info}},\mathcal{P}_{\mathrm{disc}}^{(2)}).
\]
\end{prop}

The proof is the restricted version of Theorem \ref{thm:granularity-monotonicity}.
Finer information weakly enlarges the feasible set because any binary
menu defined on a coarser information partition can be replicated
by assigning each finer block the allocation and payment of its containing
coarse block. Coarser disclosure also weakly enlarges the feasible
set because any binary menu under finer disclosure can be replicated
by offering tuples of fine-block menu options, with payments aggregated
by their probability weights. Both constructions preserve utilities,
payments, and ex-ante allocation probabilities, and both remain within
$\mathcal{M}_{\mathrm{bin}}$. Hence the corresponding binary-menu
revenue curves move monotonically in the stated directions, and the
marginal-revenue extension gives the result.

\subsection{Additive-Pricing Menus}

Our second approximation restricts payments to follow an additive
structure:
\[
\mu(\chi)=\rho_{0}+\sum_{t\in[T]}p_{t}\chi_{t}\rho_{t},
\]
where $\rho_{t}$ is a fixed price for targeting segment $t$ and
$\rho_{0}$ can be interpreted as an entry fee for participating in
the mechanism. The resulting class, $\mathcal{M}_{\mathrm{add}}^{0}\subset\mathcal{M}_{\mathrm{bin}}^{0}$,
retains the binary-allocation structure while imposing a simple, linear
pricing rule. As before, to ensure individual rationality, we normalize
the payment for the zero-allocation vector to be $\mu(\boldsymbol{0})=0$.

This structure is computationally convenient: menus need only specify
a vector $(\rho_{0},\rho_{1},\ldots,\rho_{T})$, and the resulting
mechanism is implementable in time polynomial in $T$ (Proposition
\ref{prop:additive-pricing-complexity}).

Let $\mathcal{M}_{\mathrm{add}}$ be the convex closure of $\mathcal{M}_{\mathrm{add}}^{0}$.
We denote the marginal-revenue extension of this restricted single-advertiser
class by $\mathrm{IBPA_{\mathrm{add}}}$.

Despite its simplicity, $\mathcal{\mathcal{M}_{\mathrm{add}}}$ achieves
the same revenue-approximation guarantees as $\mathcal{M}_{\mathrm{bin}}$
under both independent and correlated valuations (Proposition \ref{prop:deteministic-approx}).
In practice, $\mathcal{\mathcal{M}_{\mathrm{add}}}$ provides a highly
scalable approximation that preserves the essential mixed-bundling
logic of the full IBPA mechanism.
\begin{prop}
\label{prop:deteministic-approx} For the ex-ante constrained single-advertiser
problem, the mechanism class $\mathcal{M}_{\mathrm{bin}}$ of binary-allocation
menus and its subclass $\mathcal{\mathcal{M}_{\mathrm{add}}}$ of
additive-pricing menus each achieve a constant-factor approximation
to the optimal revenue in $\mathcal{M}$ when valuations across segments
are drawn independently (i.e., from a product distribution). More
generally, when valuations may be arbitrarily correlated across segments,
both classes achieve an $O(\log T)$ approximation to the optimal
revenue. 
\end{prop}

\begin{proof}
See Section \ref{subsec:Approx-Mechanisms-Proofs}.
\end{proof}
\begin{prop}
\label{prop:additive-pricing-complexity} For the ex-ante constrained
single-advertiser problem, optimization over the class $\mathcal{\mathcal{M}_{\mathrm{add}}}$
can be carried out, for any approximation tolerance $\epsilon>0$,
in time polynomial in both dimensions $T$ and tolerance $1/\epsilon$.
In particular, each step of an optimization routine can be implemented
in $O(T)$ operations.
\end{prop}

\begin{proof}
See Section \ref{subsec:Approx-Mechanisms-Proofs}.
\end{proof}

\subsection{Proofs}

\label{subsec:Approx-Mechanisms-Proofs}
\begin{proof}[\textbf{Proof of Proposition \ref{prop:deteministic-approx}}]
 Fix a single advertiser $a_{0}$ and the highest available slot
$s_{0}$. Because the multiplicative factor $\alpha_{s_{0}}\gamma_{a_{0}}$
is irrelevant for approximation ratios, we work with the slot-normalized
problem in Equation \eqref{eq:single-adv-slot-normalized}. A single-advertiser
mechanism is a menu of options $\{(\chi^{(k)},\mu^{(k)})\}_{k=1}^{K}$,
where $\chi^{(k)}\in[0,1]^{T}$ specifies, for each segment $t$,
the probability $\chi_{t}^{(k)}$ that the advertiser is served when
segment $t$ realizes, and $\mu^{(k)}$ is the corresponding normalized
expected payment. Let $\xi_{k}$ denote the ex-ante probability (under
the advertiser's prior) that option $k$ is chosen. For any such menu
$M$, define its normalized expected revenue and its ex-ante allocation
probability: 
\[
\mathrm{Rev}(M)\;\equiv\;\sum_{k=1}^{K}\xi_{k}\,\mu^{(k)},\qquad\mathrm{Alloc}(M)\;\equiv\;\sum_{k=1}^{K}\xi_{k}\Big(\sum_{t=1}^{T}p_{t}\,\chi_{t}^{(k)}\Big).
\]
For a mechanism class $\mathcal{C}\subseteq\mathcal{M}$, let 
\[
\mathrm{OPT}_{\mathcal{C}}(q)\;\equiv\;\sup\big\{\mathrm{Rev}(M):M\in\mathcal{C},\ \mathrm{Alloc}(M)\le q\big\}
\]
denote the optimal value at ex-ante cap $q$. Our goal is to show
that for every $q\in[0,1]$, $\mathrm{OPT}_{\mathcal{\mathcal{M}_{\mathrm{bin}}}}(q)$
and $\mathrm{OPT}_{\mathcal{\mathcal{M}_{\mathrm{add}}}}(q)$ approximate
$\mathrm{OPT}_{\mathcal{M}}(q)$ within a constant factor under independent
values and within $O(\log T)$ under arbitrary correlations.

For any mechanism class $\mathcal{C}\subseteq\mathcal{M}$ and any
multiplier $\lambda\ge0$, define the Lagrangian value 
\[
\Pi_{\mathcal{C}}(\lambda)\;\equiv\;\sup_{M\in\mathcal{C}}\big\{\mathrm{Rev}(M)-\lambda\,\mathrm{Alloc}(M)\big\}.
\]
For any $\lambda\ge0$ and any mechanism $M$ feasible at cap $q$
(i.e., $\mathrm{Alloc}(M)\le q$), 
\[
\mathrm{Rev}(M)\;\le\;\mathrm{Rev}(M)-\lambda\mathrm{Alloc}(M)+\lambda q\;\le\;\Pi_{\mathcal{M}}(\lambda)+\lambda q.
\]
Taking the supremum over all $M\in\mathcal{M}$ feasible at $q$ yields
the weak-duality bound 
\begin{equation}
\mathrm{OPT}_{\mathcal{M}}(q)\;\le\;\Pi_{\mathcal{M}}(\lambda)+\lambda q\qquad\forall\lambda\ge0.\label{eq:weak-duality}
\end{equation}

We now show that the single-advertiser problem is isomorphic to a
seller offering multiple items (one per segment $t$) to a single
buyer with additive per-item values. For a realized valuation vector
$v=(v_{1},\dots,v_{T})$, the (normalized) expected utility of option
$k$ is $U^{(k)}(v)=\sum_{t=1}^{T}p_{t}\,\beta_{t}\,v_{t}\,\chi_{t}^{(k)}\;-\;\mu^{(k)}.$
Define the transformed ``item values'' $v'_{t}\equiv p_{t}\beta_{t}v_{t}$
for $t\in[T]$. Then $U^{(k)}(v)=\sum_{t}v'_{t}\chi_{t}^{(k)}-\mu^{(k)}$,
which is exactly the utility of an \emph{additive} buyer in a multi-item
problem who receives (possibly randomized) quantities $\chi_{t}^{(k)}$
of each item $t$ and pays $\mu^{(k)}$. In particular: allowing $\chi^{(k)}\in[0,1]^{T}$
corresponds to allowing randomized bundles (the full class $\mathcal{M}$),
restricting $\chi^{(k)}\in\{0,1\}^{T}$ corresponds to deterministic
bundles (the class $\mathcal{M}_{\mathrm{bin}}$), and imposing additive
pricing $\mu(\chi)=\rho_{0}+\sum_{t}p_{t}\chi_{t}\rho_{t}$ corresponds
to item pricing (with an entry fee) for an additive buyer (the class
$\mathcal{M}_{\mathrm{add}}$). 

Fix $\lambda\ge0$. For any menu option $(\chi,\mu)$, the Lagrangian
objective contributes 
\[
\mu\;-\;\lambda\sum_{t=1}^{T}p_{t}\chi_{t}\;=\;\mu\;-\;\sum_{t=1}^{T}c_{t}\,\chi_{t},\qquad\text{where }c_{t}\equiv\lambda p_{t}.
\]

Thus $\Pi_{\mathcal{M}}(\lambda)$ is exactly the optimal expected
profit in the additive multi-item problem in which serving ``item''
$t$ incurs (expected) cost $c_{t}$ per unit of $\chi_{t}$. Equivalently,
define a shifted payment $\tilde{\mu}\equiv\mu-\sum_{t}c_{t}\chi_{t}$
and shifted item values $\tilde{v}_{t}\equiv v'_{t}-c_{t}$. Then
buyer utility is unchanged: 
\[
\sum_{t}v'_{t}\chi_{t}-\mu\;=\;\sum_{t}\tilde{v}_{t}\chi_{t}-\tilde{\mu},
\]
and seller profit becomes exactly the shifted revenue $\tilde{\mu}$.
Hence maximizing profit with costs $\{c_{t}\}$ is isomorphic to revenue
maximization for an additive buyer with (shifted) values $\{\tilde{v}_{t}\}$;
shifting by constants preserves independence/correlation structure
across items.

Known approximation results for additive single-buyer multi-item mechanism
design therefore apply: \citet{babaioff2020simple} show that when
valuations for items are drawn independently (from a product distribution),
the better of selling items separately (per-item pricing) and selling
the grand bundle achieves a universal constant-factor approximation
to optimal revenue. In particular, the optimal revenue is at most
six times the revenue from the better of these two mechanisms. When
valuations are arbitrarily correlated, \citet{chawla2019buy} show
that selling items separately guarantees an $O(\log T)$ approximation.

In our setting, both item pricing and grand-bundle pricing are contained
in $\mathcal{M}_{\mathrm{add}}$ (as special cases of additive pricing
menus). Separate good pricing corresponds to setting $\rho_{0}=0$
in the class of mechanisms with binary allocations and additive pricings,
and grand-bundle pricing corresponds to setting $\rho_{t}=0\;\forall t$.
Thus, the revenue achievable under binary allocations and additive
pricings inherits the same constant-factor and $O(\log T)$ approximation
guarantees. In particular, there exists a factor $c$ such that, for
every $\lambda\ge0$, 
\begin{equation}
\Pi_{\mathcal{M}}(\lambda)\;\le\;c\;\Pi_{\mathcal{M}_{\mathrm{add}}}(\lambda).\label{eq:lag-approx}
\end{equation}

Because $\mathcal{\mathcal{M}_{\mathrm{add}}}$ (and $\mathcal{\mathcal{M}_{\mathrm{bin}}}$)
is closed under convex combinations, the curve $q\mapsto\mathrm{OPT}_{\mathcal{\mathcal{M}_{\mathrm{add}}}}(q)$
is concave by the same argument as Proposition \eqref{prop:rev-curve-concave}.
Fix any $q\in(0,1)$ and choose any subgradient $\lambda_{q}\in\partial\mathrm{OPT}_{\mathcal{\mathcal{M}_{\mathrm{add}}}}(q)$.
Concavity implies that $q$ maximizes the function $q'\mapsto\mathrm{OPT}_{\mathcal{\mathcal{M}_{\mathrm{add}}}}(q')-\lambda_{q}q'$,
and consequently 
\begin{equation}
\Pi_{\mathcal{\mathcal{M}_{\mathrm{add}}}}(\lambda_{q})\;=\;\mathrm{OPT}_{\mathcal{\mathcal{M}_{\mathrm{add}}}}(q)-\lambda_{q}q.\label{eq:conjugate-tight}
\end{equation}

Combining \eqref{eq:weak-duality} (with $\lambda=\lambda_{q}$),
\eqref{eq:lag-approx}, and \eqref{eq:conjugate-tight} gives 
\[
\mathrm{OPT}_{\mathcal{M}}(q)\;\le\;\Pi_{\mathcal{M}}(\lambda_{q})+\lambda_{q}q\;\le\;c\,\Pi_{\mathcal{\mathcal{M}_{\mathrm{add}}}}(\lambda_{q})+\lambda_{q}q\;\le\;c\big(\Pi_{\mathcal{\mathcal{M}_{\mathrm{add}}}}(\lambda_{q})+\lambda_{q}q\big)\;=\;c\,\mathrm{OPT}_{\mathcal{\mathcal{M}_{\mathrm{add}}}}(q),
\]
where the penultimate inequality uses $c\ge1$. Thus, for every $q\in[0,1]$,
$\mathcal{\mathcal{M}_{\mathrm{add}}}$ achieves a $c$-approximation
to the \emph{ex-ante constrained} optimum in $\mathcal{M}$ (constant
$c$ under independent values and $c=O(\log T)$ under arbitrary correlations).
Since $\mathcal{\mathcal{M}_{\mathrm{add}}}\subseteq\mathcal{\mathcal{M}_{\mathrm{bin}}}$,
the same guarantee holds for $\mathcal{\mathcal{M}_{\mathrm{bin}}}$
as well. 
\end{proof}
$\;$
\begin{proof}[\textbf{Proof of Proposition \ref{prop:additive-pricing-complexity}}]
 Results such as \citet{lee2016optimizing} show that polynomial-time
global optimization is possible for broad classes of non-convex objectives
in both the number of dimensions $T$ and the tolerance parameter
$1/\epsilon$. To establish the stated result, it therefore suffices
to show that a single step of such an optimization routine---namely,
the evaluation of the objective and the constraint---can be carried
out in $O(T)$ operations.

Because the objective and the constraint are expectations over the
advertiser's valuation distribution, we approximate them using sample
averages. Replacing exact expectations with sample averages increases
the number of iterations required to reach a given tolerance level,
but does not affect the complexity of each step with respect to $T$.
Thus, it remains to show that, for any given sample of the advertiser's
valuation vector, the optimal bundle can be identified in $O(T)$
time.

Let $\sigma$ be a binary vector (excluding the all-zero vector) representing
a bundle of segments, where $\sigma_{t}=1$ indicates inclusion of
segment $t$. The advertiser's utility from bundle $\sigma$ is $U(\sigma)=-\rho_{0}+\sum_{t}p_{t}\sigma_{t}(v_{t}-\rho_{t})$.
We determine the optimal bundle by examining each segment $t$ as
follows:

\emph{Segments with large surplus.} If $v_{t}>\rho_{t}+\frac{\rho_{0}}{p_{t}}$
for some segment $t$, i.e., $-\rho_{0}+p_{t}(v_{t}-\rho_{t})>0$,
then the bundle containing only segment $t$ yields positive utility.
In addition, adding $t$ to any other bundle strictly increases utility,
so the optimal bundle must include $t$.

\emph{Segments with negative surplus.} If $v_{t}<\rho_{t}$, then
adding $t$ decreases utility in any bundle. Therefore, the optimal
bundle does not contain such segments.

\emph{Intermediate segments.} If $\rho_{t}<v_{t}<\rho_{t}+\frac{\rho_{0}}{p_{t}}$,
then segment $t$ is not profitable on its own, but if another segment
with large surplus is already included, adding $t$ increases utility.
In this case, the optimal bundle must include both the large-surplus
segment and $t$.

\emph{Residual case.} If no segment satisfies the large-surplus condition,
the only candidate optimal bundle is the set of all segment with $v_{t}>\rho_{t}$.
Denote this bundle by $\sigma$. If $-\rho_{0}+\sum_{t}p_{t}\sigma_{t}(v_{t}-\rho_{t})>0$,
then $\sigma$ is optimal; otherwise, the null bundle is optimal.

This procedure requires only a single pass through the $T$ segments,
and thus runs in $O(T)$ time. Combined with the guarantees of derivative-free
optimization, this establishes that the single-advertiser problem
with binary allocations and additive pricings can be solved in time
polynomial in $T$ and $1/\epsilon$.
\end{proof}

\clearpage{}

\section{Estimating Primitives}

\label{appendix:Primitives}

\subsection{Slot Effects}

Click-through rate (CTR) in the model is multiplicatively separable
into slot, segment, and advertiser effects: 
\[
\text{CTR}_{ast}=\alpha_{s}\beta_{t}\gamma_{a}.
\]
Following \citet{athey2010structural}, the slot effects $\{\alpha_{s}\}$
and advertiser effects $\{\gamma_{a}\}$ are identified from variation
in realized positions across advertisers over time. Because our CTR
data do not break out clicks by segment, we set $\beta_{t}=1$ for
all $t$. This normalization applies uniformly across all counterfactual
mechanisms and therefore preserves the fairness of comparisons.

Let $y_{asd}$ denote the observed CTR for advertiser $a$ in slot
$s$ on day $d$, and let $n_{asd}$ denote the corresponding number
of impressions. We estimate the multiplicative CTR model using Poisson
quasi-maximum likelihood, specifying the conditional mean as
\[
\mathbb{E}[y_{asd}]=\alpha_{s}\gamma_{a}=\exp(\log\alpha_{s}+\log\gamma_{a}).
\]

Because $y_{asd}$ is a rate estimated from $n_{asd}$ impressions,
we weight observations by $n_{asd}$ in the Poisson QMLE. We estimate
the model using slot and advertiser fixed effects, with $\{\log\alpha_{s}\}$
treated as slot fixed effects and $\{\log\gamma_{a}\}$ treated as
advertiser fixed effects. Consistent with the structure of the mechanism,
we impose the monotonicity constraint $\alpha_{1}\ge\alpha_{2}\ge\cdots\ge\alpha_{S}$
during estimation.

The specification achieves a McFadden pseudo-$R^{2}$ of $0.864$,
indicating that the assumptions of multiplicative separability and
monotonicity of slot effects provide a good approximation to the CTR
data. The resulting slot effects $\{\alpha_{s}\}$ are used both in
the estimation of distribution of advertiser valuations below and
in counterfactual mechanism comparisons.

\subsection{Distribution of Advertiser Valuations}

Because GSP is not truth-telling, advertiser bids do not equal valuations.
Nevertheless, \citet{Edelman_et_al_2007} and \citet{Varian_2007}
show that equilibrium bidding behavior in GSP, which they refer to
as\emph{ }envy-free equilibria, imposes sharp inequality constraints
that bound advertiser valuations. Let advertisers be ordered by bid-rank
($b_{a}\times\gamma_{a}$) so that advertiser $a_{s}$ occupies slot
$s$. The envy-free conditions require: 
\[
\gamma_{a_{s}}v_{a_{s}}\;\;\ge\;\;\frac{\gamma_{a_{s+1}}b_{a_{s+1}}\alpha_{s}-\gamma_{a_{s+2}}b_{a_{s+2}}\alpha_{s+1}}{\alpha_{s}-\alpha_{s+1}}\;\;\ge\;\;\gamma_{a_{s+1}}v_{a_{s+1}}.
\]
The middle term defines the \emph{incremental cost per click} (ICC)
of moving from slot $s+1$ to slot $s$: 
\[
ICC_{s,s+1}=\frac{\gamma_{a_{s+1}}b_{a_{s+1}}\alpha_{s}-\gamma_{a_{s+2}}b_{a_{s+2}}\alpha_{s+1}}{\alpha_{s}-\alpha_{s+1}}.
\]

In equilibrium these ICCs must satisfy the monotonicity condition
\[
ICC_{s,s+1}\ge ICC_{s+1,s+2}.
\]
Because observed bids may violate monotonicity, we follow \citet{Varian_2007}
and compute \emph{weighted} ICCs $\{ICC_{s,s+1}^{d^{*}}\}$ that minimize
the weighted deviation from monotonicity: 
\[
ICC_{s,s+1}^{d}=\frac{\gamma_{a_{s+1}}b_{a_{s+1}}\alpha_{s}d_{s}-\gamma_{a_{s+2}}b_{a_{s+2}}\alpha_{s+1}d_{s+1}}{\alpha_{s}d_{s}-\alpha_{s+1}d_{s+1}},
\]
with weights $d$ chosen to minimize $\sum_{s}(1-d_{s}^{2})$ subject
to the ICCs being monotone. The optimal weights $d^{*}$ yield a monotone
sequence of ICCs.

These monotone ICCs deliver valuation bounds: 
\[
\gamma_{a_{s}}v_{a_{s}}\in\big[ICC_{s,s+1}^{d^{*}},\;ICC_{s-1,s}^{d^{*}}\big],\qquad s=2,\ldots,S-1.
\]

For slot $1$, the lower bound is identified; for the upper bound
we set $\overline{v}=2b_{\max}$, where $b_{\max}$ is the largest
bid observed. For slot $S$, the upper bound is identified; the lower
bound requires the highest losing bid, which we do not observe, so
we conservatively set it to zero. 

Thus each advertiser in each auction contributes an interval $(L,U]$
containing the true valuation $v_{at}$. To estimate the valuation
distribution $F_{at}$, we pool the intervals across auctions and
compute the non-parametric maximum likelihood estimator using the
Turnbull estimator \citep{turnbull1976empirical}. Because all auctions
in our sample belong to a single product category, we assume that
advertisers are ex ante symmetric within an segment, and that valuations
are independent and identically distributed draws from a common distribution
$F_{t}$. This symmetry assumption permits pooling interval observations
across advertisers and auctions to recover a single valuation distribution
for each segment. We implement the estimator using the EM algorithm,
which yields a piecewise-uniform CDF with data-driven support.

\begin{figure}[tbph]
\centering{}\includegraphics[scale=0.7]{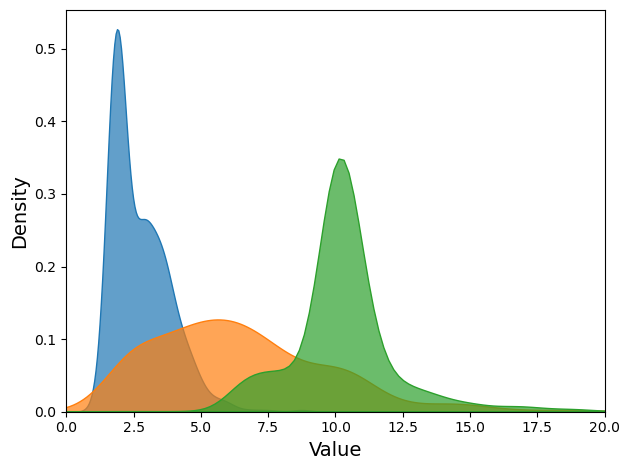}\caption{Distribution of valuations for randomly sampled segments}
\label{fig:value-dists}
\end{figure}

Figure \ref{fig:value-dists} plots estimated valuation densities
for three randomly sampled segments, illustrating substantial heterogeneity
in valuation levels and dispersion. These estimated valuation distributions,
together with the CTR primitives above, are used in the counterfactual
mechanism comparisons in Section \ref{subsec:Comparison-of-Mechanisms}.

\end{document}